\documentclass[12pt]{article}
\usepackage[english]{babel}
\usepackage[utf8]{inputenc}
\usepackage{amssymb,bm,amsthm,amsmath,xcolor,graphicx}
\evensidemargin \oddsidemargin
\newcommand{\CC}{\mathbb{C}}
\newcommand{\RR}{\mathbb{R}}

\newcommand{\NN}{\mathbb{N}}

\newcommand{\Hi}{{\cal H}}

\newcommand{\A}{{\cal A}}

\newcommand{\R}{{\cal R}}

\newcommand{\bL}{{\bf L}}
\newcommand{\bpi}{{\bm\pi}}

\newcommand{\be}{\begin{equation}}
\newcommand{\ee}{\end{equation}}
\newcommand{\bea}{\begin{eqnarray}}
\newcommand{\eea}{\end{eqnarray}}
\newcommand{\ba}{\begin{array}}
\newcommand{\ea}{\end{array}}
\def\nn{\nonumber \\}

\newtheorem{teorema}{Theorem}[section]

\newtheorem{defi}{Definition}[section]
\newtheorem{propo}{Proposition}[section]

\newtheorem{remark}{Remark}

\begin{document}
\title{$O(D)-$equivariant fuzzy hyperspheres}
\date{}

\author{Francesco Pisacane
   \\   \\  %  \and
Dip. di Matematica e applicazioni, Universit\`a di Napoli `Federico II'\\
\& INFN, Sezione di Napoli, \\
Complesso Universitario  M. S. Angelo, Via Cintia, 80126 Napoli, Italy}

\maketitle

\begin{abstract}
\noindent
Fuzzy hyperspheres $S^d_\Lambda$ of dimension $d>2$ are constructed here generalizing the procedure adopted  in \cite{FiorePisacane,{FiorePisacanePOS18}} for $d=1,2$. The starting point is an ordinary quantum particle in $\RR^D$, $D:=d+1$, subject to a rotation invariant potential well $V(r)$ with a very sharp minimum on the sphere of radius $r=1$. The subsequent imposition of a sufficiently low energy cutoff `freezes' the radial excitations, this makes only a finite-dimensional Hilbert subspace $\mathcal{H}_{\Lambda,D}$ accessible and on it the coordinates noncommutative {\it \`a la Snyder}. In addition, the coordinate operators generate the whole algebra of observables $\mathcal{A}_{\Lambda,D}$ which turns out to be realizable through a suitable irreducible vector representation of $U\bm{so}(D+1)$.
% $\mathcal{A}_{\Lambda,D}$. 
This construction is equivariant not only under $SO(D)$, but under the full orthogonal group $O(D)$, and making the cutoff and the depth of the well grow with a natural number $\Lambda$, the result is a sequence $S^d_{\Lambda}$ of fuzzy spheres converging to $S^d$ as $\Lambda\to\infty$ (where one recovers ordinary quantum mechanics on $S^d$).  
\end{abstract}
\section{Introduction}\label{intro}
The aim of this work is to apply the general procedure used in \cite{FiorePisacane,{FiorePisacanePOS18}} to the $D$-dimensional case, when $D>3$; in this way the fuzzy hypersphere constructed is equivariant not only under $SO(D)$, but under the full orhogonal group $O(D)$, obtaining then an $O(D)-$equivariant fuzzy hypersphere. Furthermore, the algebra of observables $\mathcal{A}_{\Lambda,D}$ is realized through a suitable irreducible vector representation of $U\bm{so}\left(D+1\right)$, and then there is the proof of the convergence (in a certain sense) of this new fuzzy hypersphere to ordinary quantum mechanics on the hypersphere $S^{d}$ (here and on $d:=D-1$). 

{
First of all, a fuzzy space is a sequence $\{\mathcal{A}_n\}_{n\in\NN}$
of {\it finite-dimensional} and {\it noncommutative algebras} such that
 $\mathcal{A}_n\overset{n\rightarrow\infty}\longrightarrow\mathcal{A}$, which is the algebra 
of regular functions on an ordinary manifold, with \ dim$(\mathcal{A}_n)\overset{n\rightarrow\infty}\longrightarrow\infty$. 

The first and seminal fuzzy space is the Fuzzy 2-Sphere (FS) of Madore and Hoppe \cite{Madore,HopdeWNic}, where ${\cal A}_n\simeq M_n(\CC)$ is the algebra of complex $n\times n$ matrices and it is generated by coordinate operators $\left\{x_h\right\}_{h=1}^{3}$ fulfilling
\be
[x_h,x_j]=\frac {2i}{\sqrt{n^2\!-\!1}}\varepsilon^{hjk}x_k \qquad\mbox{and}\qquad
%r^2:=
\sum_{h=1}^3 x_hx_h\equiv 1,                     \label{FS}
\ee
with $n\in\mathbb{N}\setminus \{1\}$.

In fact, these operators are obtained by the rescaling 
\be
x_h=\frac{2L_h}{\sqrt{n^2\!-\!1}},\quad h=1,2,3                   \label{rescale}
\ee
of the elements $L_h$ of the standard basis of $\bm{so}(3)$ in the irrep $(\pi_l,V_{l,3})$ characterized by $\bL^2:=\sum_{h=1}^3L_hL_h\equiv l(l\!+\!1) I$,  or equivalently the one of dimension $n=2l\!+\!1$.

The relations (\ref{FS}) are covariant under $SO(3)$, but not under the whole $O(3)$,  
in particular not under parity $x_i\mapsto -x_i$; this is in contrast with the $O(3)$-covariance of both the ordinary sphere $S^2$ [where the right-hand side 
of (\ref{FS})$_1$ is zero] and our $S^2_\Lambda$ [where the right-hand side of (\ref{FS})$_1$ depends on the angular momentum components, as in Snyder \cite{Snyder} commutation relations]; in addition, the coordinate operators $\left\{\overline{x}^i\right\}_{i=1}^D$ of our fuzzy spaces generate also the whole algebra of observables $\mathcal{A}_{\Lambda,D}$, as for the FS.

Moreover, while the Hilbert space 
$V_{l,3}$ of the FS carries an irreducible representation of $SO(3)$,  that ${\cal L}^2(S^d)$
 of a quantum particle on $S^d$ decomposes as the direct sum of {\it all} the irreducible representations of $SO(D)$:
\be
\mathcal{L}^2(S^d)=\bigoplus\limits_{l=0}^\infty V_{l,D}.
\ee
Furthermore, it turns out that the one $\mathcal{H}_{\Lambda,D}$ of  $S^d_\Lambda$ decomposes as  the direct sum \
$\mathcal{H}_{\Lambda,D}=\bigoplus\nolimits_{l=0}^\Lambda V_{l,D}$, \ and therefore also in this aspect $S^d_\Lambda$ better approximates the configuration space $S^d$  in the limit $\Lambda\to\infty$.}

The aforementioned general procedure does not strictly depend on the dimension of the carrier space, but one has to replace all the $2$-dimensional and $3$-dimensional objects by the corresponding $D$-dimensional ones; for instance, the $D$-dimensional spherical harmonics are needed, together with the action on them of the $D$-dimensional angular momentum operator components.

For this reason, let
\be\label{defiLhj}
L_{h,j}:=\frac{1}{i}\left(x_h\frac{\partial}{\partial x_j}-x_j\frac{\partial}{\partial x_h}\right)\quad\mbox{with}\quad h,j\in\{1,2,\cdots,D\}
\ee
be a component of the quantum angular momentum in $\mathbb{R}^D$, and
\begin{equation}\label{C_p}
C_p:=\sum_{1\leq h<j\leq p}L_{h,j}^2\quad \mbox{with }p\in\left\{2,3,\cdots,D\right\}
\end{equation}
be the realization of the quadratic Casimir of $U\bm{so}(p)$; in particular, $C_D\equiv\bm{L}^2$ is the opposite of the Laplace-Beltrami operator $\Delta_{S^{d}}$ on the sphere $S^{d}$. This and the fact that the action of $C_{\widetilde{D}}$ in $S^{\widetilde{D}-1}$ coincides with the one in $S^d$ (see section \ref{picche_fiori}) imply that $C_p$ is the opposite of the Laplace-Beltrami operator $\Delta_{S^{p-1}}$ on the sphere $S^{p-1}$ in every dimension $D\geq p\geq 2$, and its eigenvalues (see \cite{Shubin}, p. 169, theorem 22.1) are 
\be\label{eigenvaluesC_p}
l_{p-1}(l_{p-1}+p-2),\quad \mbox{with}\quad l_1\in\mathbb{Z}\quad\mbox{and}\quad l_{p-1}\in\mathbb{N}_0\quad \forall p>2.
\ee
Following \cite{FiorePisacane,{FiorePisacanePOS18}}, start with a quantum particle in $\mathbb{R}^D$ subject to a confining potential well $V(r)$, which has a very deep minimum in $r=1$ [$\Rightarrow V'(1)=0$, $ V'(1)=:4k_D$, with $k_D\gg 0$]; assume that, when $r\approx1$, it can be approximated with the potential of a one-dimensional harmonic oscillator, in symbols $V(r)\simeq V_0+2k_D(r-1)^2$, where $V_0:=V(1)$ and $k_D$ plays the role of a confining parameter.

This choice of $V(r)$ ensures that, in the limit $k_D\rightarrow+\infty$, the quantum particle is forced to stay on the unit $d$-dimensional sphere $S^d$, and this leads to prove also the convergence of this new fuzzy space to ordinary quantum mechanics on the sphere, in that limit.

Once introduced this $V(r)$, one has to study the eigenvalue equation 
\be\label{eqHpsi=Epsi}
H\bm{\psi}=\left[-\frac{1}{2}\Delta +V(r) \right]\bm{\psi}=E\bm{\psi},
\ee
which is a PDE in the unknowns $\bm{\psi},E$ and its resolution provides a basis of the Hilbert space of quantum states $\mathcal{H}_D$; in addition, from
\be\label{commutHC_p} 
\begin{split}
\left[H,L_{i,j}\right]=0 \quad&\forall 1\leq i<j\leq D,\\
 \left[L_{1,2},C_{p_2}\right]=\left[C_{p_1},C_{p_2}\right]=0,&\quad\forall p_1, p_2\in\left\{2,3,\cdots,D\right\}
\end{split}
\ee
it follows that $H$, $L_{1,2}$ and all these $C_p$ operators can be simultaneously diagonalized in the resolution of (\ref{eqHpsi=Epsi}).

In order to do this, let's look for an eigenfunction $\bm{\psi}$ in the form
\be\label{Ansatz}
\bm{\psi}=f(r)Y(\theta_{d},\theta_{d-1},\cdots,\theta_1),
\ee
where $Y$ is a common eigenfunction of the CSCO (Complete set of commuting observables, i.e. a set of commuting operators whose set of eigenvalues completely specify the state of a system) $L_{1,2}$, $C_2,\cdots, C_d$ and $\bm{L}^2$; while $r,\theta_{d},\theta_{d-1},\cdots,\theta_1$ are polar coordinates. It is obvious that, in order to have $\bm{\psi}\in\mathcal{L}^2\left(\mathbb{R}^D\right)$, it is necessary that $r^d f\in\mathcal{L}^2\left(\mathbb{R}_+\right)$ and $Y\in\mathcal{L}^2\left(S^d\right)$.

The Ansatz (\ref{Ansatz}) transforms the PDE $H\bm{\psi}=E\bm{\psi}$ into an ODE in the unknown $f$, which is solved in section \ref{section_determination_f}; while in section \ref{section_D-dimSA} an orthonormal basis of $\mathcal{L}^2\left(S^d\right)$ of eigenfunctions of $\bm{L}^2$ is determined, in particular I prove that every basis-function $Y$ is uniquely determined by a collection of $d$ indices $\bm{l}:=(l_{d},\cdots,l_2,l_1)$, fulfilling
$$
C_p Y_{\bm{l}}=l_{p-1}(l_{p-1}+p-2) Y_{\bm{l}}\quad,\quad l_{d}\geq \cdots\geq l_2\geq |l_1|\quad\mbox{and}\quad l_i\in\mathbb{Z}\quad\forall i.
$$
Then, it turns out that an orthonormal basis $\mathcal{B}_D$ of the space of quantum states $\mathcal{H}_D$ is (here and later on $l:=l_{d}$)
$$
\mathcal{B}_D=\left\{f_{n,l,D}(r)Y_{\bm{l}}(\theta_{d},\theta_{d-1},\cdots,\theta_1)\vert n\in\mathbb{N}_0,l\geq l_{d-1}\geq \cdots\geq l_2\geq |l_1|,l_i\in\mathbb{Z} \forall i\right\}.
$$
Furthermore, the consequence of the imposition of a sufficiently low energy cutoff $E\leq\overline{E}$ (see section \ref{sec_cutoff}) is that the Hilbert space of `admitted' states $\mathcal{H}_{\overline{E},D}\subset\mathcal{H}_D$ becomes finite-dimensional and spanned by all the $H$-eigenstates having eigenvalues $E\leq\overline{E}$. I also replace every observable $A$ by the corresponding projected one $\overline{A}:=P_{\overline{E},D}AP_{\overline{E},D}$ (here and later on $P_{\overline{E},D}$ is the projection on $\mathcal{H}_{\overline{E},D}$) and I give to $\overline{A}$ the same physical interpretation; in this way I have only states and operators that are `physical'. 

The condition $\overline{E}<2\sqrt{2k_D}$ implies that the Hamiltonian operator $H$ can be seen, in a first approximation, as the square angular momentum operator $\bm{L}^2$ (in other words radial excitations are `frozen'), while two crucial steps, necessary to obtain a fuzzy space, are the choice of a $\Lambda$-dependent energy cutoff $\overline{E}:=\overline{E}\left(\Lambda\right)$ so that $\overline{E}\left(\Lambda\right)$ diverges with $\Lambda\in\mathbb{N}$, and the assumption that also $k_D$ depends on $\Lambda$ in a way such that $\overline{E}\left(\Lambda\right)<2\sqrt{2k_D\left(\Lambda\right)}$. This implies that the Hilbert space of admitted states can be definitively re-labeled as $\mathcal{H}_{\Lambda,D}$, and the corresponding algebra of observables $End\left(\mathcal{H}_{\Lambda,D}\right)$ as $\mathcal{A}_{\Lambda,D}$; then the sequence $\left\{\mathcal{A}_{\Lambda,D}\right\}_{\Lambda\in\mathbb{N}}$ is made of finite-dimensional algebras, which become infinite dimensional in the limit $\Lambda\rightarrow+\infty$.

In order to calculate the algebraic relations between the generators of $\mathcal{A}_{\Lambda,D}$ I need to determine the action of every $\overline{L}_{h,j}\equiv L_{h,j}$ and $\overline{x}_h$ on a basis of $\mathcal{H}_{\Lambda,D}$. Because of (\ref{Ansatz}), it is possible to use the knowledge of the action of $L_{h,j}$ on the spherical harmonics $Y_{\bm{l}}$ obtained from the above CSCO, to recover the one on $\bm{\psi}_{\bm{l},D}$; since I have not found the action of this in the literature when $D>3$, I have explicitely calculated it in \ref{Section_sphharm}, while in section \ref{Section_Algebra} I compute the action of coordinate operators $\overline{x}_h$.

As in \cite{FiorePisacane,{FiorePisacanePOS18}}, in section \ref{commrel+R^2} it is shown that $\overline{x}_h,\overline{x}_j$ fulfill \emph{Snyder} commutation relations, in other words their commutator is proportional to the component $L_{h,j}$ of the $D$-dimensional angular momentum operator, up to a scalar operator depending on $\bm{L}^2$. Then there is a list of all the relations involving the projectors of $\mathcal{A}_{\Lambda,D}$ which show that the $\overline{x}$s generate the whole algebra of observables, for instance every component of the angular momentum operator can be written as a ordered polynomial in the $\overline{x}_{h}$. The square distance from the origin operator $\bm{x}^2:=\sum_h\overline{x}_h\overline{x}_h$ is not identically $1$, but a function of $\bm{L}^2$ such that nevertheless its spectrum is very close to $1$ and collapses to $1$ in the $k_D\to\infty$ limit. 

Furthermore, in section \ref{realso(D+1)} some tools of Lie algebra theory are used in order to realize the algebra of observables $\mathcal{A}_{\Lambda,D}$ through a suitable irreducible vector representation $\pi_{\Lambda,D+1}$ of $U\bm{so}\left(D+1\right)$; this is suggested by the fact that the dimension of $\mathcal{H}_{\Lambda,D}$ coincides with the one of the representation space $V_{\Lambda,D+1}$ of $\pi_{\Lambda,D+1}$; then (up to isomorphisms)
$$
\mathcal{H}_{\Lambda,D}=\bigoplus_{l=0}^{\Lambda}V_{l,D}=V_{\Lambda,D+1}.
$$
That realization is $O(D)$-equivariant and the algebra isomorphism $\Phi$ fulfills
\be
\left[\Phi\left(A\right)\right]^{\dag}=\Phi\left(A^{\dag}\right).\label{adjoint}
\ee

The proof of the aforementioned convergence is a sort of certification of the goodness of this approximation of quantum mechanics on the sphere $S^d$, and this job is done in section \ref{converge}; this is inspired by the behavior of the potential $V(r)$ in the limit $k_D\rightarrow+\infty$, where it forces the quantum particle to stay on the unit $d$-dimensional sphere $S^d$. The `projected' spherical harmonics are firstly identified as a basis of a space of all spherical harmonics, $\mathcal{A}_{\Lambda,D}$ as a subalgebra of $\mathcal{B}\left(S^{d}\right)$, the algebra of bounded functions on $S^d$ [or $\mathcal{C}\left(S^{d}\right)$, the algebra of polynomial functions on $S^d$], and then I prove the convergence (in a certain sense) of the operators in $\mathcal{A}_{\Lambda,D}$ to the corresponding ones in $\mathcal{B}\left(S^{d}\right)$ [$\mathcal{C}\left(S^{d}\right)$, respectively]; furthermore, I use a $k_{D}(\Lambda)$ growing faster with $\Lambda\in\mathbb{N}$ to prove this result.

Section \ref{conclu} contains final remarks, conclusions and a comparison with literature; in {section \ref{appendix}} (appendix) there are lengthy computations and proofs.

{The construction of this work is valid for any $D$, but an explicit exhibition of the resulting $S_{\Lambda}^3$ and $S_{\Lambda}^4$  (see next section) may help to better understand how to move from the cases $D=1,2$ of \cite{FiorePisacane,{FiorePisacanePOS18}} to the dimensions $D>3$, and also to see analogies and differences with other fuzzy $3$- and $4$-spheres.}
\section{Two particolar cases: $D=4$ and $D=5$}
\subsection{$S_{\Lambda}^3$} 
When $D=4$ the choices $\overline{E}=\overline{E}(\Lambda):=\Lambda(\Lambda+2)$ and $k_4=k_4(\Lambda)\geq \left[\Lambda(\Lambda+2)\right]^2$ imply that the Hilbert space of admitted states $\mathcal{H}_{\Lambda,4}$ is generated by all the functions (see sections \ref{section_determination_f} and  \ref{section_D-dimSA})
$$
\bm{\psi}_{l,l_2,l_1,4}=\bm{\psi}_{l,l_2,l_1,4}(r,\theta_1,\theta_2,\theta_3):=f_{l,4}(r)Y_{l,l_2,l_1}(\theta_1,\theta_2,\theta_3),\quad l\leq \Lambda
$$
hence
$$
\dim \mathcal{H}_{\Lambda,4}\overset{(\ref{dimHLD})}={{\Lambda+2}\choose{\Lambda-1}}\frac{2\Lambda+3}{\Lambda}=\frac{2\Lambda^3+9\Lambda^2+13\Lambda+6}{6}=\frac{1}{3}(\Lambda+1)(\Lambda+2)\left(\Lambda+\frac{3}{2}\right).
$$
The angular momentum components are $\{L_{h,j}:1\leq h<j\leq 4\}$, $L_{\pm,4}:=L_{2,4}\mp iL_{1,4}$ and they act as follows (see definition \ref{defiL} in section \ref{Section_sphharm}): 
\begin{equation*}
\begin{split}
L_{1,2} \bm{\psi}_{l,l_2,l_1}=&l_1 \bm{\psi}_{l,l_2,l_1},\\
L_{1,3}\bm{\psi}_{l,l_2,l_1}=\frac{1}{i}&\left[\frac{d_{l_2,l_1,3}}{2}\bm{\psi}_{l,l_2,l_1-1}-\frac{d_{l_2,l_1+1,3}}{2}\bm{\psi}_{l,l_2,l_1+1} \right],\\
L_{2,3}\bm{\psi}_{l,l_2,l_1}=\frac{1}{i}&\left[-\frac{d_{l_2,l_1,3}}{2i}\bm{\psi}_{l,l_2,l_1-1}-\frac{d_{l_2,l_1+1,3}}{2i}\bm{\psi}_{l,l_2,l_1+1} \right],\\
L_{1,4}\bm{\psi}_{l,l_2,l_1}=\frac{1}{i}&\left[\frac{d_{l,l_2,4}B(l_2,l_1,2)}{2}\bm{\psi}_{l,l_2-1,l_1+1} + \frac{d_{l,l_2,4}D(l_2,l_1,2)}{2}\bm{\psi}_{l,l_2-1,l_1-1}\right.\\
&\left. -\frac{d_{l,l_2+1,4}A(l_2,l_1,2)}{2}\bm{\psi}_{l,l_2+1,l_1+1} - \frac{d_{l,l_2+1,4}C(l_2,l_1,2)}{2}\bm{\psi}_{l,l_2+1,l_1-1}  \right],\\
L_{2,4}\bm{\psi}_{l,l_2,l_1}=\frac{1}{i}&\left[\frac{d_{l,l_2,4}B(l_2,l_1,2)}{2i}\bm{\psi}_{l,l_2-1,l_1+1} - \frac{d_{l,l_2,4}D(l_2,l_1,2)}{2i}\bm{\psi}_{l,l_2-1,l_1-1}\right.\\
&\left. -\frac{d_{l,l_2+1,4}A(l_2,l_1,2)}{2i}\bm{\psi}_{l,l_2+1,l_1+1} + \frac{d_{l,l_2+1,4}C(l_2,l_1,2)}{2i}\bm{\psi}_{l,l_2+1,l_1-1}  \right],
\end{split}
\end{equation*}
\begin{equation*}
\begin{split}
L_{3,4}\bm{\psi}_{l,l_2,l_1}=\frac{1}{i}&\left[d_{l,l_2,4}G(l_2,l_1,2)\bm{\psi}_{l,l_2-1,l_1} - d_{l,l_2+1,4}F(l_2,l_1,2)\bm{\psi}_{l,l_2+1,l_1}\right],\\
L_{+,4}\bm{\psi}_{l,l_2,l_1}=&-d_{l,l_2,4}B(l_2,l_1,2)\bm{\psi}_{l,l_2-1,l_1+1} +d_{l,l_2+1,4}A(l_2,l_1,2)\bm{\psi}_{l,l_2+1,l_1+1},\\
L_{-,4}\bm{\psi}_{l,l_2,l_1}=&d_{l,l_2,4}D(l_2,l_1,2)\bm{\psi}_{l,l_2-1,l_1-1}-d_{l,l_2+1,4}C(l_2,l_1,2)\bm{\psi}_{l,l_2+1,l_1-1},
\end{split}
\end{equation*}
where $d_{l,l_2,4}=\sqrt{(l-l_2+1)(l+l_2+1)}$.

They fulfill
$$
\left[L_{h,j},L_{p,s}\right]\bm{\psi}_{l,l_2,l_1}=i\left(\delta_{h,p}L_{j,s}+\delta_{j,s}L_{h,p}-\delta_{h,s}L_{j,p}-\delta_{j,p}L_{h,s}\right)\bm{\psi}_{l,l_2,l_1},
$$
$$
\bm{L}^2 \bm{\psi}_{l,l_2,l_1}=l(l+2)\bm{\psi}_{l,l_2,l_1},\quad C_3\psi_{l,l_2,l_1}=l_2(l_2+1)\bm{\psi}_{l,l_2,l_1}\quad\mbox{and}\quad C_2\psi_{l,l_2,l_1}=l_1^2\psi_{l,l_2,l_1}.
$$
The coordinate operators are $\overline{x}_1,\overline{x}_2,\overline{x}_3,\overline{x}_4$ $\overline{x}_{\pm}:=\overline{x}_1\pm i\overline{x}_2$, and they act on $\mathcal{H}_{\Lambda,4}$ as follows (see section \ref{Section_Algebra}):
\begin{equation*}
\begin{split}
\overline{x}_1 \bm{\psi}_{l,l_2,l_1}=&\left[\frac{c_{l,4}D(l,l_2,3) D(l_2,l_1,2)}{2}\bm{\psi}_{l-1,l_2-1,l_1-1} +\frac{c_{l,4}D(l,l_2,3) B(l_2,l_1,2)}{2}\bm{\psi}_{l-1,l_2-1,l_1+1}\right.\\
&+\frac{c_{l,4}B(l,l_2,3) C(l_2,l_1,2)}{2}\bm{\psi}_{l-1,l_2+1,l_1-1} +\frac{c_{l,4}B(l,l_2,3) A(l_2,l_1,2)}{2}\bm{\psi}_{l-1,l_2+1,l_1+1} \\
&+\frac{c_{l+1,4}C(l,l_2,3) D(l_2,l_1,2)}{2}\bm{\psi}_{l+1,l_2-1,l_1-1} +\frac{c_{l+1,4}C(l,l_2,3) B(l_2,l_1,2)}{2}\bm{\psi}_{l+1,l_2-1,l_1+1} \\
&\left. +\frac{c_{l+1,4}A(l,l_2,3) C(l_2,l_1,2)}{2}\bm{\psi}_{l+1,l_2+1,l_1-1} +\frac{c_{l+1,4}A(l,l_2,3) A(l_2,l_1,2)}{2}\bm{\psi}_{l+1,l_2+1,l_1+1}  \right],\\
\overline{x}_2 \bm{\psi}_{l,l_2,l_1}=&\left[\frac{c_{l,4}D(l,l_2,3) D(l_2,l_1,2)}{2i}\bm{\psi}_{l-1,l_2-1,l_1-1} -\frac{c_{l,4}D(l,l_2,3) B(l_2,l_1,2)}{2i}\bm{\psi}_{l-1,l_2-1,l_1+1}\right.\\
&+\frac{c_{l,4}B(l,l_2,3) C(l_2,l_1,2)}{2i}\bm{\psi}_{l-1,l_2+1,l_1-1} -\frac{c_{l,4}B(l,l_2,3) A(l_2,l_1,2)}{2i}\bm{\psi}_{l-1,l_2+1,l_1+1} \\
&+\frac{c_{l+1,4}C(l,l_2,3) D(l_2,l_1,2)}{2i}\bm{\psi}_{l+1,l_2-1,l_1-1} -\frac{c_{l+1,4}C(l,l_2,3) B(l_2,l_1,2)}{2i}\bm{\psi}_{l+1,l_2-1,l_1+1} \\
&\left. +\frac{c_{l+1,4}A(l,l_2,3) C(l_2,l_1,2)}{2i}\bm{\psi}_{l+1,l_2+1,l_1-1} -\frac{c_{l+1,4}A(l,l_2,3) A(l_2,l_1,2)}{2i}\bm{\psi}_{l+1,l_2+1,l_1+1}  \right],\\
\overline{x}_3 \bm{\psi}_{l,l_2,l_1}=&\left[c_{l,4}D(l,l_2,3) G(l_2,l_1,2)\bm{\psi}_{l-1,l_2-1,l_1} +c_{l,4}B(l,l_2,3) F(l_2,l_1,2)\bm{\psi}_{l-1,l_2+1,l_1}\right.\\
&\left. +c_{l+1,4}C(l,l_2,3) G(l_2,l_1,2)\bm{\psi}_{l+1,l_2-1,l_1} +c_{l+1,4}A(l,l_2,3) F(l_2,l_1,2)\bm{\psi}_{l+1,l_2+1,l_1}  \right],\\
\overline{x}_4 \bm{\psi}_{l,l_2,l_1}=&c_{l,4}G(l,l_2,3)\bm{\psi}_{l-1,l_2,l_1} + c_{l+1,4}F(l,l_2,3)\bm{\psi}_{l+1,l_2,l_1},\\
\overline{x}_+ \bm{\psi}_{l,l_2,l_1}=&\left[c_{l,4}D(l,l_2,3) B(l_2,l_1,2)\bm{\psi}_{l-1,l_2-1,l_1+1}+c_{l,4}B(l,l_2,3) A(l_2,l_1,2)\bm{\psi}_{l-1,l_2+1,l_1+1}\right.\\
&\left.+c_{l+1,4}C(l,l_2,3) B(l_2,l_1,2)\bm{\psi}_{l+1,l_2-1,l_1+1}+c_{l+1,4}A(l,l_2,3) A(l_2,l_1,2)\bm{\psi}_{l+1,l_2+1,l_1+1} \right],\\
\overline{x}_- \bm{\psi}_{l,l_2,l_1}=&\left[c_{l,4}D(l,l_2,3) D(l_2,l_1,2)\bm{\psi}_{l-1,l_2-1,l_1-1} +c_{l,4}B(l,l_2,3) C(l_2,l_1,2)\bm{\psi}_{l-1,l_2+1,l_1-1} \right.\\
&\left.+c_{l+1,4}C(l,l_2,3) D(l_2,l_1,2)\bm{\psi}_{l+1,l_2-1,l_1-1} +c_{l+1,4}A(l,l_2,3) C(l_2,l_1,2)\bm{\psi}_{l+1,l_2+1,l_1-1}\right],
\end{split}
\end{equation*}
where
$$
c_{l,4}\overset{(\ref{valueclD})}=\begin{cases}
\sqrt{1+\frac{l^2+l+\frac{1}{4}}{k_4}}\quad\mbox{if } 1\leq l\leq \Lambda ,\\
\quad\quad\quad 0\quad\quad \quad\mbox{otherwise},
\end{cases}
$$
and, according to (\ref{ABCDFG}),
\begin{equation*}
\begin{split}
A(l,l_2,3)&=\sqrt{\frac{(l+l_2+2)(l+l_2+3)}{(2l+2)(2l+4)}},\\
B(l,l_2,3)&=-\sqrt{\frac{(l-l_2-1)(l-l_2)}{(2l+2)(2l)}},\\
C(l,l_2,3)&=-\sqrt{\frac{(l-l_2+2)(l-l_2+1)}{(2l+2)(2l+4)}},
\end{split}
\end{equation*}
\begin{equation*}
\begin{split}
D(l,l_2,3)&=\sqrt{\frac{(l+l_2+1)(l+l_2)}{(2l+2)(2l)}},\\
F(l,l_2,3)&=\sqrt{\frac{(l+l_2+2)(l-l_2+1)}{(2l+2)(2l+4)}},\\
G(l,l_2,3)&=\sqrt{\frac{(l-l_2)(l+l_2+1)}{(2l+2)(2l)}}.
\end{split}
\end{equation*}
They fulfill (see section \ref{commrel+R^2})
$$
\left[\overline{x}_h,\overline{x}_j\right]=\left[-\frac{I}{k_4}+\left(\frac{1}{k_4}+\frac{\left(c_{\Lambda,4}
\right)^2}{2\Lambda+2} \right)\widehat{P}_{\Lambda,4}
\right]
\overline{L}_{h,j},\quad\left[L_{h,s},\overline{x}_j\right]=\frac{1}{i}\left(\delta_j^s \overline{x}_h -\delta_j^h \overline{x}_s\right),
$$
\begin{equation*}
\begin{split}
\bm{x}^2:=\sum_{h=1}^4\overline{x}_h\overline{x}_h=\left\{1+\frac{4\bm{L}^2+9}{4k_4(\Lambda)}-\left[\left(1+\frac{4\Lambda^2+12\Lambda+9}{4k_4(\Lambda)} \right)\frac{\Lambda+2}{2\Lambda+2} \right]\widehat{P}_{\Lambda,4} \right\}
\end{split}
\end{equation*}
and
\begin{equation}
\begin{split}
\prod_{l=0}^{\Lambda}\left[\bm{L}^2-l(l+2)I\right]=0\quad,&\quad \prod_{l_{2}=0}^{l}\left[{C}_{3}-l_{2}(l_{2}+1)I\right]\widetilde{P}_{1,l}=0,\\
 \prod_{l_{1}=-l_2}^{l_2}\left[{L}_{1,2}-l_{1}I\right]\widetilde{P}_{2,l_2}=0,&\quad \left(\overline{x}_{\pm}\right)^{2\Lambda+1}=0,\mbox{ and }\left(L_{\nu,\pm}\right)^{2\Lambda+1}=0,\forall \nu\geq 3,
\end{split}
\end{equation}
where $\widetilde{P}_{h,j}$ is the projector on the eigenspace of ${C}_{4-h}$ corresponding to $l_{4-h}\equiv j$.

According to this, the algebra of observables is generated by the coordinate operators, in fact every projector can be written as a ordered polynomial in the $\overline{x}_{\nu}$.

Furthermore, the $SO(5)$-irrep $\pi_{\Lambda,5}$, the one characterized by $C_5\equiv \Lambda(\Lambda+3)I$ with representation space
$$
V_{\Lambda,5}:=span\left\{Y_{\Lambda,l,l_2,l_1}(\theta_4,\theta_3,\theta_2,\theta_1):\Lambda\geq l\geq l_2\geq |l_1|, l_i\in\mathbb{Z}\forall i \right\},
$$
can be used to identify $\bm{\psi}_{l,l_2,l_1}\equiv Y_{\Lambda,l,l_2,l_1}$, and also the operators
\begin{equation}
\begin{split}
L_{h,j}\equiv L_{h,j}\quad\mbox{for}\quad 1\leq h<j\leq 4\quad\mbox{and}\quad\overline{x}_s\equiv p_4(\lambda)L_{s,5}p_{4}(\lambda),
\end{split}
\end{equation}
where
$$
\lambda:=\frac{-2+\sqrt{4+4\bm{L}^2}}{2}=\sqrt{1+\bm{L}^2}-1,
$$
while $p_{4}(\lambda)$ is an analytic function and the values $p_{4}(l)$, when $l\in\mathbb{N}_0$, can be obtained recursively from (\ref{condition_p}) starting from $p_{4}(0):=1$.

Furthermore, in order to prove the convergence of $S_{\Lambda}^3$ to ordinary quantum mechanics on $S^3$, it is convenient to identify $\bm{\psi}_{l,l_2,l_1}\equiv Y_{l,l_2,l_1}$ and then to consider their fuzzy counterparts $\widehat{Y}_{l,l_2,l_1}$ [see (\ref{fuzzyY})], which can be used to approximate a generic $f\in B(S^3)$ or $f\in C(S^3)$; this is possible because the $Y_{l,l_2,l_1}$ are an orthonormal basis of $\mathcal{L}^2(S^3)$, and also homogeneous polynomials in the $t_h:=x_h/r$ variables. Then,
$$
\widehat{f}_{\Lambda}:= \sum_{l=0}^{2\Lambda}
\sum_{l_2=0 }^{l}\sum_{l_1=-l_2}^{l_2}
f_{l,l_2,l_1}\widehat{Y}_{l,l_2,l_1},\quad\mbox{where}\quad f_{l,l_2,l_1}:=\left\langle Y_{l,l_2,l_1},f\right\rangle,
$$
is an approximation of $f$ because of the following two theorems (see section \ref{converge})
\begin{teorema}
Every projected coordinate operator $\overline{x}_h$ converges strongly to the corresponding $t_{h}$ as $\Lambda\to\infty$ if
\begin{equation*}
\begin{split}
k_4\left(\Lambda\right)&\geq \Lambda \frac{1}{9}(\Lambda+1)^2(\Lambda+2)^2\left(\Lambda+\frac{3}{2}\right)^2 \frac{4\Lambda(\Lambda+2)+3}{4}\\
&\Lambda\frac{1}{9}(\Lambda+1)^2(\Lambda+2)^2\left(\Lambda+\frac{3}{2}\right)^3 \left(\Lambda+\frac{1}{2}\right).
\end{split}
\end{equation*}
\end{teorema}
\begin{teorema} 
Choosing $k_4\left(\Lambda\right)$ fulfilling (\ref{kineq2}) for $D=4$, then for all $f,g\in B(S^3)$ the following strong limits as $\Lambda\rightarrow \infty$ hold: $\hat{f}_{\Lambda}\rightarrow f\cdot,\widehat{\left(fg\right)}_{\Lambda}\rightarrow fg\cdot$ and $\hat{f}_{\Lambda}\hat{g}_{\Lambda}\rightarrow fg\cdot$. 
\end{teorema}
\subsection{$S_{\Lambda}^4$}
When $D=5$ the choices $\overline{E}=\overline{E}(\Lambda):=\Lambda(\Lambda+3)$ and $k_5=k_5(\Lambda)\geq \left[\Lambda(\Lambda+3)\right]^2$ imply that the Hilbert space of admitted states $\mathcal{H}_{\Lambda,5}$ is generated by all the functions
$$
\bm{\psi}_{l,l_3,l_2,l_1,5}=\bm{\psi}_{l,l_3,l_2,l_1,4}(r,\theta_1,\theta_2,\theta_3,\theta_4):=f_{l,5}(r)Y_{l,l_3,l_2,l_1}(\theta_1,\theta_2,\theta_3,\theta_4),\quad l\leq \Lambda
$$
hence
$$
\dim \mathcal{H}_{\Lambda,5}= {{\Lambda+3}\choose{\Lambda-1}}\frac{2\Lambda+4}{\Lambda}=\frac{1}{12}(\Lambda+1)(\Lambda+2)^2(\Lambda+3).
$$
The angular momentum components are $\{L_{h,j}:1\leq h<j\leq 5\}$, $L_{\pm ,5}:=L_{2,5}\mp i L_{1,5}$ and they act as follows: 
\begin{equation*}
\begin{split}
L_{1,2} \bm{\psi}_{l,l_3,l_2,l_1}=&l_1 \bm{\psi}_{l,l_3,l_2,l_1},\\
L_{1,3}\bm{\psi}_{l,l_3,l_2,l_1}=\frac{1}{i}&\left[\frac{d_{l_2,l_1,3}}{2}\bm{\psi}_{l,l_3,l_2,l_1-1}-\frac{d_{l_2,l_1+1,3}}{2}\bm{\psi}_{l,l_3,l_2,l_1+1} \right],\\
L_{2,3}\bm{\psi}_{l,l_3,l_2,l_1}=\frac{1}{i}&\left[-\frac{d_{l_2,l_1,3}}{2i}\bm{\psi}_{l,l_3,l_2,l_1-1}-\frac{d_{l_2,l_1+1,3}}{2i}\bm{\psi}_{l,l_3,l_2,l_1+1} \right],\\
L_{1,4}\bm{\psi}_{l,l_3,l_2,l_1}=\frac{1}{i}&\left[\frac{d_{l_3,l_2,4}B(l_2,l_1,2)}{2}\bm{\psi}_{l,l_3,l_2-1,l_1+1} + \frac{d_{l_3,l_2,4}D(l_2,l_1,2)}{2}\bm{\psi}_{l,l_3,l_2-1,l_1-1}\right.\\
&\left. -\frac{d_{l_3,l_2+1,4}A(l_2,l_1,2)}{2}\bm{\psi}_{l,l_3,l_2+1,l_1+1} - \frac{d_{l_3,l_2+1,4}C(l_2,l_1,2)}{2}\bm{\psi}_{l,l_3,l_2+1,l_1-1}  \right],\\
L_{2,4}\bm{\psi}_{l,l_3,l_2,l_1}=\frac{1}{i}&\left[\frac{d_{l_3,l_2,4}B(l_2,l_1,2)}{2i}\bm{\psi}_{l,l_3,l_2-1,l_1+1} - \frac{d_{l_3,l_2,4}D(l_2,l_1,2)}{2i}\bm{\psi}_{l,l_3,l_2-1,l_1-1}\right.\\
&\left. -\frac{d_{l_3,l_2+1,4}A(l_2,l_1,2)}{2i}\bm{\psi}_{l,l_3,l_2+1,l_1+1} + \frac{d_{l_3,l_2+1,4}C(l_2,l_1,2)}{2i}\bm{\psi}_{l,l_3,l_2+1,l_1-1}  \right],\\
L_{3,4}\bm{\psi}_{l,l_3,l_2,l_1}=\frac{1}{i}&\left[d_{l_3,l_2,4}G(l_2,l_1,2)\bm{\psi}_{l,l_3,l_2-1,l_1} - d_{l_3,l_2+1,4}F(l_2,l_1,2)\bm{\psi}_{l,l_3,l_2+1,l_1}\right],\\
\end{split}
\end{equation*}
\begin{equation*}
\begin{split}
L_{1,5}\bm{\psi}_{l,l_3,l_2,l_1}=\frac{1}{i}&\left[\frac{d_{l,l_3,5}D(l_3,l_2,3) D(l_2,l_1,2)}{2}\bm{\psi}_{l,l_3-1,l_2-1,l_1-1}\right.\\
&+\frac{d_{l,l_3,5}D(l_3,l_2,3) B(l_2,l_1,2)}{2}\bm{\psi}_{l,l_3-1,l_2-1,l_1+1}\\
&+\frac{d_{l,l_3,5}B(l_3,l_2,3) C(l_2,l_1,2)}{2}\bm{\psi}_{l,l_3-1,l_2+1,l_1-1}\\
&+\frac{d_{l,l_3,5}B(l_3,l_2,3) A(l_2,l_1,2)}{2}\bm{\psi}_{l,l_3-1,l_2+1,l_1+1} \\
&-\frac{d_{l,l_3+1,5}C(l_3,l_2,3) D(l_2,l_1,2)}{2}\bm{\psi}_{l,l_3+1,l_2-1,l_1-1}\\
&-\frac{d_{l,l_3+1,5}C(l_3,l_2,3) B(l_2,l_1,2)}{2}\bm{\psi}_{l,l_3+1,l_2-1,l_1+1} \\
&-\frac{d_{l,l_3+1,5}A(l_3,l_2,3) C(l_2,l_1,2)}{2}\bm{\psi}_{l,l_3+1,l_2+1,l_1-1}\\
&\left. -\frac{d_{l,l_3+1,5}A(l_3,l_2,3) A(l_2,l_1,2)}{2}\bm{\psi}_{l,l_3+1,l_2+1,l_1+1}  \right],
\end{split}
\end{equation*}
\begin{equation*}
\begin{split}
L_{2,5} \bm{\psi}_{l,l_3,l_2,l_1}=\frac{1}{i}&\left[\frac{d_{l,l_3,5}D(l_3,l_2,3) D(l_2,l_1,2)}{2i}\bm{\psi}_{l,l_3-1,l_2-1,l_1-1}\right.\\
& -\frac{d_{l,l_3,5}D(l_3,l_2,3) B(l_2,l_1,2)}{2i}\bm{\psi}_{l,l_3-1,l_2-1,l_1+1}\\
&+\frac{d_{l,l_3,5}B(l_3,l_2,3) C(l_2,l_1,2)}{2i}\bm{\psi}_{l,l_3-1,l_2+1,l_1-1}\\
&-\frac{d_{l,l_3,5}B(l_3,l_2,3) A(l_2,l_1,2)}{2i}\bm{\psi}_{l,l_3-1,l_2+1,l_1+1} \\
&-\frac{d_{l,l_3+1,5}C(l_3,l_2,3) D(l_2,l_1,2)}{2i}\bm{\psi}_{l,l_3+1,l_2-1,l_1-1}\\
& +\frac{d_{l,l_3+1,5}C(l_3,l_2,3) B(l_2,l_1,2)}{2i}\bm{\psi}_{l,l_3+1,l_2-1,l_1+1} \\
& -\frac{d_{l,l_3+1,5}A(l_3,l_2,3) C(l_2,l_1,2)}{2i}\bm{\psi}_{l,l_3+1,l_2+1,l_1-1}\\
&\left. +\frac{d_{l,l_3+1,5}A(l_3,l_2,3) A(l_2,l_1,2)}{2i}\bm{\psi}_{l,l_3+1,l_2+1,l_1+1}  \right],
\end{split}
\end{equation*}
\begin{equation*}
\begin{split}
L_{3,5} \bm{\psi}_{l,l_3,l_2,l_1}=\frac{1}{i}&\left[d_{l,l_3,5}D(l_3,l_2,3) G(l_2,l_1,2)\bm{\psi}_{l,l_3-1,l_2-1,l_1}\right. \\
&+d_{l,l_3,5}B(l_3,l_2,3) F(l_2,l_1,2)\bm{\psi}_{l,l_3-1,l_2+1,l_1}\\
& -d_{l,l_3+1,5}C(l_3,l_2,3) G(l_2,l_1,2)\bm{\psi}_{l,l_3+1,l_2-1,l_1} \\
&\left.-d_{l,l_3+1,5}B(l_3,l_2,3) A(l_2,l_1,2)\bm{\psi}_{l,l_3+1,l_2+1,l_1}  \right],\\
L_{4,5} \bm{\psi}_{l,l_3l_2,l_1}=\frac{1}{i}&\left[d_{l,l_3,5}G(l_3,l_2,3)\bm{\psi}_{l,l_3-1,l_2,l_1} - d_{l,l_3+1,5}F(l_3,l_2,3)\bm{\psi}_{l,l_3+1,l_2,l_1}\right],\\
\end{split}
\end{equation*}
\begin{equation*}
\begin{split}
L_{+,5} \bm{\psi}_{l,l_3,l_2,l_1}=&d_{l,l_3,5}D(l_3,l_2,3) B(l_2,l_1,2)\bm{\psi}_{l,l_3-1,l_2-1,l_1+1}\\
&+d_{l,l_3,5}B(l_3,l_2,3) A(l_2,l_1,2)\bm{\psi}_{l,l_3-1,l_2+1,l_1+1} \\
& -d_{l,l_3+1,5}C(l_3,l_2,3) B(l_2,l_1,2)\bm{\psi}_{l,l_3+1,l_2-1,l_1+1} \\
&-d_{l,l_3+1,5}A(l_3,l_2,3) A(l_2,l_1,2)\bm{\psi}_{l,l_3+1,l_2+1,l_1+1} ,
\end{split}
\end{equation*}
\begin{equation*}
\begin{split}
L_{-,5} \bm{\psi}_{l,l_3,l_2,l_1}=&-d_{l,l_3,5}D(l_3,l_2,3) D(l_2,l_1,2)\bm{\psi}_{l,l_3-1,l_2-1,l_1-1}\\
&-d_{l,l_3,5}B(l_3,l_2,3) C(l_2,l_1,2)\bm{\psi}_{l,l_3-1,l_2+1,l_1-1}\\
&+d_{l,l_3+1,5}C(l_3,l_2,3) D(l_2,l_1,2)\bm{\psi}_{l,l_3+1,l_2-1,l_1-1}\\
& +d_{l,l_3+1,5}A(l_3,l_2,3) C(l_2,l_1,2)\bm{\psi}_{l,l_3+1,l_2+1,l_1-1},\\
\end{split}
\end{equation*}
where $d_{l,l_3,5}=\sqrt{(l-l_3+1)(l+l_3+2)}$.

They fulfill
$$
\left[L_{h,j},L_{p,s}\right]\bm{\psi}_{l,l_3,l_2,l_1}=i\left(\delta_{h,p}L_{j,s}+\delta_{j,s}L_{h,p}-\delta_{h,s}L_{j,p}-\delta_{j,p}L_{h,s}\right)\bm{\psi}_{l,l_3,l_2,l_1},
$$
$$
\bm{L}^2 \bm{\psi}_{l,l_3,l_2,l_1}=l(l+3)\bm{\psi}_{l,l_3,l_2,l_1},\quad C_4\psi_{l,l_3,l_2,l_1}=l_3(l_3+2)\bm{\psi}_{l,l_3,l_2,l_1},
$$
$$
C_3\psi_{l,l_3,l_2,l_1}=l_2(l_2+1)\bm{\psi}_{l,l_3,l_2,l_1}\quad\mbox{and}\quad C_2\psi_{l,l_3,l_2,l_1}=l_1^2\psi_{l,l_3,l_2,l_1}.
$$
The coordinate operators are $\overline{x}_1,\overline{x}_2,\overline{x}_3,\overline{x}_4,\overline{x}_5$, $\overline{x}_{\pm}:=\overline{x}_1\pm i \overline{x}_2$, and they act on $\mathcal{H}_{\Lambda,5}$ as follows:
\begin{equation*}
\begin{split}
\overline{x}_1 \bm{\psi}_{l,l_3,l_2,l_1}=&\frac{c_{l,5} D(l,l_3,4) D(l_3,l_2,3)D(l_2,l_1,2)}{2} \bm{\psi}_{l-1,l_3-1,l_2-1,l_1-1}\\
&+\frac{c_{l+1,5} C(l,l_3,4) D(l_3,l_2,3)D(l_2,l_1,2)}{2} \bm{\psi}_{l+1,l_3-1,l_2-1,l_1-1}\\
&+\frac{c_{l,5} B(l,l_3,4) C(l_3,l_2,3)D(l_2,l_1,2)}{2} \bm{\psi}_{l-1,l_3+1,l_2-1,l_1-1}\\
&+\frac{c_{l+1,5} A(l,l_3,4) C(l_3,l_2,3)D(l_2,l_1,2)}{2} \bm{\psi}_{l+1,l_3+1,l_2-1,l_1-1}\\
&+\frac{c_{l,5} D(l,l_3,4) B(l_3,l_2,3)C(l_2,l_1,2)}{2} \bm{\psi}_{l-1,l_3-1,l_2+1,l_1-1}\\
&+\frac{c_{l+1,5} C(l,l_3,4) B(l_3,l_2,3)C(l_2,l_1,2)}{2} \bm{\psi}_{l+1,l_3-1,l_2+1,l_1-1}\\
&+\frac{c_{l,5} B(l,l_3,4) A(l_3,l_2,3)C(l_2,l_1,2)}{2} \bm{\psi}_{l-1,l_3+1,l_2+1,l_1-1}\\
&+\frac{c_{l+1,5} A(l,l_3,4) A(l_3,l_2,3)C(l_2,l_1,2)}{2} \bm{\psi}_{l+1,l_3+1,l_2+1,l_1-1}\\
&+\frac{c_{l,5} D(l,l_3,4) D(l_3,l_2,3)B(l_2,l_1,2)}{2} \bm{\psi}_{l-1,l_3-1,l_2-1,l_1+1}\\
&+\frac{c_{l+1,5} C(l,l_3,4) D(l_3,l_2,3)B(l_2,l_1,2)}{2} \bm{\psi}_{l+1,l_3-1,l_2-1,l_1+1}\\ 
&+\frac{c_{l,5} B(l,l_3,4) C(l_3,l_2,3)B(l_2,l_1,2)}{2} \bm{\psi}_{l-1,l_3+1,l_2-1,l_1+1}\\
&+\frac{c_{l+1,5} A(l,l_3,4) C(l_3,l_2,3)B(l_2,l_1,2)}{2} \bm{\psi}_{l+1,l_3+1,l_2-1,l_1+1}\\
&+\frac{c_{l,5} D(l,l_3,4) B(l_3,l_2,3)A(l_2,l_1,2)}{2} \bm{\psi}_{l-1,l_3-1,l_2+1,l_1+1}\\
&+\frac{c_{l+1,5} C(l,l_3,4) B(l_3,l_2,3)A(l_2,l_1,2)}{2} \bm{\psi}_{l+1,l_3-1,l_2+1,l_1+1}\\
&+\frac{c_{l,5} B(l,l_3,4) A(l_3,l_2,3)A(l_2,l_1,2)}{2} \bm{\psi}_{l-1,l_3+1,l_2+1,l_1+1}\\
&+\frac{c_{l+1,5} A(l,l_3,4) A(l_3,l_2,3)A(l_2,l_1,2)}{2} \bm{\psi}_{l+1,l_3+1,l_2+1,l_1+1},\\
\end{split}
\end{equation*}
\begin{equation*}
\begin{split}
\overline{x}_2 \bm{\psi}_{l,l_3,l_2,l_1}=&\frac{c_{l,5} D(l,l_3,4) D(l_3,l_2,3)D(l_2,l_1,2)}{2i} \bm{\psi}_{l-1,l_3-1,l_2-1,l_1-1}\\
&+\frac{c_{l+1,5} C(l,l_3,4) D(l_3,l_2,3)D(l_2,l_1,2)}{2i} \bm{\psi}_{l+1,l_3-1,l_2-1,l_1-1}\\
&+\frac{c_{l,5} B(l,l_3,4) C(l_3,l_2,3)D(l_2,l_1,2)}{2i} \bm{\psi}_{l-1,l_3+1,l_2-1,l_1-1}\\
&+\frac{c_{l+1,5} A(l,l_3,4) C(l_3,l_2,3)D(l_2,l_1,2)}{2i} \bm{\psi}_{l+1,l_3+1,l_2-1,l_1-1}
\end{split}
\end{equation*}
\begin{equation*}
\begin{split}
&+\frac{c_{l,5} D(l,l_3,4) B(l_3,l_2,3)C(l_2,l_1,2)}{2i} \bm{\psi}_{l-1,l_3-1,l_2+1,l_1-1}\\
&+\frac{c_{l+1,5} C(l,l_3,4) B(l_3,l_2,3)C(l_2,l_1,2)}{2i} \bm{\psi}_{l+1,l_3-1,l_2+1,l_1-1}\\
&+\frac{c_{l,5} B(l,l_3,4) A(l_3,l_2,3)C(l_2,l_1,2)}{2i} \bm{\psi}_{l-1,l_3+1,l_2+1,l_1-1}\\
&+\frac{c_{l+1,5} A(l,l_3,4) A(l_3,l_2,3)C(l_2,l_1,2)}{2i} \bm{\psi}_{l+1,l_3+1,l_2+1,l_1-1}\\
&-\frac{c_{l,5} D(l,l_3,4) D(l_3,l_2,3)B(l_2,l_1,2)}{2i} \bm{\psi}_{l-1,l_3-1,l_2-1,l_1+1}\\
&-\frac{c_{l+1,5} C(l,l_3,4) D(l_3,l_2,3)B(l_2,l_1,2)}{2i} \bm{\psi}_{l+1,l_3-1,l_2-1,l_1+1}\\ 
&-\frac{c_{l,5} B(l,l_3,4) C(l_3,l_2,3)B(l_2,l_1,2)}{2i} \bm{\psi}_{l-1,l_3+1,l_2-1,l_1+1}\\
&-\frac{c_{l+1,5} A(l,l_3,4) C(l_3,l_2,3)B(l_2,l_1,2)}{2i} \bm{\psi}_{l+1,l_3+1,l_2-1,l_1+1}\\
&-\frac{c_{l,5} D(l,l_3,4) B(l_3,l_2,3)A(l_2,l_1,2)}{2i} \bm{\psi}_{l-1,l_3-1,l_2+1,l_1+1}\\
&-\frac{c_{l+1,5} C(l,l_3,4) B(l_3,l_2,3)A(l_2,l_1,2)}{2i} \bm{\psi}_{l+1,l_3-1,l_2+1,l_1+1}\\
&-\frac{c_{l,5} B(l,l_3,4) A(l_3,l_2,3)A(l_2,l_1,2)}{2i} \bm{\psi}_{l-1,l_3+1,l_2+1,l_1+1}\\
&-\frac{c_{l+1,5} A(l,l_3,4) A(l_3,l_2,3)A(l_2,l_1,2)}{2i} \bm{\psi}_{l+1,l_3+1,l_2+1,l_1+1},
\end{split}
\end{equation*}
\begin{equation*}
\begin{split}
\overline{x}_3 \bm{\psi}_{l,l_3,l_2,l_1}=&c_{l,5}D(l,l_3,4) D(l_3,l_2,3)G(l_2,l_1,2)\bm{\psi}_{l-1,l_3-1,l_2-1,l_1}\\
&+c_{l+1,5} C(l,l_3,4) D(l_3,l_2,3)G(l_2,l_1,2)\bm{\psi}_{l+1,l_3-1,l_2-1,l_1}\\
&+c_{l,5} B(l,l_3,4) C(l_3,l_2,3)G(l_2,l_1,2)\bm{\psi}_{l-1,l_3+1,l_2-1,l_1}\\
&+c_{l+1,5} A(l,l_3,4) C(l_3,l_2,3)G(l_2,l_1,2)\bm{\psi}_{l+1,l_3+1,l_2-1,l_1} \\
&+c_{l,5} D(l,l_3,4) B(l_3,l_2,3)F(l_2,l_1,2)\bm{\psi}_{l-1,l_3-1,l_2+1,l_1}\\
&+c_{l+1,5} C(l,l_3,4) B(l_3,l_2,3)F(l_2,l_1,2)\bm{\psi}_{l+1,l_3-1,l_2+1,l_1} \\
&+c_{l,5} B(l,l_3,4) A(l_3,l_2,3)F(l_2,l_1,2)\bm{\psi}_{l-1,l_3+1,l_2+1,l_1}\\
&+c_{l+1,5} A(l,l_3,4) A(l_3,l_2,3)F(l_2,l_1,2)\bm{\psi}_{l+1,l_3+1,l_2+1,l_1},
\end{split}
\end{equation*}
\begin{equation*}
\begin{split}
\overline{x}_4 \bm{\psi}_{l,l_3,l_2,l_1}=&c_{l,5}D(l,l_3,4) G(l_3,l_2,3)\bm{\psi}_{l-1,l_3-1,l_2,l_1} +c_{l,5}B(l,l_3,4) F(l_3,l_2,3)\bm{\psi}_{l-1,l_3+1,l_2,l_1}\\
&\left. +c_{l+1,5}C(l,l_3,4) G(l_3,l_2,3)\bm{\psi}_{l+1,l_3-1,l_2,l_1} +c_{l+1,5}A(l,l_3,4) F(l_3,l_2,3)\bm{\psi}_{l+1,l_3+1,l_2,l_1}  \right.,\\
\overline{x}_5 \bm{\psi}_{l,l_3,l_2,l_1}=&c_{l,5}G(l,l_3,4)\bm{\psi}_{l-1,l_3,l_2,l_1} + c_{l+1,5}F(l,l_3,4)\bm{\psi}_{l+1,l_3,l_2,l_1},\\
\end{split}
\end{equation*}
\begin{equation*}
\begin{split}
\overline{x}_+ \bm{\psi}_{l,l_3,l_2,l_1}=&c_{l,5} D(l,l_3,4) D(l_3,l_2,3)B(l_2,l_1,2) \bm{\psi}_{l-1,l_3-1,l_2-1,l_1+1}\\
&+c_{l+1,5} C(l,l_3,4) D(l_3,l_2,3)B(l_2,l_1,2) \bm{\psi}_{l+1,l_3-1,l_2-1,l_1+1}\\ 
&+c_{l,5} B(l,l_3,4) C(l_3,l_2,3)B(l_2,l_1,2) \bm{\psi}_{l-1,l_3+1,l_2-1,l_1+1}\\
&+c_{l+1,5} A(l,l_3,4) C(l_3,l_2,3)B(l_2,l_1,2) \bm{\psi}_{l+1,l_3+1,l_2-1,l_1+1}\\
&+c_{l,5} D(l,l_3,4) B(l_3,l_2,3)A(l_2,l_1,2) \bm{\psi}_{l-1,l_3-1,l_2+1,l_1+1}\\
&+c_{l+1,5} C(l,l_3,4) B(l_3,l_2,3)A(l_2,l_1,2) \bm{\psi}_{l+1,l_3-1,l_2+1,l_1+1}\\
&+c_{l,5} B(l,l_3,4) A(l_3,l_2,3)A(l_2,l_1,2) \bm{\psi}_{l-1,l_3+1,l_2+1,l_1+1}\\
&+c_{l+1,5} A(l,l_3,4) A(l_3,l_2,3)A(l_2,l_1,2) \bm{\psi}_{l+1,l_3+1,l_2+1,l_1+1},\\
\end{split}
\end{equation*}
\begin{equation*}
\begin{split}
\overline{x}_- \bm{\psi}_{l,l_3,l_2,l_1}=&c_{l,5} D(l,l_3,4) D(l_3,l_2,3)D(l_2,l_1,2)\bm{\psi}_{l-1,l_3-1,l_2-1,l_1-1}\\
&+c_{l+1,5} C(l,l_3,4) D(l_3,l_2,3)D(l_2,l_1,2)\bm{\psi}_{l+1,l_3-1,l_2-1,l_1-1}\\
&+c_{l,5} B(l,l_3,4) C(l_3,l_2,3)D(l_2,l_1,2) \bm{\psi}_{l-1,l_3+1,l_2-1,l_1-1}\\
&+c_{l+1,5} A(l,l_3,4) C(l_3,l_2,3)D(l_2,l_1,2) \bm{\psi}_{l+1,l_3+1,l_2-1,l_1-1}\\
&+c_{l,5} D(l,l_3,4) B(l_3,l_2,3)C(l_2,l_1,2) \bm{\psi}_{l-1,l_3-1,l_2+1,l_1-1}\\
&+c_{l+1,5} C(l,l_3,4) B(l_3,l_2,3)C(l_2,l_1,2) \bm{\psi}_{l+1,l_3-1,l_2+1,l_1-1}\\
&+c_{l,5} B(l,l_3,4) A(l_3,l_2,3)C(l_2,l_1,2) \bm{\psi}_{l-1,l_3+1,l_2+1,l_1-1}\\
&+c_{l+1,5} A(l,l_3,4) A(l_3,l_2,3)C(l_2,l_1,2) \bm{\psi}_{l+1,l_3+1,l_2+1,l_1-1},\\
\end{split}
\end{equation*}
where
$$
c_{l,5}\overset{(\ref{valueclD})}=\begin{cases}
\sqrt{1+\frac{l^2+2l+1}{k_5}}\quad\mbox{if } 1\leq l\leq \Lambda ,\\
\quad\quad\quad 0\quad\quad \quad\mbox{otherwise},
\end{cases}
$$
and, according to (\ref{ABCDFG}),
\begin{equation*}
\begin{split}
A(l,l_3,4)&=\sqrt{\frac{(l+l_3+3)(l+l_3+4)}{(2l+3)(2l+5)}},\\
B(l,l_3,4)&=-\sqrt{\frac{(l-l_3-1)(l-l_3)}{(2l+3)(2l+1)}},\\
C(l,l_3,4)&=-\sqrt{\frac{(l-l_3+2)(l-l_3+1)}{(2l+3)(2l+5)}},\\
D(l,l_3,4)&=\sqrt{\frac{(l+l_3+2)(l+l_3+1)}{(2l+3)(2l+1)}},\\
F(l,l_3,4)&=\sqrt{\frac{(l+l_3+3)(l-l_3+1)}{(2l+3)(2l+5)}},\\
G(l,l_3,4)&=\sqrt{\frac{(l-l_3)(l+l_3+2)}{(2l+3)(2l+1)}}.
\end{split}
\end{equation*}
They fulfill (see section \ref{commrel+R^2})
$$
\left[\overline{x}_h,\overline{x}_j\right]=\left[-\frac{I}{k_5}+\left(\frac{1}{k_5}+\frac{\left(c_{\Lambda,5}
\right)^2}{2\Lambda+3} \right)\widehat{P}_{\Lambda,5}
\right]
\overline{L}_{h,j},\quad\left[L_{h,s},\overline{x}_j\right]=\frac{1}{i}\left(\delta_j^s \overline{x}_h -\delta_j^h \overline{x}_s\right),
$$
\begin{equation*}
\begin{split}
\bm{x}^2:=\sum_{h=1}^5\overline{x}_h\overline{x}_h=\left\{1+\frac{2\bm{L}^2+8}{4k_5(\Lambda)}-\left[\left(1+\frac{ 2\Lambda^2+8\Lambda+8 }{4k_5(\Lambda)} \right)\frac{\Lambda+3}{2\Lambda+3} \right]\widehat{P}_{\Lambda,4} \right\}
\end{split}
\end{equation*}
and
\begin{equation}
\begin{split}
\prod_{l=0}^{\Lambda}\left[\bm{L}^2-l(l+2)I\right]=0\quad,&\quad \prod_{l_{2}=0}^{l}\left[{C}_{3}-l_{2}(l_{2}+1)I\right]\widetilde{P}_{1,l}=0,\\
 \prod_{l_{1}=-l_2}^{l_2}\left[{L}_{1,2}-l_{1}I\right]\widetilde{P}_{2,l_2}=0,&\quad \left(\overline{x}_{\pm}\right)^{2\Lambda+1}=0,\mbox{ and }\left(L_{\nu,\pm}\right)^{2\Lambda+1}=0,\forall \nu\geq 3,
\end{split}
\end{equation}
where $\widetilde{P}_{h,j}$ is the projector on the eigenspace of ${C}_{5-h}$ corresponding to $l_{5-h}\equiv j$.

According to this, the algebra of observables is generated by the coordinate operators, in fact every projector can be written as a ordered polynomial in the $\overline{x}_{\nu}$.

Furthermore, the $SO(6)$-irrep $\pi_{\Lambda,6}$, the one characterized by $C_6\equiv \Lambda(\Lambda+4)I$ with representation space
$$
V_{\Lambda,6}:=span\left\{Y_{\Lambda,l,l_3,l_2,l_1}(\theta_5,\theta_4,\theta_3,\theta_2,\theta_1):\Lambda\geq l\geq l_3\geq l_2\geq |l_1|, l_i\in\mathbb{Z}\forall i \right\},
$$
can be used to identify $\bm{\psi}_{l,l_3,l_2,l_1}\equiv Y_{\Lambda,l,l_3,l_2,l_1}$, and also the operators
\begin{equation}
\begin{split}
L_{h,j}\equiv L_{h,j}\quad\mbox{for}\quad 1\leq h<j\leq 5\quad\mbox{and}\quad\overline{x}_s\equiv p_5(\lambda)L_{s,6}p_{5}(\lambda),
\end{split}
\end{equation}
where
$$
\lambda:=\frac{-3+\sqrt{9+4\bm{L}^2}}{2},
$$
while $p_{5}(\lambda)$ is an analytic function and the values $p_{5}(l)$, when $l\in\mathbb{N}_0$, can be obtained recursively from (\ref{condition_p}) starting from $p_{5}(0):=1$.

Furthermore, in order to prove the convergence of $S_{\Lambda}^4$ to ordinary quantum mechanics on $S^4$, it is convenient to identify $\bm{\psi}_{l,l_3,l_2,l_1}\equiv Y_{l,l_3,l_2,l_1}$ and then to consider their fuzzy counterparts $\widehat{Y}_{l,l_3,l_2,l_1}$, which can be used to approximate a generic $f\in B(S^4)$ or $f\in C(S^4)$; this is possible because the $Y_{l,l_3,l_2,l_1}$ are an orthonormal basis of $\mathcal{L}^2(S^4)$, and also homogeneous polynomials in the $t_h:=x_h/r$ variables. Then,
$$
\widehat{f}_{\Lambda}:= \sum_{l=0}^{2\Lambda}\sum_{l_3=0}^{l}
\sum_{l_2=0 }^{l_3}\sum_{l_1=-l_2}^{l_2}
f_{l,l_3,l_2,l_1}\widehat{Y}_{l,l_3,l_2,l_1},\quad\mbox{where}\quad f_{l,l_3,l_2,l_1}:=\left\langle Y_{l,l_3,l_2,l_1},f\right\rangle,
$$
is an approximation of $f$ because of the following two theorems (see section \ref{converge})
\begin{teorema}
Every projected coordinate operator $\overline{x}_h$ converges strongly to the corresponding $t_{h}$ as $\Lambda\to\infty$ if
$$
k_5\left(\Lambda\right)\geq \Lambda \frac{1}{144}(\Lambda+1)^2(\Lambda+2)^4(\Lambda+3)^2\frac{4\Lambda(\Lambda+3)+8}{4}=
\frac{1}{144}(\Lambda+1)^3(\Lambda+2)^5(\Lambda+3)^2.
$$
\end{teorema}
\begin{teorema} 
Choosing $k_5\left(\Lambda\right)$ fulfilling (\ref{kineq2}) for $D=5$, then for all $f,g\in B(S^4)$ and $C(S^4)$ the following strong limits as $\Lambda\rightarrow \infty$ hold: $\hat{f}_{\Lambda}\rightarrow f\cdot,\widehat{\left(fg\right)}_{\Lambda}\rightarrow fg\cdot$ and $\hat{f}_{\Lambda}\hat{g}_{\Lambda}\rightarrow fg\cdot$. 
\end{teorema}
\section{General setting}\label{General_setting}
As mentioned before, consider a quantum particle in $\mathbb{R}^D$, with a Hamiltonian operator
$$
H:=-\frac 12\Delta + V(r)
$$
such that the potential $V(r)$ has a very sharp minimum at $r=1$ with a very large $k_D:= V'(1)/4>0$, and fix 
$V_0:= V(1)$ so that the ground state has zero energy, i.e. $E_0=0$. In addition, impose here that the energy cutoff $\overline{E}$ is chosen so that
\be
V(r)\simeq V_0+2k_D (r-1)^2\quad \mbox{if }r\mbox{ fulfills}\quad V(r)\leq  \overline{E},
%|r\!-\!1|\le \sqrt{\frac{\overline{E}\!-\!V_0}{2k_D}},
\label{cond1}
\ee
then one can neglect terms of order higher than $2$ in the Taylor expansion of $V(r)$ around $r=1$ and approximate the potential with a harmonic one in the classical region $b_{\overline{E}}\subset \mathbb{R}^D$ compatible with the energy cutoff \ $V(r)\leq \overline{E}$. \ 
  \begin{figure}[htbp]
  \begin{center}
          \includegraphics[scale=0.23]{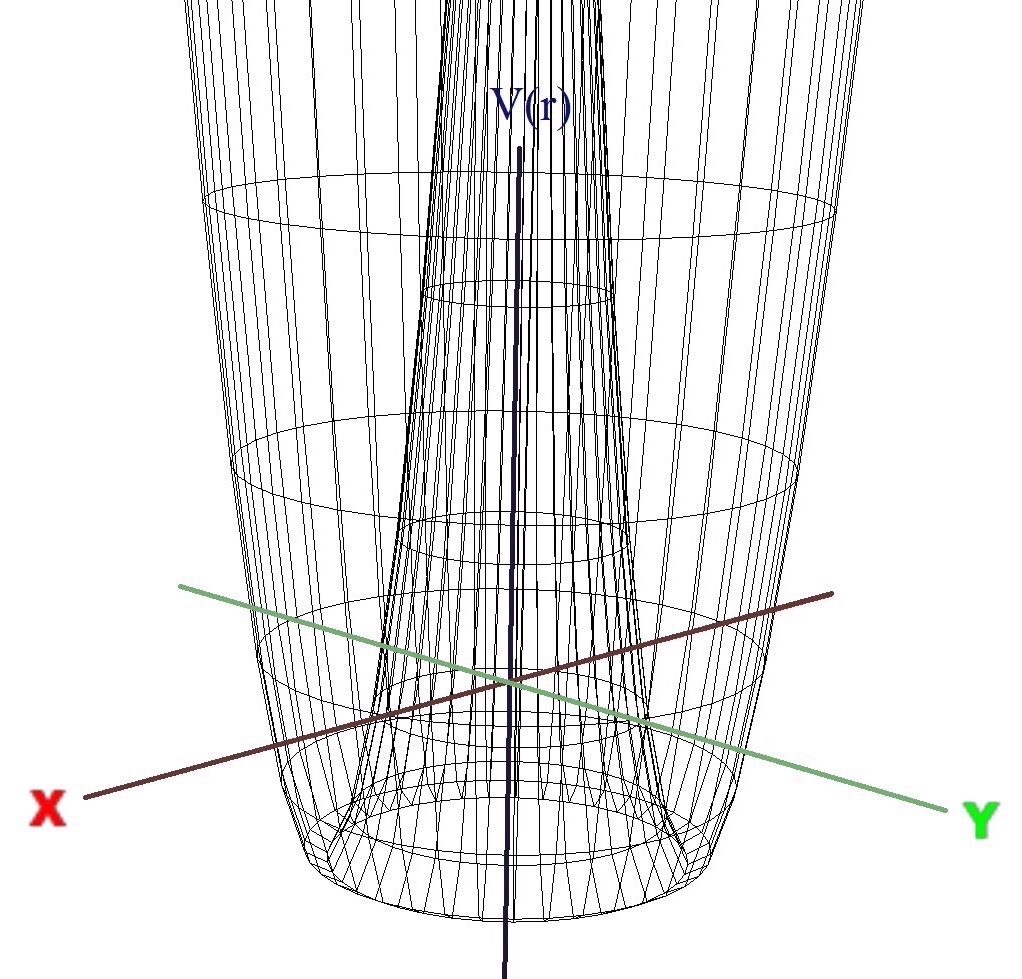}
          \caption{Three-dimensional plot of $V(r)$}
          \end{center}
      \end{figure}
The equality $\bm{L}^2 Y(\theta_{d},\theta_{d-1},\cdots,\theta_1)=l(l+D-2)Y(\theta_{d},\theta_{d-1},\cdots,\theta_1)$ and the Ansatz (\ref{Ansatz}) are used to simplify the resolution of the PDE $H\bm{\psi}=E\bm{\psi}$, in fact this problem is consequently split in two:
\begin{enumerate}
\item Solve the corresponding ODE for $f(r)$;
\item Determine all the eigenfunctions of $\bm{L}^2$, which will be also square-integrable because $S^2$ is compact and $\bm{L}^2$ is regular.
\end{enumerate}
In addition, it is also necessary to verify if $H$ is a self-adjoint operator on the Hilbert space $\mathcal{H}_D$ of pure quantum states.
\subsection{Resolution of $H\bm{\psi}=E\bm{\psi}$ $-$ Step 1}\label{section_determination_f}
The ODE for $f(r)$ turns out to be equivalent to equation (9) in \cite{FiorePisacane,{FiorePisacanePOS18}}; this means that one has to solve
\be
\left[-\partial_r^2-(D-1)\frac 1r\partial_r+\frac 1{r^2}l\left(l+D-2\right)+V(r)\right] f(r)=Ef(r).\label{eqpolarD}
\ee
In section \ref{section_reg_f} it is shown that the hypothesis $r^2V(r)\overset{r\rightarrow 0^+}\longrightarrow T\in\mathbb{R}^+$ [which is obviously compatible with (\ref{cond1})] and the request that $\bm{\psi}\in D(H)\equiv D(H^*)$ (self-adjointness of $H$) imply that $f(r)$ is regular at $r=0$, and then the same applies to the function $g(r):=f(r)r^{\frac{D-1}{2}}$. Consequently, (\ref{eqpolarD}) becomes
\be
-g''(r)+g(r)\frac{\left[D^2-4D+3+4l(l+D-2) \right]}{4r^2}+V(r)g(r)=Eg(r).
\label{eqpolarD2}
\ee
For the purposes of this thesis, the solution of this last equation is interesting only around $r=1$; this means that one can use the equalities (at leading order)
$$
\frac{1}{r^2}=1-2(r-1)+3(r-1)^2,\quad V(r)=V_0+2k_D(r-1)^2,
$$
which lead to this $1$-dimensional harmonic oscillator equation
\be
-g''(r)+g(r)k_{l,D}\left(r-\widetilde{r}_{l,D}\right)^2=\widetilde{E}_{l,D}g(r),
\label{eqpolarD3}
\ee
where 
\begin{equation}\label{definizioni1}
\begin{split}
&b(l,D):=\frac{D^2-4D+3+4l(l+D-2)}{4},\quad k_{l,D}:=3b(l,D)+2k_D,\\
&\widetilde{r}_{l,D}:=\frac{4b(l,D)+2k_D}{3b(l,D)+2k_D},\quad \widetilde{E}_{l,D}:=E-V_0-\frac{2b(l,D)\left[k_D+b(l,D)\right]}{3b(l,D)+2k_D};
\end{split}
\end{equation}
so at leading order the lowest eigenvalues $E$ are those of the $1$-dimensional harmonic oscillator approximation of (\ref{eqpolarD2}).

The (Hermite) square-integrable solutions of (\ref{eqpolarD3}) are ($M_{n,l,D}$ is a suitable normalization constant)
\be\label{valueg}
g_{n,l,D}(r)=M_{n,l,D}\hspace{0.15cm}e^{-\frac{\sqrt{k_{l,D}}}{2}\left(r-\widetilde{r}_{l,D}\right)^2}\cdot H_n\left((r-\widetilde{r}_{l,D})\sqrt[4]{k_{l,D}}\right)\quad\mbox{ with }n\in\mathbb{N}_0,
\ee
implying
\be\label{valuef}
f_{n,l,D}(r)=\frac{M_{n,l,D}}{r^{\frac{D-1}{2}}}\hspace{0.15cm}e^{-\frac{\sqrt{k_{l,D}}}{2}\left(r-\widetilde{r}_{l,D}\right)^2}\cdot H_n\left((r-\widetilde{r}_{l,D})\sqrt[4]{k_{l,D}}\right)\quad\mbox{ with }n\in\mathbb{N}_0.
\ee
The corresponding `eigenvalues' in (\ref{eqpolarD3}) are  $\widetilde{E}_{n,l,D}=(2n+1)\sqrt{k_{l,D}}$ and this leads to energies
\be
E_{n,l,D}=(2n+1)\sqrt{k_{l,D}}+V_0+\frac{2b(l,D)[k_D+b(l,D)]}{3b(l,D)+2k_D}
.\label{energiesD}
\ee
As mentioned before, $V_0$ is fixed requiring that the lowest energy level, which corresponds to $n=l=0$, is $E_{0,0,D}=0$; this implies
\be 
V_0=-\sqrt{k_{0,D}}-\frac{2b(0,D)\left[k_D+b(0,D)\right]}{3b(0,D)+2k_D};
\ee
while the expansions of $\widetilde{r}_{l,D}$ and $E_{n,l,D}$ at leading order in $k_D$ are the following ones:
\begin{equation}
\begin{split}
\widetilde{r}_{l,D}&=1+\frac{b(l,D)}{2k_D}-\frac{3b(l,D)^2}{4k_D^2}+O\left(k_D^{-3}\right),\\
V_0&=-\sqrt{2k_D}-b(0,D)-\frac{3b(0,D)}{2\sqrt{2k_D}}+\frac{b(0,D)^2}{2k_D}+\frac{9b(0,D)^2}{8\left(2k_D\right)^{\frac{3}{2}}}-\frac{3b(0,D)^3}{4k_D^2}+O\left(k_D^{-\frac{5}{2}}\right),\\
E_{n,l,D}&=(2n+1)\sqrt{2k_D}+V_0+b(l,D)+(2n+1)\frac{3b(l,D)}{2\sqrt{2k_D}}\\
&-\frac{b(l,D)^2}{2k_D}-(2n+1)\frac{9b(l,D)^2}{16k\sqrt{2k_D}}+\frac{3b(l,D)^3}{4k_D^2}+O\left(k_D^{-\frac{5}{2}}\right)\\
&=2n\sqrt{2k_D}+l(l+D-2)+\frac{1}{\sqrt{2k_D}}\left[3nb(l,D)+\frac{3}{2}l(l+D-2) \right]\\
&+\frac{1}{2k_D}\left[-l(l+D-2)\right]\left[\frac{2D^2-8D+6+4l(l+D-2)}{4} \right]+O\left(k_D^{-\frac{3}{2}}\right).
\end{split}
\end{equation}
\subsection{Resolution of $H\bm{\psi}=E\bm{\psi}$ $-$ Step 2}\label{section_D-dimSA}
In section \ref{D-dimsa} it is shown that an orthonormal basis of $\mathcal{L}^2(S^d)$, made up of $\bm{L}^2$-eigenfunctions, is the collection of all the
\be\label{explicitY}
Y=Y_{\bm{l}}(\theta_{d},\cdots,\theta_2,\theta_1)=\frac{e^{il_1\theta_1}}{\sqrt{2\pi}}\left[\prod_{n=2}^{d}{}_n \overline{P}_{l_{n}}^{l_{n-1}}(\theta_n) \right],\quad \bm{l}=(l_d,\cdots,l_2,l_1),
\ee
where
\be
{}_j \overline{P}_{L}^{M}(\theta):=\sqrt{\frac{2L+j-1}{2}}\sqrt{\frac{(L+M+j-2)!}{(L-M)!}}\left[\sin{\theta}\right]^{\frac{2-j}{2}}P_{L+\frac{j-2}{2}}^{-\left(M+\frac{j-2}{2}\right)}(\cos{\theta}),
\ee
$l_d\geq\cdots\geq l_2\geq |l_1|$, $l_i\in\mathbb{Z}$ $\forall i$ and $P_l^m$ is the associated Legendre function of first kind (see \cite{Bateman} for a summary about these special functions).

They fulfill
\begin{equation}\label{ArmSfe}
\begin{split}
L_{1,2}Y_{\bm{l}}=l_1 Y_{\bm{l}}\Rightarrow C_2 Y_{\bm{l}}=l_1^2 Y_{\bm{l}},&\quad  C_p Y_{\bm{l}}=l_{p-1}(l_{p-1}+p-2) Y_{\bm{l}},\\
\mbox{and}\qquad&\int_{S^d}Y_{\bm{l}}Y^*_{\bm{l'}}d\alpha=\delta_{\bm{l}}^{\bm{l'}},
\end{split}
\end{equation}
where $d\alpha$ is the usual measure on $S^d$,
$$
d\alpha=\left[\sin^{d-1}{(\theta_{d})} \sin^{d-2}{(\theta_{d-1})} \cdots \sin{(\theta_2)}\right] d\theta_1 d\theta_2\cdots d\theta_{d}.
$$

According to these last equations, every $\overline{l}\in\mathbb{N}_0$ identifies a
\be\label{defV_lD}
V_{\overline{l},D}:=span\left\{Y_{\overline{\bm{l}}}:\overline{\bm{l}}:=\left(\overline{l},l_{d-1},\cdots,l_2,l_1\right),\overline{l}\geq l_{d-1}\geq\cdots\geq l_2\geq  |l_1|, l_i\in\mathbb{Z}\forall i \right\},
\ee
which is the representation space of an irrep of $U\bm{so}(D)$, and $\left\{L_{1,2},C_2,\cdots,C_{D}\right\}$ is a CSCO of this irrep, where CSCO stands for \emph{complete set of commuting observables}, i.e. a set of commuting operators whose set of eigenvalues completely specify elements of a basis of $\mathcal{H}_{\Lambda,D}$.

In addition, in section \ref{Y_lbasisL^2} it is shown that
\begin{equation*}
\begin{split}
&V_{l,D}\quad\mbox{is isomorphic to}\quad \bigoplus_{m=0}^{l}V_{m,d} \quad\mbox{if }D>3,\\
\mbox{ while }&V_{l,3}\quad\mbox{is isomorphic to}\quad \bigoplus_{m=-l}^{l}V_{m,2};
\end{split}
\end{equation*}
this decomposition can be also applied to $\mathcal{H}_{\Lambda,D}$, up to isomorphisms, and this job is done in section \ref{realso(D+1)}. 

So, the pure quantum states (the elements of an orthonormal basis of $\mathcal{H}_{D}$) are the following ones:
\be\label{pure_psi}
\bm{\psi}_{n,\bm{l},D}(r,\theta_{d},\cdots,\theta_2,\theta_1):=f_{n,l,D}(r)Y_{\bm{l}}(\theta_{d},\cdots,\theta_2,\theta_1),
\ee
with $n\in\mathbb{N}_0$, $l\equiv l_d\geq\cdots\geq l_2\geq |l_1|$, $l_i\in\mathbb{Z}$ $\forall i$.
\section{The imposition of the cutoff}\label{sec_cutoff}
As mentioned before, a low enough energy cutoff $E\leq\overline{E}$ is imposed in a way such that it excludes all the states with $n>0$; according to this, it must be $\overline E<2\sqrt{2k_D}$, which (from the physical point of view) means that radial oscillations are `frozen' ($\Rightarrow n=0$, as wanted), so that all corresponding classical trajectories are circles; the energies $E$ below $\overline E$ will therefore depend only on $l$ and $D$, and are consequently denoted by $E_{l,D}$.

 \begin{figure}[htbp]
\begin{center}          \includegraphics[scale=0.5]{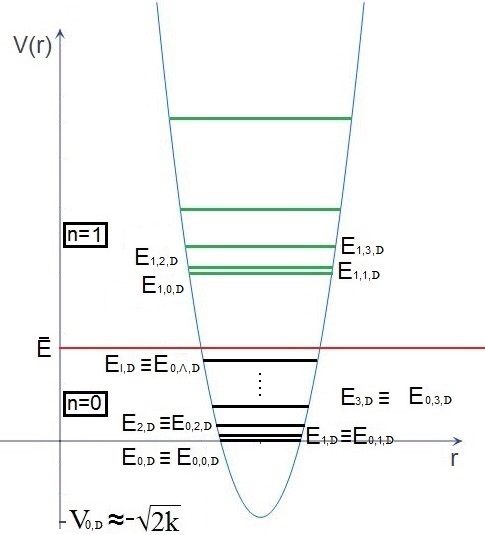}
          \caption{Two-dimensional plot of $V(r)$ including the energy-cutoff}
\end{center}      
      \end{figure}

The Hilbert space of `admitted' states is $\mathcal{H}_{\overline{E},D}\subset\mathcal{H}$, it is finite-dimensional and spanned by the states $\bm{\psi}$ fulfilling the cutoff condition; on the other hand, one has also to replace every observable $A$ by $\overline A:= P_{\overline{E},D}AP_{\overline{E},D}$,  where $P_{\overline{E},D}$ is the projection on $\Hi_{\overline{E},D}$, and I give to $\overline{A}$ the same physical interpretation.

Then, at leading orders in $1/\sqrt{k_D}$,
\be
\begin{split}
H=E_{l,D}&=l(l+D-2)+O\left(\frac 1{\sqrt{k_D}}\right);\\
\bm{\psi}_{\bm{l},D}(r,\theta_{D-1},\cdots,\theta_1)&=\frac{M_{l,D}}{r^{\frac{D-1}{2}}} e^{-\frac{\sqrt{k_{l,D}}}{2}\left(r-\widetilde{r}_{l,D}\right)^2}Y_{\bm{l}}\left(\theta_{d},\cdots,\theta_1\right),
\end{split}\label{statopsi}
\ee
where the normalization factor $M_{l,D}$ is fixed so that $M_{l,D}>0$ and all $\bm{\psi}_{\bm{l},D}$ have unit norm in $\mathcal{L}^2\left(\mathbb{R}^D\right)$ (see {section \ref{squintpsi}}). 

The choice of a $\Lambda$-dependent energy cutoff $\overline{E}=\overline{E}\left(\Lambda\right):=\Lambda(\Lambda+D-2)$, implies that the condition $E\leq \overline E$ becomes equivalent to the projection of the theory onto the Hilbert subspace $\mathcal{H}_{{{\Lambda}},D}\equiv \mathcal{H}_{\overline E,D}$ spanned by all the states $\bm{\psi}_{\bm{l},D}$ with  $l(l+D-2)\leq \Lambda(\Lambda+D-2)\Leftrightarrow l\leq\Lambda$. For consistency it must be
\be
\Lambda(\Lambda+D-2)<2\sqrt{2k_D},\label{consistency}
\ee
and for instance one can define $k_D\left(\Lambda\right)\geq\left[\Lambda(\Lambda+D-2) \right]^2$, while in {section \ref{convergproof}} a larger $k_D(\Lambda)$ is used in order to prove the convergence to ordinary quantum mechanics on $S^d$. According to this first choice of $k_D(\Lambda)$, all $E_{l,D}$ are smaller than the energy levels corresponding to $n>0$;
this is also sufficient to guarantee that $k_{l,D}%>2(k_D\!-\!2\sqrt{2k_D})
\gg 1$ for all $l\leq {{\Lambda}}$ \ [by the way,
$k_{l,D}>0$ is a  necessary condition for (\ref{eqpolarD3}) to be the eigenvalue equation of a harmonic oscillator]; furthermore, the spectrum of $\overline{H}$ becomes the whole spectrum $\{l(l+D-2)\}_{l\in\mathbb{N}_0}$ of   $\bm{L}^2$ 
in the commutative limit, i.e. $\Lambda \to\infty$.

\section{The algebra $\mathcal{A}_{\Lambda,D}$}
\subsection{The action of angular momentum components on $Y_{\bm{l}}$}\label{Section_sphharm}
In the next lines there are the $R$ coefficients, they are determined in {section \ref{Appendix_defi}} and used in the following definition (which is given by induction) of the action of a generic $L_{h,j}$ on a spherical harmonic $Y_{\bm{l}}$.
\begin{defi}\label{defiL}
For $D=2$ there is only one angular momentum component, $L_{1,2}$, and its action is $L_{1,2}Y_{l_1}=l_1 Y_{l_1}$. For $D>2$, let 
$$d_{L,l,D}:=\sqrt{(L+1)(L+D-3)-l(l+D-4)}=\sqrt{(L-l+1)(L+l+D-3)}$$
and
$$
R_{h,D}\left(\bm{l},\bm{l'}\right):=\left\langle Y_{\bm{l'}},
t_h Y_{\bm{l}}\right\rangle;
$$
the action of the angular momentum operators is defined in this way:
\be\label{azL1}
L_{\nu,D} Y_{\bm{l}}:=\frac{1}{i}\sum_{
\substack{
{l'}_j:|l_j-{l'}_j|=1\\
\mbox{for }j=\nu-1,\cdots,d-1
}}\left\{d_{l,l_{d-1},D}R_{\nu,d}\left(\bm{l},\widetilde{\bm{l'}}_\nu\right)Y_{\widetilde{\bm{l'}}_{\nu}}-d_{l,l_{d-1}+1,D}R_{\nu,d}\left(\bm{l},\widehat{\bm{l'}}_\nu\right)Y_{\widehat{\bm{l'}}_{\nu}}\right\},\\
\ee
where
\begin{equation*}
\begin{split}
\widetilde{\bm{l'}}_\nu&:=\left(l,l_{d-1}-1,l'_{d-2},\cdots,{l'}_{\nu-1},l_{\nu-2},\cdots,{l}_1\right), \\
\widehat{\bm{l'}}_\nu&:=\left(l,l_{d-1}+1,l'_{d-2},\cdots,{l'}_{\nu-1},l_{\nu-2},\cdots,{l}_1\right),
\end{split}
\end{equation*}
for $\nu\in\{1,\cdots,d-2\}$, $l_0\equiv l_1$ and
$$
\widetilde{\bm{l'}}_{d-1}:=\left(l,l_{d-1}-1,l_{d-2},\cdots,{l}_1\right),\quad \widehat{\bm{l'}}_{d-1}:=\left(l,l_{d-1}+1,l_{d-2},\cdots,{l}_1\right).
$$

Furthermore, 
$$
L_{D,j}:=-L_{j,D}\quad,\quad L_{\pm,\nu}:=\frac{L_{2,\nu}\mp iL_{1,\nu}}{\sqrt{2}}\quad\forall\nu\geq 3
$$
and the action of $L_{h,\widetilde{D}}$ on a $D$-dimensional spherical harmonic, when $h<\widetilde{D}<D$, is defined as the same of $L_{h,\widetilde{D}}$ on a $\widetilde{D}$-spherical harmonic in $\mathbb{R}^{\widetilde{D}}$; then it, when acts in $\mathbb{R}^D$, does not `affect' the indices $l,l_{d-1},\cdots l_{\widetilde{D}-1}$.
\end{defi}
Summarizing,
\begin{itemize}
\item In {section \ref{Appendix_defi}} the action in $\mathbb{R}^D$ of the coordinate operators $t_\nu:=\frac{x_\nu}{r}$ on the $D$-dimensional spherical harmonics $Y_{\bm{l}}$ is calculated, this action essentially defines the aforementioned $R_{\nu,D}$ coefficients; 
\item This implies that one can easily derive the action in $\mathbb{R}^{D-1}$ of coordinate operators $t_h$ on a generic $(D-1)$-dimensional spherical harmonic $Y_{l_{d-1},\cdots,l_1}$, which consequently uses the $R_{h,d}$ coefficients;
\item So, in {definition \ref{defiL}} the action of $L_{\nu,D}$ on $Y_{\bm{l}}$ is the same, up to the $\frac{d_{l,l_{d-1},D}}{i}$ and $-\frac{d_{l,l_{d-1}+1,D}}{i}$ coefficients, of $t_\nu$ on $Y_{l_{d-1},\cdots,l_1}$; this is also in agreement with the Wigner-Eckart theorem, because 
$$
\left\langle Y_{\bm{l'}},L_{\nu,D} Y_{\bm{l}}\right\rangle=
\begin{cases}
\frac{1}{i}d_{l,l_{d-1},D}R_{\nu,d}\left(\bm{l},\widetilde{\bm{l'}}_\nu\right)\quad\mbox{ if }l'_{d-1}=l_{d-1}-1,\\
-\frac{1}{i}d_{l,l_{d-1}+1,D}R_{\nu,d}\left(\bm{l},\widetilde{\bm{l'}}_\nu\right)\quad\mbox{ if }l'_{d-1}=l_{d-1}+1,\\
0\quad\mbox{otherwise,}
\end{cases}
$$
where the first factor depends only on the index $l_{d-1}$, which identifies the $SO(d)$ irrep, while the second one is a Clebsch-Gordan coefficient.
\end{itemize}
In {sections \ref{proofLij1}} and {\ref{proofLij2}} the following relations are explicitly checked for the reader's convenience:
\be\label{propLij}
\begin{split}
\bm{L}^2 Y_{\bm{l}}&=\sum_{1\leq h<j\leq D}L_{h,j}^2Y_{\bm{l}}=l\left(l+D-2 \right)Y_{\bm{l}},\\
\left[L_{h,j},L_{p,s}\right]&=i\left(\delta_{h,p}L_{j,s}+\delta_{j,s}L_{h,p}-\delta_{h,s}L_{j,p}-\delta_{j,p}L_{h,s}\right).
\end{split}
\ee

\subsection{The action of `projected' operators on $\mathcal{H}_{\Lambda,D}$}\label{Section_Algebra}
The Hilbert space of admitted states $\mathcal{H}_{\Lambda,D}$, constructed in {section \ref{sec_cutoff}}, is spanned by all the states $\bm{\psi}_{\bm{l},D}$ fulfilling $l\leq \Lambda$. In the following lines I do a complete study of the action of the `projected' angular momentum operators $\overline{L}_{h,j}$ and of the `projected' coordinate operators $\overline{x}_h$ on the pure quantum states. The {definition \ref{defiL}} implies $\overline{L}_{h,j}\bm{\psi}_{\bm{l},D}={L}_{h,j}\bm{\psi}_{\bm{l},D}$, which is a consequence of the invariance of $H$ (and therefore $P_{\overline{E},D}$) with respect to rotations (i.e. they commute with every $L_{h,j}$); from this and the fact that the action of every $L_{h,j}$ does not `affect' the index $l$ it follows that the action of $L_{h,j}$ on a $\bm{\psi}_{\bm{l},D}$ essentially coincides with the one of $Y_{\bm{l}}$. Then
\be\label{azLpsi}
L_{\nu,D} \bm{\psi}_{\bm{l},D}:=\frac{1}{i}\sum_{
\substack{
{l'}_j:|l_j-{l'}_j|=1\\
\mbox{for }j=\nu-1,\cdots,d-1
}}\left\{d_{l,l_{d-1},D}R_{\nu,d}\left(\bm{l},\widetilde{\bm{l'}}_\nu\right)\bm{\psi}_{\widetilde{\bm{l'}}_{\nu},D}-d_{l,l_{d-1}+1,D}R_{\nu,d}\left(\bm{l},\widehat{\bm{l'}}_\nu\right)\bm{\psi}_{\widehat{\bm{l'}}_{\nu},D}\right\};
\ee
$$
L_{D,j} \bm{\psi}_{\bm{l},D}:=-L_{j,D} \bm{\psi}_{\bm{l},D}\quad,\quad L_{\pm,\nu} \bm{\psi}_{\bm{l},D}:=\left(L_{2,\nu}\mp iL_{1,\nu}\right)\bm{\psi}_{\bm{l},D}\quad\forall\nu\geq 3,
$$
and the action of $L_{h,\widetilde{D}}$ on a $\bm{\psi}_{\bm{l},D}$, when $\widetilde{D}<D$, is essentially the same of $L_{h,\widetilde{D}}$ on a $\widetilde{D}$-spherical harmonic in $\mathbb{R}^{\widetilde{D}}$, as for (\ref{azL1}) and (\ref{azLpsi}).

On the other hand, the action of $\overline{x}_h$ on a state $\bm{\psi}_{\bm{l},D}$ can be obtained from the one of the multiplication operator $t_h\cdot$ on a $D$-dimensional spherical harmonic $Y_{\bm{l}}$ (see section \ref{Appendix_defi}), while sometimes it is useful to consider the operators
$$
\overline{x}_{\pm}:=\overline{x}_1\pm i\overline{x}_2.
$$
It is easy to see that the action of projected coordinate operators `affect' the index $l$, for this reason further calculations are needed, because in this case the integral 
$$
\int_{0}^{+\infty}r f_{l,D}(r) f_{l',D}(r) dr
$$
is not trivial, unlike what happens for the action of $L_{h,j}$.

According to this,
\begin{equation}
\overline{x}_h\psi_{\bm{l},D}=\sum_{\substack{
|l_j-{l'}_j|=1\\
j\in\{h-1,\cdots,d-1\}
}}\left[c_{l,D}
R_{h,D}\left(\bm{l},\widetilde{\widetilde{\bm{l'}}}_h\right)\bm{\psi}_{\widetilde{\widetilde{\bm{l'}}}_h,D}+c_{l+1,D}R_{h,D}\left(\bm{l},\widehat{\widehat{\bm{l'}}}_h\right)\bm{\psi}_{\widehat{\widehat{\bm{l'}}}_h,D}\right],
\end{equation}
where
\begin{equation}\label{defc}
\begin{split}
\widetilde{\widetilde{\bm{l'}}}_h:=\left(l-1,{l'}_{d-1},\cdots,{l'}_{h-1},l_{h-2}\cdots,l_1\right)&,\hspace{0.2cm}\widehat{\widehat{\bm{l'}}}_h:=\left(l+1,{l'}_{d-1},\cdots,{l'}_{h-1},l_{h-2}\cdots,l_1\right),\\
c_{l,D}:=\int_{0}^{+\infty}r f_{l,D}(r) f_{l-1,D}(r) dr&,\quad c_{l+1,D}:=\int_{0}^{+\infty}r f_{l,D}(r) f_{l+1,D}(r) dr,\\
c_{-\Lambda,2}= c_{\Lambda+1,2}:=0\quad&\mbox{and}\quad c_{0,D}= c_{\Lambda+1,D}:=0\quad\forall D\geq 3;
\end{split}
\end{equation}
the explicit values of $c_{l,D}$ are calculated in {section \ref{scalprod}} and
\be\label{valueclD}
c_{l,D}\overset{(\ref{valuescalprod})}=\sqrt{1+\frac{\left[b(l,D)+b(l-1,D)\right]}{2k_D}}\quad\mbox{up to terms}\quad O\left(\frac{1}{(k_D)^{\frac{3}{2}}}\right).
\ee
\subsection{The commutation relations and the action of $\bm{x}^2$}\label{commrel+R^2}
The calculations of {section \ref{proofLDaLDb}} can be used to determine the action of $\left[\overline{x}_h,\overline{x}_j\right]$ on $\mathcal{H}_{\Lambda,D}$, this because the action of $\overline{x}_h$ on $\bm{\psi}_{\bm{l},D}$ is essentially the same of $L_{h,D+1}$ on $Y_{l_D,\bm{l}}$; the only difference is the replacement of $-\frac{1}{i}d_{l_D,l+1,D+1}$ with $c_{l+1,D}$ and $\frac{1}{i}d_{l_D,l,D+1}$ with $c_{l,D}$, respectively. These arguments and (\ref{diffc}) are sufficient to prove that (see {section \ref{proofcommx-ax-b}} for explicit calculations)
\begin{equation}\label{commx-ax-b}
\left[\overline{x}_h,\overline{x}_j\right]=
\left[-\frac{I}{k_D}+\left(\frac{1}{k_D}+\frac{\left(c_{\Lambda,D}
\right)^2}{2\Lambda+D-2} \right)\widehat{P}_{\Lambda,D}
\right]
\overline{L}_{h,j},
\end{equation}
where $\widehat{P}_{\Lambda,D}$ is the projector on the $\Lambda(\Lambda+D-2)$-eigenspace of $\bm{L}^2$.

On the other hand, it is obvious that $\widehat{P}_{\Lambda,D}:=\widehat{P}_{\overline{E},D}$ commutes with $L_{h,j}$, for all $1\leq h<j\leq D$; this and 
\be\label{commLhjxp}
\left[L_{h,s},{x}^j\right]\overset{(\ref{defiLhj})}=\frac{1}{i}\left(\delta_j^s {x}^h -\delta_j^h {x}^s\right)
\ee
imply
\begin{equation*}
\begin{split}
\left[\overline{L}_{h,s},\overline{x}_j\right]&= \widehat{P}_{\Lambda,D}L_{h,s} \widehat{P}_{\Lambda,D} \widehat{P}_{\Lambda,D}x_j \widehat{P}_{\Lambda,D}-\widehat{P}_{\Lambda,D}x_h \widehat{P}_{\Lambda,D} \widehat{P}_{\Lambda,D}L_{h,s}\widehat{P}_{\Lambda,D}\\
&= \widehat{P}_{\Lambda,D} L_{h,s}x_j \widehat{P}_{\Lambda,D}-\widehat{P}_{\Lambda,D}x_j L_{h,s} \widehat{P}_{\Lambda,D}\\
&= \widehat{P}_{\Lambda,D}\left[L_{h,s},x_j\right] \widehat{P}_{\Lambda,D}\\
&=\frac{1}{i}\left(\delta_j^s \overline{x}_h -\delta_j^h \overline{x}_s\right).
\end{split}
\end{equation*}
Furthermore, if one defines $\bm{x}^2:=\sum_h \overline{x}_h\overline{x}_h$, then the calculations of {section \ref{proofLij1}} can be used to prove that [see {section \ref{proofR^2value}} for the explicit calculations, while here the $b(l,D)$ coefficients are the ones defined in (\ref{definizioni1})]
\be\label{R^2value}
\begin{split}
\bm{x}^2 \bm{\psi}_{\bm{l},D}=&\left\{1+\frac{b(l,D)+[b(l+1,D)]\frac{l+D-2}{2l+D-2}+[b(l-1,D)]\frac{l}{2l+D-2}}{2k_D(\Lambda)}\right.\\
&\left.-\left[\left(1+\frac{b(\Lambda,D)+b(\Lambda+1,D)}{2k_D(\Lambda)} \right)\frac{\Lambda+D-2}{2\Lambda+D-2} \right]\widehat{P}_{\Lambda,D} \right\}\bm{\psi}_{\bm{l},D}.
\end{split}
\ee
In addition (here $\widetilde{P}_{h,j}$ is the projector on the eigenspace of ${C}_{D-h}$ corresponding to $l_{D-h}\equiv j$),
\begin{equation}\label{relaz_proj}
\begin{split}
\prod_{l=0}^{\Lambda}\left[\bm{L}^2-l(l+D-2)I\right]=0\quad,&\quad \prod_{l_{d-1}=0}^{l}\left[{C}_{D-1}-l_{d-1}(l_{d-1}+D-3)I\right]\widetilde{P}_{1,l}=0,\\
\cdots\quad,\quad \prod_{l_{1}=-l_2}^{l_2}\left[{L}_{1,2}-l_{1}I\right]\widetilde{P}_{D-2,l_2}=0,&\quad \left(\overline{x}_{\pm}\right)^{2\Lambda+1}=0,\mbox{ and }\left(L_{\nu,\pm}\right)^{2\Lambda+1}=0,\forall \nu\geq 3
\end{split}
\end{equation}
The relations (\ref{commx-ax-b})-(\ref{relaz_proj}) imply that the coordinate operators generate the whole algebra of observables $\mathcal{A}_{\Lambda,D}$, in fact every $L_{h,j}$ can be written in terms of $\left[\overline{x}_h,\overline{x}_j\right]$ and therefore every projector $\widetilde{P}_{h,j}$ can be written as a non-ordered polynomial in the $\overline{x}_p$.
\section{Realization of $\mathcal{A}_{\Lambda,D}$ through $U\bm{so}(D+1)$}\label{realso(D+1)}
Let $\Lambda\in\mathbb{N}$, $\pi_{\Lambda,D+1}$ be the irreducible representation of $U\bm{so}(D+1)$ having $l_D\equiv \Lambda$ and $V_{\Lambda,D+1}$ be the corresponding representation space [see (\ref{defV_lD})]. First of all, in section \ref{Y_lbasisL^2} it is shown that $\dim{\mathcal{H}_{\Lambda,D}}=\dim{V_{\Lambda,D+1}}$, and if one identifies $\bm{\psi}_{\bm{l},D}\equiv Y_{\Lambda,\bm{l}}\in V_{\Lambda,D+1}$, then the operators on $\mathcal{H}_{\Lambda,D}$, in particular $\overline{L}_{h,j}$ and $\overline{x}_h$, are naturally realized in $\pi_{\Lambda,D+1} [U\bm{so}(D+1)]$.

In fact one has [here the ${L}_{h,j}$s are seen as basis elements of $\bm{so}(D+1)$]
\be\label{defpi}
\begin{split}
\overline{L}_{h,j}={L}_{h,j}& \quad \mbox{if }h<j<D+1 \quad \mbox{and}\quad\overline{x}_h=p_D^{*}\left(\lambda\right) L_{h,D+1} p_D\left(\lambda\right),\\
&\mbox{where}\quad\lambda:=\frac{2-D+\sqrt{(D-2)^2+4\bm{L}^2}}{2}.
\end{split}
\ee
It turns out that the function $p_D$ has to fulfill
\be 
\begin{split}\label{condition_p}
p_D^{*}\left(l+1\right)p_D\left(l\right)&=\frac{1}{i}\frac{c_{l+1,D}}{d_{\Lambda,l+1,D+1}}=\frac{1}{i}\frac{\sqrt{1+\frac{b(l,D)+b(l+1,D)}{4k_D(\Lambda)}}}{\sqrt{(\Lambda-l)(\Lambda+l+D-1)}},\\
p_D^{*}\left(l-1\right)p_D\left(l\right)&=i\frac{c_{l,D}}{d_{\Lambda,l,D+1}}=i\frac {\sqrt{1+\frac{b(l,D)+b(l-1,D)}{4k_D(\Lambda)}}}{\sqrt{(\Lambda-l+1)(\Lambda+l+D-2)}};
\end{split}
\ee
it can be determined recursively, starting from $p_{D}(0):=1$ and then using the last formulas.

This means that
\begin{teorema}
Formulas (\ref{defpi}), (\ref{condition_p}) and section define an $O(D)$-equivariant $*$-algebra isomorphism between the algebra $\A_{\Lambda}=End(\Hi_{{{\Lambda}}})$  of observables (endomorphisms) on $\Hi_{{{\Lambda}}}$  and the  $C_{D+1}=\Lambda\left[\Lambda+(D+1)-2\right]$ irreducible representation $\pi_{\Lambda,D+1}$ of  $U\bm{so}(D+1)$:
\be
\begin{split}
\A_{\Lambda}&:=End(\Hi_{{{\Lambda}}})\simeq M_N(\CC)\simeq\bpi_\Lambda[U\bm{so}(D+1)],\\
\mbox{where}\quad\dim{\mathcal{H}_{\Lambda,D}}&\equiv N:\overset{(\ref{dimHLD})} =\binom{\Lambda+D-2}{\Lambda-1}\frac{2\Lambda+D-1}{\Lambda}.            
\end{split}
\ee
\end{teorema}
\medskip

As already recalled, the group of  $*$-automorphisms of $M_N(\CC)\simeq \A_{\Lambda}$ is inner and isomorphic
to $SU(N)$, i.e. of the type 
$$
a\mapsto g\, a \, g^{-1}, \qquad a\in \A_{\Lambda} ,   
$$
with $g$ an unitary $N\times N$ matrix with unit determinant. A special role is played 
by the subgroup $SO(D+1)$ acting in the representation $\bpi_\Lambda$, namely $g=\bpi_\Lambda\left[e^{i\alpha}\right]$,
where $\alpha\in \bm{so}(D+1)$.
In particular, choosing $\alpha=\alpha_{h,j}L_{h,j} $ ($\alpha_{h,j}\in\RR$ and $h<j\leq D$)  the automorphism amounts to 
  a $SO(D)\subset SO(D+1)$ transformation (a rotation in $D$-dimensional space). Parity $(L_{h,j},L_{p,D+1})\mapsto (L_{h,j},-L_{p,D+1})$, is an $O(D)\subset SO(D+1)$ transformation
with determinant  $-1$ in the $L_{p,D+1}$ space, and therefore also
in the $\bar x_p$ space. This shows that (\ref{defpi}) is equivariant
under $O(D)$, which plays the role of isometry group of this fuzzy sphere.
\section{Convergence to $O(D)$-equivariant quantum mechanics on $S^d$}\label{converge}
Here it is explained how this new fuzzy space converges to $O(D)$-equivariant quantum mechanics on the sphere $S^{d}$ as $\Lambda\to\infty$.  

The fuzzy analogs of the vector spaces $B(S^d),C(S^d)$ are defined as [see (\ref{fuzzyY}) for the explicit definition of $\widehat{Y}_{\bm{l}}$]
\be
{\cal C}_{\Lambda,D}:=span_{\mathbb{C}}\left\{\widehat{Y}_{\bm{l}}:2\Lambda\geq l\equiv l_d\geq \cdots\geq l_2\geq |l_1|,l_i\in\mathbb{Z}\forall i \right\}
\subset\A_{\Lambda,D}\subset B[{\cal L}^2(S^d)],
\label{def_CLambda3D}
\ee
and here the highest $l$ is $2\Lambda$ because $\widehat{Y}_{2\Lambda,2\Lambda,\cdots,2\Lambda}$ is the 
`highest' multiplying operator acting nontrivially on $\mathcal{H}_{\Lambda,D}$ (it does 
not annihilate $\bm{\psi}_{\Lambda,\Lambda,\cdots,-\Lambda,D}$).

So
\be
{\cal C}_{\Lambda,D}=\bigoplus\limits_{l=0}^{2\Lambda} V_{l,D}\hspace{0.1cm}\label{deco2}
\ee
is the decomposition
of ${\cal C}_{\Lambda,D}$ into irreducible components under $O(D)$; furthermore, $V_{l,D}$ is trace-free
for all $l>0$, i.e. its projection on the single component $V_{0,D}$ is zero and it is easy to see that (\ref{deco2}) becomes the
decomposition of $B(S^d),C(S^d)$ in the limit $\Lambda\to\infty$. 

In addition, the fuzzy analog of $f\in B(S^d)$ is
\be
\hat f_\Lambda:=\sum_{l=0}^{2\Lambda}
\sum_{\substack{
l_{d-1}\leq l\\
l_{h-1}\leq l_h \mbox{ for }h=d-1,\cdots,3\\
|l_1|\leq l_2
}}
f_{\bm{l}}\widehat{Y}_{\bm{l}}
\in\mathcal{A}_{\Lambda,D}\subset B[{\cal L}^2(S^2)];
\ee
while the $\bm{\psi}_{\bm{l},D}\in\mathcal{H}_{\Lambda,D}$ are the fuzzy analogs of the spherical
harmonics $Y _{\bm{l}} $  considered just as elements  of an orthonormal
basis of the Hilbert space ${\cal L}^2(S^d)$; for this reason, consider the $O(D)$-covariant embedding ${\cal I}:\mathcal{H}_{\Lambda,D}\hookrightarrow {\cal L}^2(S^d)$ 
% $\Hi_\Lambda\subset {\cal L}^2(S)$ 
defined by 
$$
{\cal I}\left(\sum_{l=0}^{\Lambda}\sum_{\substack{
l_{d-1}\leq l\\
l_{j-1}\leq l_j \mbox{ for }j=d-1,\cdots,3\\
|l_1|\leq l_2
}}
\phi_{\bm{l}} \bm{\psi}_{\bm{l},D}
\right)= 
\sum_{l=0}^{\Lambda}
\sum_{\substack{
l_{d-1}\leq l\\
l_{j-1}\leq l_j \mbox{ for }j=d-1,\cdots,3\\
|l_1|\leq l_2
}}
\phi_{\bm{l}} Y_{\bm{l}},
$$
and below the symbol ${\cal I}$ is dropped and then simply identified $\bm{\psi}_{\bm{l},D}\equiv Y_{\bm{l}}$.

The decomposition
of $\mathcal{H}_{\Lambda,D}$ into irreducible components under $O(D)$ reads
\be
\mathcal{H}_{\Lambda,D}=\bigoplus\limits_{l=0}^{\Lambda} V_l,\quad V_l:=
\left\{\sum_{\substack{
l_{d-1}\leq l\\
l_{j-1}\leq l_j \mbox{ for }j=d-1,\cdots,3\\
|l_1|\leq l_2
}}
\phi_{\bm{l}} \bm{\psi}_{\bm{l},D}: \phi_{\bm{l}}\in\CC
\right\}, \label{deco1}
\ee
and (\ref{deco1})$_1$ becomes the decomposition of ${\cal L}^2(S^d)$ in the limit $\Lambda\to\infty$. 

For all $\phi\in {\cal L}^2(S^d)$ let 
$$\phi_ \Lambda:=\sum_{l=0}^{\Lambda} \sum_{\substack{
l_{d-1}\leq l\\
l_{j-1}\leq l_j \mbox{ for }j=d-1,\cdots,3\\
|l_1|\leq l_2
}}
\phi_{\bm{l}} \bm{\psi}_{\bm{l},D},
$$
where $\phi_{\bm{l}} $ are the coefficients of the decomposition of $\phi$ in the orthonormal basis of spherical harmonics; clearly $\phi_ \Lambda\to\phi$
in the ${\cal L}^2(S^d)$-norm $\Vert\,\Vert$, and in this sense  $\mathcal{H}_{\Lambda,D}$ invades ${\cal L}^2(S^d)$
 as $\Lambda\to\infty$.

Let $B\left[\mathcal{L}^2\left(S^d\right)\right]$ be the algebra of bounded operators on $\mathcal{L}^2\left(S^d\right)$, the embedding $\mathcal{I}$ induces the one $\mathcal{J}:\mathcal{A}_{\Lambda,D}\hookrightarrow B\left[\mathcal{L}^2\left(S^d\right)\right]$ and by construction $\mathcal{A}_{\Lambda,D}$ annihilates $\mathcal{H}_{\Lambda,D}^{\perp}$; the operators $L_{h,j},\overline{L}_{h,j}$ coincide on $\mathcal{H}_{\Lambda,D}$, while one can easily check that $\overline{L}_{h,j}\to L_{h,j}$
strongly as $\Lambda\to\infty$ on the domain $D\left(L_{h,j}\right)\subset \mathcal{L}^2\left(S^d\right)$
\footnote{The strict inclusion it follows from the fact that
$L_{h,j}$ is unbounded; for example $\phi\in D(L_{1,2})$ only if 
$
\sum_{l=0}^{+\infty} \sum_{l_{d-1}\leq l,\cdots}
|l_1|^2|\phi_{\bm{l}}|^2<+\infty.
$
} and, similarly, 
$f(\overline{L}_{h,j})\to f(L_{h,j})$ strongly on $D[f(L_{h,j})]$ for all measurable function $f(s)$.

Bounded (in particular, continuous) functions $f$ on the sphere $S^d$, acting as multiplication operators 
$f\cdot:\phi\in{\cal L}^2(S^d)\mapsto f\phi\in{\cal L}^2(S^d)$, make up a subalgebra  $B(S^d)$  [resp. $C(S^d)$]
of $B\left[{\cal L}^2(S^d)\right]$. An element of $B(S^d)$ is actually an equivalence class $[f]$ of bounded functions differing from $f$ only on a set of zero measure, because this ensures that for any $f_1,f_2\in[f]$, and $\phi\in{\cal L}^2(S^d)$, $f_1\phi$ and $f_2\phi$ differ only on a set of zero measure, and therefore are two equivalent representatives of the same element of ${\cal L}^2(S^d)$.
Since $f$ belongs also to ${\cal L}^2(S^d)$, then 
$$f_N(\theta_{d},\cdots,\theta_1):=\sum_{l=0}^N \sum_{\substack{
l_{d-1}\leq l\\
l_{j-1}\leq l_j \mbox{ for }j=d-1,\cdots,3\\
|l_1|\leq l_2
}}
f_{\bm{l}} Y_{\bm{l}}\left(\theta_{d},\cdots,\theta_1\right)$$ converges to $f(\theta_{d},\cdots,\theta_1)$ in the ${\cal L}^2(S^d)$ norm as $N\to \infty$.

In {section \ref{convergproof}} it is shown that every projected coordinate operator $\overline{x}_h$ converges strongly to the corresponding $t_{h}$ as $\Lambda\to\infty$ if
$$
k_D\left(\Lambda\right)\geq \Lambda \left[\dim{\mathcal{H}_{\Lambda,D}}\right]^2b(\Lambda,D).
$$
Again, since for all $\Lambda\!>\!0$ the operator $\overline{x}_h$ annihilates $\Hi_{\Lambda,D}^\perp$, 
$\overline{x}_h$  {\it does not} converge  to $t_h$ {\it in operator norm}.
It is possible to prove also this more general result:
\begin{teorema} \label{propoD}
Choosing 
\be\label{kineq2}
k_D\left(\Lambda\right)\geq \Lambda^2 [\dim{\mathcal{H}_{2\Lambda,D}}]^3 \left[(2\Lambda)!\right]^D 2^{\Lambda D}\left[(2\Lambda+1)!!\right]^{2D}b(\Lambda,D)\sqrt{\dim{\mathcal{H}_{\Lambda,D}}},
\ee
then for all $f,g\in B(S^d)$ the following strong limits as $\Lambda\rightarrow \infty$ hold: $\hat{f}_{\Lambda}\rightarrow f\cdot,\widehat{\left(fg\right)}_{\Lambda}\rightarrow fg\cdot$ and $\hat{f}_{\Lambda}\hat{g}_{\Lambda}\rightarrow fg\cdot$. 
\end{teorema}
\noindent
In other words, the product in $\mathcal{A}_{\Lambda,D}$ between the approximations $\widehat{f}_{\Lambda}$ and $\widehat{g}_{\Lambda}$ goes to the product in $B\left[\mathcal{L}^2(S^d)\right]$ between $f\cdot$ and $g\cdot$ [although $\left(\widehat{fg} \right)_{\Lambda}\neq \widehat{f}_{\Lambda}\widehat{g}_{\Lambda}$].

\section{Conclusions, outlook and comparison with literature}\label{conclu}
The construction of the $O(D)$-equivariant fuzzy sphere in the second section has been done through the imposition of a sufficiently low (and $\Lambda$-dependent, with $\Lambda\in\mathbb{N}$) energy cutoff $\overline{E}:=\Lambda\left(\Lambda+D-2\right)$ on the quantum mechanics of a particle subject to a rotation-invariant potential $V(r)$ having a very deep minimum in $r=1$, and regulated by a confining parameter $k(\Lambda)\geq \left[\Lambda(\Lambda+D-2)\right]^2$, which expresses the sharpness of that minimum.

The output is a sequence $\left\{\mathcal{A}_{\Lambda,D}\right\}_{\Lambda\in\mathbb{N}}$ of finite-dimensional algebras. Every operator $A\in\mathcal{A}_{\Lambda,D}$ acts on the corresponding Hilbert space of admitted states $\mathcal{H}_{\Lambda,D}$, which is also finite-dimensional and can be realized using an irreducible representation of $U\bm{so}(D+1)$ (the one having $l_D\equiv \Lambda$), but also a reducible representation of $U\bm{so}(D)$; in fact it can be decomposed through the irreps of $U\bm{so}(D)$ having $0\leq l\leq\Lambda$.

The algebraic relations involving $\overline{L}_{h,j},\overline{x}_p$ are invariant under parity, as well as under any $O(D)$-transformation of the coordinates, and this was expected because of the application of a rotation-invariant energy-cutoff to a theory having the same covariance; then, as shown, the projected theory has inherited that symmetry. It is also important to underline that these relations are nothing but the generalizations, to the $D$-dimensional case, of the ones calculated for $D=2$ and $D=3$. 

The focal point is the {definition \ref{defiL}}. It is inspired by the action of a generic coordinate $t_h$ on a spherical harmonic, and it allowed to repeat (in the generic $D$-dimensional case) what was done in \cite{FiorePisacane,{FiorePisacanePOS18}}; in fact, in almost all the proof it was fundamental that the action of $L_{h,D}$ on $Y_{\bm{l}}$ coincides, more or less, with the one of the coordinate $t_h$ on $Y_{l_{d-1},\cdots,l_1}$ this is also in agreement with the Wigner-Eckart theorem, and this is also in agreement with the Wigner-Eckart theorem, because both $t_h$ and $L_{h,D}$ transform in the same way under $SO(d)$.

Another crucial point of this section is the research of all the eigenfunctions of $\bm{L}^2$ (section \ref{D-dimsa}) on $S^d$, for this reason the goal was the determination of an orthonormal basis of eigenfunctions $\left\{Y_{\bm{l}}\right\}_{\bm{l}}$ for $\bm{L}^2$ in $\mathcal{L}^2\left(S^d\right)$; this returned an orthonormal basis of $\mathcal{H}_{\Lambda, D}$ and then the subsequent possibility of calculating explicitely the action of $\overline{x}_p$ and $\overline{L}_{h,j}$ on every state $\psi$ (section \ref{Section_Algebra}).

On the other hand, every space (here $\overline{l}$ is a fixed number of $\mathbb{N}_0$)
$$
span\left\{Y_{\bm{l}}(\theta_d,\cdots,\theta_1):\overline{l}\equiv l_d\geq\cdots\geq l_2\geq |l_1|, l_i\in\mathbb{Z}\forall i \right\}
$$
is the representation space of an irrep of $\bm{so}(D)$, the one corresponding to $\bm{L}^2\equiv \overline{l}\left(\overline{l}+D-2\right)I$; it is important to underline that the Cartan subalgebra is too small to be a CSCO, which means that the sets of their eigenvalues do not univocally identify all spherical harmonics; then, in this case, one is not able to write down explicitely an orthonormal basis of $\bm{L}^2$-eigenfunctions in $\mathcal{L}^2\left(S^d\right)$ [as for (\ref{pure_psi})] and, consequently, to calculate the action of $\overline{x}_h$ and $\overline{L}_{h,j}$ on every quantum state $\psi$.

The aforementioned definition of the components ${L}_{h,j}$ of the $D$-dimensional angular momentum operator was also fundamental to realize the algebra of observables $\mathcal{A}_{\Lambda,D}$ with a suitable irreducible representation of $U\bm{so}(D+1)$. In fact, in that realization the `projected' coordinate operator $\overline{x}_h$ is identified with $L_{h,D+1}$ up to some scalar left and right factors.

Finally, I do the proof of the convergence of this new fuzzy hypersphere to quantum mechanics on $S^d$ in the commutative limit $\Lambda\rightarrow+\infty$, this was also expected because in that limit the potential $V(r)$ forces the particle to stay on the unit sphere, which (from the mathematical point of view) is represented by $c_{l,D}\rightarrow 1$, and then that every operator $\overline{x}_h$ converges to the corresponding $t_h\cdot$.

I now compare our fuzzy spheres with with other ones appeared in the literature; in \cite{Dolan:2003kq} the authors build their two fuzzy versions of $S^3$:
\begin{itemize}
\item In the first case, from $\mathbb{C}P_F^3$ they firstly obtain a fuzzy $S_F^4$ using the fact that  $\mathbb{C}P^3$ is a $S^2$ bundle over $S^4$ and that there is a well defined matrix approximation of $\mathbb{C}P^3\simeq \frac{SU(4)}{U(3)}$, then they construct $S_F^3$ from this $S_F^4$.
\item In the second case, they obtain $S_F^3$ starting from the orthogonal Grassmanian $\frac{SO(5)}{SO(3)\times SO(3)}$ and then using the existence of a well defined matrix approximation of the algebra of functions on this Grassmanian, in other words they consider fuzzy orthogonal Grassmanians.
\end{itemize} 

A well-known fuzzy $4$-sphere is built in \cite{Ramgoolam}, and it essentially coincides with \cite{Castelino}; there the author considers the Dirac $\Gamma$ matrices, which form the $4$-dimensional spin representation of $\bm{so}(5)$, and are used in the $n$-fold symmetric tensor representation of $\Gamma$ (here $Sym$ means the restriction to the completely symmetrized tensor product space)
$$
G_i^{(n)}:=\left(\Gamma_i\otimes\mathbb{I}\otimes\cdots\otimes \mathbb{I}+\mathbb{I}\otimes\Gamma_i\otimes\cdots\otimes \mathbb{I} +\cdots+\mathbb{I}\otimes\cdots\otimes \mathbb{I}\otimes \Gamma_i \right)_{Sym},
$$
for $i=1,\cdots,5$. The $G_i^{(n)}$ defined above are $N\times N$ matrices, with
$$
N=\frac{(n+1)(n+2)(n+3)}{6},
$$
and they fulfill
$$
\sum_{i} \left[G_i^{(n)} \right]^2=n(n+4)\mathbb{I}_{N}.
$$
Then, from
$$
X_i:=\frac{r}{n}G_i^{(n)}\quad\mbox{it follows}\quad 
\sum_{i}X_i^2=r^2\mathbb{I}_{N}+O\left(\frac{1}{n} \right).
$$

The representations of $Spin(5)$ [or equivalently $Sp(2)$] are considered in \cite{Medina-Connor} in order to build another fuzzy $S^4$; in particular, the irrep $\left(\frac{L}{2},\frac{L}{2}\right)$ contains the $5$ Dirac matrices $J_a$, $a=1,\cdots,5$, which can be realized as the symmetrization of $L$ copies of the $\Gamma$ matrices in the $Spin(5)$ fundamental representation:
$$
J_a:=\left(\Gamma_a\otimes\mathbb{I}\otimes\cdots\otimes \mathbb{I}+\mathbb{I}\otimes\Gamma_a\otimes\cdots\otimes \mathbb{I} +\cdots+\mathbb{I}\otimes\cdots\otimes \mathbb{I}\otimes \Gamma_a \right)_{Sym},
$$
where $Sym$ means the projection in the totally symmetrized irreducible representation.

The $J_a$ fulfill $J_aJ_a=L(L+4)\mathbb{I}$, then from $X_a:=\frac{R}{\sqrt{L(L+4)}}J_a$, it follows $X_aX_A=R^2\mathbb{I}$ and that in the limit $L\rightarrow+\infty$ the algebra becomes commutative.

In \cite{Dolan:2003th} the authors approximate the sphere $S^N\cong \frac{SO(N+1)}{SO(N)}$ starting from the cartesian co-coordinates $X^a$, the angular momentum components $L_{a,b}$ in $\mathbb{R}^{N+1}$, with $a,b\in\left\{1,2,\cdots,N+1\right\}$, and then also the $L_{A,B}$ in $\mathbb{R}^{N+2}$, with $A,B\in\left\{1,2,\cdots,N+2\right\}$. The definition $X_a:=\mu L_{a,N+2}$, with $\mu\in\mathbb{R}$, returns Snyder-type commutation relations
$$
\left[X^a,X^b\right]=-i\mu^2 L_{a,b},
$$ 
and also that (here $C_2^{N'}$ is the square angular momentum in $\mathbb{R}^{N'}$)
$$
X_aX_a=\mu^2\left[C_2^{N+2}-C_2^{N+1}\right],
$$
which is central in the fundamental spinor representation of $Spin(N+2)$
$$
X_aX_a=\frac{\mu^2(N+1)}{2}\mathbb{I}.
$$
In agreement with the above construction, Sperling and Steinacker \cite{Ste16,Ste17} build their approximation $S^4_N$ of $S^4$ with a reducible representation of $U\bm{so}(5)$ (as for the above $S_{\Lambda}^4$) on a Hilbert space $V$ obtained decomposing an irreducible representation $\pi$ of $U\bm{so}(6)$ characterized by a triple of highest weights
$(n_1,n_2,N)$; so $End(V)\simeq  \pi[U\bm{so}(6)]$, in analogy with our scheme. 
The elements $X^a:=r\mathcal{M}^{a6}$ \ play the role of noncommutative cartesian coordinates and they fulfill \emph{Snyder}-type commutation relations (as for the above $S_{\Lambda}^4$).
As a consequence the $O(5)$-scalar $\R^2=X^aX^a$ is no longer central, but its spectrum is still very close to $1$ if $N\gg n_1,n_2$ [because then the decomposition of $V$ contains few irreducible representations under $SO(5)$].

On the other hand, if $n_1=n_2=0$, the representation of $U\bm{so}(5)$ turns out to be irreducible (unlike the above $S_{\Lambda}^4$) [the highest weight is $(0,0,N)$], and one obtain the basic fuzzy $4$-sphere $S_N^4$, which is essentially the same of \cite{Ramgoolam,Dolan:2003th}, but in the case $N\equiv4$:
$$
X^aX_a=\mathcal{R}^2=\frac{1}{4}N(N+4)\mathbb{I},
$$
so the coordinates can be trivially `normalized'; furthermore, from $\bm{su}(4)\simeq \bm{so}(6)$ it follows
$$
\mathcal{H}_{\Lambda}=(0,0,N)_{\bm{su}(4)}=(0,N)_{\bm{so}(5)}.
$$
The authors fuzzy approximate the quantum mechanics on the $4$-sphere with the algebra $End\left(\mathcal{H}_N\right)$, and it fulfills
$$
End\left(\mathcal{H}_N\right)=\left(0,0,N\right)\otimes\left(N,0,0\right)=\bigoplus_{n=0}^N(n,0,n),
$$
which is its decomposition in the $\bm{su}(4)$ harmonics.

In turn, every $(n,0,n)$ decomposes in this way in the $\bm{so}(5)$ harmonics:
$$
(n,0,n)=\bigoplus_{m=0}^n(n-m,2m).
$$
So, in $End\left(\mathcal{H}_N\right)$, there are
$$
\bigoplus_{n=0}^N(n,0),
$$
which corresponds to the algebra $\mathcal{A}_{N,D}$ when $D=5$, but there are also `further modes', i.e. the representations $(n,2s)$ with $s\geq 1$, that can be seen as higher spin algebras in the Vasiliev theory.

Their physical interpretation of  $End(V)$ is that it represents a fuzzy approximation of some fiber bundle on a sphere $S^4$ 
(rather than of the algebra  of observables of a quantum particle on a $S^4$).
\subsubsection*{Acknowledgments}

I am grateful to G. Fiore and H. Steinacker for useful discussions.
\section{Appendix}\label{appendix}
\subsection{The action of $C_{\widetilde{D}}$ in $\mathbb{R}^D$ and in $\mathbb{R}^{\widetilde{D}}$, when $2\leq\widetilde{D}<D$}\label{picche_fiori}
Let $(x_1,\cdots,x_D)$ be the rectangular coordinates in $\mathbb{R}^D$ and $\left(r,\theta_{d},\cdots,\theta_1\right)$ the spherical ones:\begin{equation}\label{cambiocoord}
\begin{split}
x_1&=r\sin{\theta_{d}}\sin{\theta_{d-1}}\cdots\sin{\theta_{2}}\cos{\theta_1},\\
x_2&=r\sin{\theta_{d}}\sin{\theta_{d-1}}\cdots\sin{\theta_{2}}\sin{\theta_1},\\
x_3&=r\sin{\theta_{d}}\sin{\theta_{d-1}}\cdots\cos{\theta_{2}},\\
\vdots\\
x_{d}&=r\sin{\theta_{d}}\cos{\theta_{d-1}},\\
x_{D}&=r\cos{\theta_{d}},
\end{split}
\end{equation}
with $r\geq0$, $\theta_1\in[0,2\pi[$ and $\theta_2,\cdots,\theta_{d}\in\left[0,\pi\right]$.

First of all, in both $\mathbb{R}^{\widetilde{D}}$ and $\mathbb{R}^D$ the equality
$$
C_{\widetilde{D}}\overset{(\ref{C_p})}=\sum_{1\leq j<h\leq \widetilde{D}}L_{j,h}^2
$$
holds, but the crucial difference is that the expression of $x_p$ in polar coordinates (\ref{cambiocoord}) changes when one passes from $\mathbb{R}^{\widetilde{D}}$
$$
\widehat{x}_p:=r'\sin{\theta_{\widetilde{D}-1}}\cdots\sin{\theta_p}\cos{\theta_{p-1}},
$$
to $\mathbb{R}^D$
$$
x_p:=r\sin{\theta_d}\cdots\sin{\theta_{\widetilde{D}}}\sin{\theta_{\widetilde{D}-1}}\cdots\sin{\theta_p}\cos{\theta_{p-1}}=\frac{r}{r'}\sin{\theta_d}\cdots\sin{\theta_{\widetilde{D}}}\widehat{x}_p,
$$
where
$$
r':=\sqrt{\sum_{p=1}^{\widetilde{D}} \widehat{x}_p^2}\quad\mbox{and}\quad r:=\sqrt{\sum_{p=1}^D x_p^2}.
$$
This means that, in order to understand the difference between the action of the operator $C_{\widetilde{D}}$ in the two ambient spaces, one can focus the attention only on the differences between the action $(\clubsuit)$ of $L_{j,h}$ in $\mathbb{R}^{\widetilde{D}}$ and the one $(\spadesuit)$ in $\mathbb{R}^D$. 

According to this, if $f(x_D,\cdots,x_1)$ is a differentiable function on $\mathbb{R}^D$ and $r'=r$, then
$$
\widehat{x}_j\frac{\partial}{\partial \widehat{x}_h}f(x_D,\cdots,x_1)=\widehat{x}_j \sin{\theta_d}\cdots\sin{\theta_{\widetilde{D}}}\frac{\partial f}{\partial x_h}(x_D,\cdots,x_1)=x_j \frac{\partial f}{\partial x_h}(x_D,\cdots,x_1);
$$
which implies that the action $\clubsuit$ on the sphere $S^{\widetilde{D}-1}_r$ coincides with the one $\spadesuit$ on $S^d_r$, in particular they coincide on the corresponding unit spheres.

\subsection{About the regularity of $f(r)$ in (\ref{eqpolarD})}\label{section_reg_f}
{In the case of a second order linear ODE
\be\label{ODEordine2}
y''(z)+P(z) y'(z)+Q(z)y(z)=0,
\ee
a point $z_0\in\mathbb{C}$ is singular for the equation if $P(z)$ and $Q(z)$ have an isolated singularity at $z=z_0$; $z_0$ is a fuchsian point if $P(z)$ has a pole of order at most $1$ in $z=z_0$ and $Q(z)$ has a pole of order at most $2$ in $z=z_0$.

Fuchs theorem states that in the neighborhood of a fuchsian point every solution of (\ref{ODEordine2}) is a combination of the two independent ones having the following behavior:
$$
y_1(z)=(z-z_0)^{\alpha_1}w_1(z)\quad\mbox{and}\quad y_2(z)=(z-z_0)^{\alpha_2}w_2(z),
$$
where $\alpha_i$ are the solutions of the algebraic equation
$$
x^2+(p_0-1)x+q_0=0,
$$
$w_i(z)$ are holomorphic functions which do not vanish in $z=z_0$,
$$
p_0=\lim_{z\rightarrow z_0}(z-z_0)P(z)\quad\mbox{and}\quad q_0=\lim_{z\rightarrow z_0}(z-z_0)^2Q(z).
$$

From this last theorem, applied to (\ref{eqpolarD}) under the hypothesis
\be\label{ineqT}
\lim_{r\rightarrow 0^+}r^2 V(r)=T\in\mathbb{R}^+,
\ee
it follows
$$
p_0=D-1\quad \mbox{and}\quad q_0=-\left[l(l+D-2)+T\right],
$$
then
\begin{equation*}
\begin{split}
\alpha_1&=\frac{2-D+\sqrt{(D-2)^2+4\left[l(l+D-2)+T\right]}}{2}\overset{T> 0}\geq\frac{2-D+\sqrt{(D-2)^2}}{2}=0,\\
\alpha_2&=\frac{2-D-\sqrt{(D-2)^2+4\left[l(l+D-2)+T\right]}}{2}\overset{T>0}< 0.
\end{split}
\end{equation*}
Hence
$$
f(r)= \gamma r^{\alpha_1}w_1(r)+\delta r^{\alpha_2}w_2(r)\quad \mbox{when}\quad r\rightarrow 0;
$$
in addition, according to the self-adjointness of $H$, it must be
$$
\psi\in D(H)\equiv D(H^*)=\left\{\psi\in\mathcal{L}^2\left(\mathbb{R}^D \right):\psi\mbox{ is twice differentiable and }H\psi\in\mathcal{L}^2\left(\mathbb{R}^D\right) \right\},
$$
which implies $\delta\equiv 0$ and then $f(0)=0$.}
\subsection{The $D$-dimensional spherical harmonics}\label{D-dimsa}
In this section it is explained how to determine an orthonormal basis of $\mathcal{L}^2\left(S^d\right)$ made up of eigenfunctions $Y$ of $\bm{L}^2$ in $\mathbb{R}^D$.
\subsubsection{The resolution of $\bm{L}^2 Y=l(l+D-2)Y$ by separation of variables}
First of all, from (\ref{commutHC_p}) it follows that $L_{1,2}$ and all these $C_p$ operators can be simultaneously diagonalized; in addition, in section \ref{picche_fiori} there is the proof that $C_p$ coincides with the opposite of the Laplace-Beltrami operator $\Delta_{S^{p-1}}$ on the sphere $S^{p-1}$ in every dimension $D$, then from \cite{Morimoto} p. 21, it follows
\begin{equation}\label{decomp_Lapla}
\begin{split}
\Delta=&\frac{\partial^2}{\partial r^2}+(D-1)\frac{1}{r}\frac{\partial}{\partial r}-\frac{1}{r^2} \bm{L^2},\\
\bm{L}^2=&-(1-t^2)\frac{\partial^2}{\partial t^2}+(D-1)t\frac{\partial}{\partial t}+\frac{1}{1-t^2}C_{d},
\end{split}
\end{equation}
where $t=\cos{\theta_{d}}$.

Furthermore, when $\theta\in[0,\pi]$,
$$
\frac{\partial}{\partial \cos{\theta}}=\frac{\partial \theta}{\partial \cos{\theta}}\frac{\partial}{\partial\theta}=-\frac{1}{\sin{\theta}}\frac{\partial}{\partial \theta}
$$
and
$$
\frac{\partial^2}{\partial\cos^2{\theta}}=-\frac{1}{\sin{\theta}}\frac{\partial}{\partial\theta}\left(-\frac{1}{\sin{\theta}}\frac{\partial}{\partial\theta} \right)=\frac{1}{\sin{\theta}}\left(-\frac{\cos{\theta}}{\sin^2{\theta}}\frac{\partial}{\partial\theta}+\frac{1}{\sin{\theta}}\frac{\partial^2}{\partial \theta^2}        \right)=-\frac{\cos{\theta}}{\sin^3{\theta}}\frac{\partial}{\partial\theta}+\frac{1}{\sin^2{\theta}}\frac{\partial^2}{\partial \theta^2}.  
$$
According to this,
\begin{equation}\label{cambiovar}
\begin{split}
C_D=\bm{L}^2&=-(1-t^2)\frac{\partial^2}{\partial t^2}+(D-1)t\frac{\partial}{\partial t}+\frac{1}{1-t^2}C_{d}\\
&=-\frac{\partial^2}{\partial\theta_{d}^2}+\frac{\cos{\theta_{d}}}{\sin{\theta_{d}}}\frac{\partial}{\partial\theta_{d}}-(D-1)\frac{\cos{\theta_{d}}}{\sin{\theta_{d}}}\frac{\partial}{\partial\theta_{d}}+\frac{1}{\sin^2{\theta_{d}}}C_{d}\\
&=-\frac{\partial^2}{\partial\theta_{d}^2}-(D-2)\frac{\cos{\theta_{d}}}{\sin{\theta_{d}}}\frac{\partial}{\partial\theta_{d}}+\frac{1}{\sin^2{\theta_{d}}}C_{d}\\
&=-\frac{1}{\sin^{d-1}{\theta_{d}}}\frac{\partial}{\partial \theta_{d}}\left(\sin^{d-1}{\theta_{d}}\frac{\partial}{\partial\theta_{d}} \right)+\frac{1}{\sin^2{\theta_{d}}}C_{d}.
\end{split}
\end{equation}
The aforementioned proof of (\ref{decomp_Lapla})$_2$ and also (\ref{cambiovar}) can be trivially generalized to every dimension, which means that, when $n\in\{3,\cdots,D\}$,
\begin{equation}\label{Sum_Lapl}
\begin{split}
C_n=-\frac{1}{\sin^{n-2}{\theta_{n-1}}}&\frac{\partial}{\partial \theta_{n-1}}\left(\sin^{n-2}{\theta_{n-1}}\frac{\partial}{\partial\theta_{n-1}} \right)+\frac{1}{\sin^2{\theta_{n-1}}}C_{n-1},\\
&\mbox{while}\quad L_{1,2}=\frac{1}{i}\frac{\partial}{\partial \theta_1}\Rightarrow C_2=-\frac{\partial^2}{\partial \theta_1^2}.
\end{split}
\end{equation}

Section \ref{picche_fiori} and (\ref{Sum_Lapl}) suggest to apply a separation of variables in the resolution of $C_p Y=l_{p-1}(l_{p-1}+p-2)Y$ for $p=2,\cdots D$; then $Y=Y_1(\theta_{d},\cdots,\theta_2) g_1(\theta_1)$, (\ref{eigenvaluesC_p}), (\ref{Sum_Lapl})$_2$ and $C_2 Y=L_{1,2}^2Y=l_1^2 Y$ with $l_1\in\mathbb{Z}$ imply $g_1(\theta_1)=C e^{i l_1\theta_1}$, with $l_1\in\mathbb{Z}$. 

The constant $C$ can be fixed by requiring that
$$
\int_{0}^{2\pi}g_1 g_1^* d\theta_1=1,
$$
which implies $C=\frac{1}{\sqrt{2\pi}}$.

Furthermore
$$
C_3=-\frac{1}{\sin{\theta_{2}}}\frac{\partial}{\partial \theta_{2}}\left(\sin{\theta_{2}}\frac{\partial}{\partial\theta_{2}} \right)+\frac{1}{\sin^2{\theta_{2}}}C_{2}\quad\mbox{and}\quad C_3 Y\overset{(\ref{eigenvaluesC_p})}=l_2(l_2+1)Y,
$$
while $L_{h,j}^{\dag}=L_{h,j}$ and the fact that every operator $B:=A^{\dag} A$ has positive spectrum imply
$$
\left\langle Y,C_3 Y \right\rangle\geq \left\langle Y,C_2 Y\right\rangle\Longleftrightarrow l_2^2+l_2-l_1^2\geq0 \quad\mbox{with }l_1,l_2\in\mathbb{Z},
$$
and this is possible if and only if $l_2\geq |l_1|$.

The separation of variables 
$$
Y_1(\theta_{D-1},\cdots,\theta_2)=Y_2(\theta_{D-1},\cdots,\theta_3) g_2(\theta_2),
$$
returns
$$
l_2(l_2+1)g_2=-\frac{1}{\sin{\theta_{2}}}\frac{\partial }{\partial \theta_{2}}\left(\sin{\theta_{2}}\frac{\partial g_2}{\partial\theta_{2}} \right)+\frac{1}{\sin^2{\theta_{2}}}l_1^2 g_2,
$$
and setting $z=\cos{\theta_2}$, then
$$
\frac{\partial}{\partial \theta_2}=\frac{\partial z}{\partial \theta_2}\frac{\partial}{\partial z}=-\sqrt{1-z^2}\frac{\partial}{\partial z},
$$
so one has to solve
$$
l_2(l_2+1) g_2=\frac{\partial }{\partial z}\left(-(1-z^2)\frac{\partial g_2}{\partial z} \right)+\frac{l_1^2}{1-z^2} g_2,
$$
which is equivalent to
$$
\left[(1-z^2)\frac{\partial^2}{\partial z^2}-2z \frac{\partial}{\partial z} +l_2(l_2+1)-\frac{l_1^2}{1-z^2} \right]g_2=0.
$$
This last equation is the general Legendre differential equation (see \cite{Stegun} formula 8.1.1 p. 332) and the solution is the associated Legendre function of first kind:
$$
g_2(z)= C P_{l_2}^{l_1}(z)\Longrightarrow g_2(\theta_2)= C P_{l_2}^{l_1}(\cos{\theta_2}).
$$
The constant $C$ can be determined by requiring that
$$
|C|^2 \int_0^{\pi} P_{l_2}^{l_1}(\cos{\theta_2})\left[P_{l'_2}^{l'_1}(\cos{\theta_2})\right]^* \sin{\theta_2}d\theta_2=\delta_{l_2}^{l'_2}\delta_{l_1}^{l'_1},
$$
and after the replacement $z=\cos{\theta_2}$ it becomes
$$
|C|^2 \int_{-1}^{1} P_{l_2}^{l_1}(z)\left[P_{l'_2}^{l'_1}(z)\right]^*dz=\delta_{l_2}^{l'_2}\delta_{l_1}^{l'_1}.
$$

The equalities
\begin{equation}\label{condition_ortho_Plm}
\int_{-1}^1 P_n^m(x) P_l^m(x)dx=0 \quad (l\neq n)\quad\mbox{and}\quad\int_{-1}^1 \left[P_n^m(x)\right]^2 dx=\frac{2}{2n+1}\frac{(n+m)!}{(n-m)!}
\end{equation}
from \cite{Stegun} formulas 8.14.11, 8.14.13 p. 338 and $P_l^m(x)\in\mathbb{R}$ $\forall x\in\mathbb{R}$ imply
\be\label{constant_C}
|C|=\sqrt{\frac{2l_2+1}{2}}\sqrt{\frac{(l_2-l_1)!}{(l_2+l_1)!}},
\ee
then
$$
g_2(\cos{\theta_2})=\sqrt{\frac{2l_2+1}{2}}\sqrt{\frac{(l_2-l_1)!}{(l_2+l_1)!}}P_{l_2}^{l_1}(\cos{\theta_2}).
$$
On the other hand, $l_2\geq |l_1|$ and $l_1,l_2\in\mathbb{Z}$ imply that the formula 8.2.5 in \cite{Stegun} (here $Q_{\nu}^{\mu}$ is the associated Legendre function of second kind)
$$
P_{\nu}^{-\mu}(z)=\frac{\Gamma(\nu-\mu+1)}{\Gamma(\nu+\mu+1)}\left[P_{\nu}^{\mu}(z)-\frac{2}{\pi}e^{-i\mu\pi}\sin{(\mu\pi)}Q_{\nu}^{\mu}\right]
$$
becomes
\be\label{da+ma-m}
P_{\nu}^{\mu}(z)=\frac{(\nu+\mu)!}{(\nu-\mu)!}P_{\nu}^{-\mu}(z);
\ee
then
$$
g_2(\cos{\theta_2})=\sqrt{\frac{2l_2+1}{2}}\sqrt{\frac{(l_2+l_1)!}{(l_2-l_1)!}}P_{l_2}^{-l_1}(\cos{\theta_2}).
$$
This last procedure can be repeated for the angular variables $\theta_3,\cdots,\theta_{d}$, because (\ref{Sum_Lapl})$_1$ links every $C_n$ with $C_{n-1}$, for this reason one can now work with a generic $C_n$.

From (\ref{Sum_Lapl})$_1$,
$$
C_{n-1} Y\overset{(\ref{eigenvaluesC_p})}=l_{n-2}(l_{n-2}+n-3)Y\quad\mbox{and}\quad C_n Y\overset{(\ref{eigenvaluesC_p})}=l_{n-1}(l_{n-1}+n-2)Y
$$
it follows $l_{n-1}\geq l_{n-2}$ and
$$
l_{n-1}(l_{n-1}+n-2) g_{n-1}=\left[-\frac{\partial^2}{\partial \theta_{n-1}^2}-(n-2)\frac{\cos{\theta_{n-1}}}{\sin{\theta_{n-1}}}\frac{\partial}{\partial \theta_{n-1}}+\frac{l_{n-2}(l_{n-2}+n-3)}{\sin^2{\theta_{n-1}}}\right]g_{n-1}.
$$
The replacement $z=\cos{\theta_{n-1}}$ implies
$$
\frac{\partial}{\partial \theta_{n-1}}=\frac{\partial z}{\partial \theta_{n-1}}\frac{\partial}{\partial z}=-\sqrt{1-z^2}\frac{\partial}{\partial z},
$$
and then the last ODE becomes
$$
l_{n-1}(l_{n-1}+n-2) g_{n-1}=\frac{\partial }{\partial z}\left[-(1-z^2)\frac{\partial g_{n-1}}{\partial z} \right]+\frac{l_{n-2}(l_{n-2}+n-3)}{1-z^2} g_{n-1},
$$
which is equivalent to
$$
\left[(1-z^2)\frac{\partial^2}{\partial z^2}-(n-1)z \frac{\partial}{\partial z} +l_{n-1}(l_{n-1}+n-2)-\frac{l_{n-2}(l_{n-2}+n-3)}{1-z^2} \right]g_{n-1}=0.
$$
Assume that
\be\label{dagadf}
g_{n-1}(z)=\left(1-z^2\right)^{\frac{3-n}{4}}f_{n-1}(z)
\ee
and that the function $f_{n-1}(z)$ (which is determined in the following lines) has a zero in $z=1$ of order higher than $\frac{n-3}{4}$ (see section \ref{GALF} for the proof of this); then
$$
g_{n-1}'(z)=\frac{z(n-3)}{2}\left(1-z^2\right)^{-\frac{1+n}{4}}f_{n-1}(z)+\left(1-z^2\right)^{\frac{3-n}{4}}f'_{n-1}(z)
$$
and
\begin{equation*}
\begin{split}
g_{n-1}''(z)=&\frac{n-3}{2}\left(1-z^2\right)^{-\frac{1+n}{4}}f_{n-1}(z)+\frac{z(n-3)}{2}\frac{z(n+1)}{2}\left(1-z^2\right)^{-\frac{5+n}{4}}f_{n-1}(z)\\
&+z(n-3)\left(1-z^2\right)^{-\frac{1+n}{4}}f'_{n-1}(z)+\left(1-z^2\right)^{\frac{3-n}{4}}f''_{n-1}(z),
\end{split}
\end{equation*}
which implies
$$
-(n-1)zg_{n-1}'(z)=-(n-1)z\left[\frac{z(n-3)}{2}\left(1-z^2\right)^{-\frac{1+n}{4}}f_{n-1}(z)+\left(1-z^2\right)^{\frac{3-n}{4}}f'_{n-1}(z)\right]
$$
$$
=\left(1-z^2\right)^{\frac{3-n}{4}}\left[-\frac{z(n^2-4n+3)}{2(1-z^2)}f_{n-1}(z)-(n-1)z f'_{n-1}(z) \right]
$$
and, similarly,
\begin{equation*}
\begin{split}
(1-z^2)g''_{n-1}=\left(1-z^2\right)^{\frac{3-n}{4}}\left[\frac{n-3}{2}f_{n-1}(z)+\frac{z^2(n^2-2n-3)}{4(1-z^2)}f_{n-1}(z)\right.\\
\left.+z(n-3)f'_{n-1}(z)+(1-z^2)f''_{n-1}(z) \right].
\end{split}
\end{equation*}
Furthermore,
\begin{equation*}
\begin{split}
&-\frac{l_{n-2}(l_{n-2}+n-3)}{1-z^2}+\frac{z^2(n^2-2n+3)}{4(1-z^2)}-\frac{z^2(n^2-4n+3)}{2(1-z^2)}\\
&=-\frac{l_{n-2}(l_{n-2}+n-3)}{1-z^2}+\frac{z^2(n^2-2n-3-2n^2+8n-6)}{4(1-z^2)}\\
&=-\frac{l_{n-2}(l_{n-2}+n-3)}{1-z^2}+\frac{z^2(-n^2+6n-9)}{4(1-z^2)}\\
&=\frac{1}{4}(n^2-6n+9)-\frac{l_{n-2}(l_{n-2}+n-3)+\frac{1}{4}(n^2-6n+9)}{1-z^2}.
\end{split}
\end{equation*}
At this point, the first term of the ODE for $f_{n-1}$ [after deleting the common factor $\left(1-z^2\right)^{\frac{3-n}{4}}$] is
$$
(1-z^2)f''_{n-1}(z),
$$
the second term is
$$
z(n-3)f'_{n-1}(z)-z(n-1)f'_{n-1}(z)=-2zf'_{n-1}(z),
$$
the third term is
\begin{equation*}
\begin{split}
&l_{n-1}(l_{n-1}+n-2)f_{n-1}(z)+\frac{n-3}{2}f_{n-1}(z)+\frac{n^2-6n+9}{4}f_{n-1}(z)\\
=&\left(l_{n-1}^2+l_{n-1}n-2l_{n-1}+\frac{n^2-4n+3}{4} \right)f_{n-1}(z)\\
=&\left(l_{n-1}+\frac{n-3}{2}\right)\left(l_{n-1}+\frac{n-3}{2}+1\right)f_{n-1}(z)\\
=&l'(l'+1)f_{n-1}(z),
\end{split}
\end{equation*}
with $l':=l_{n-1}+\frac{n-3}{2}$,
and the last term is
$$
\frac{l_{n-2}^2+l_{n-2}n-3l_{n-2}+n^2-6n+9}{1-z^2}f_{n-1}(z)=\frac{\left(l_{n-2}+\frac{n-3}{2} \right)^2}{1-z^2}f_{n-1}(z)=\frac{\left(m'\right)^2}{1-z^2}f_{n-1}(z),
$$
with $m':=l_{n-2}+\frac{n-3}{2}$.

This means that there is another associated Legendre equation:
$$
\left[(1-z^2)\frac{\partial^2}{\partial z^2}-2z \frac{\partial}{\partial z} +l'(l'+1)-\frac{\left(m'\right)^2}{1-z^2} \right]f_{n-1}(z)=0,
$$
and then the solution is [here the constant $C$ is fixed as done in (\ref{constant_C})]
\begin{equation*}
\begin{split}
f_{n-1}(\cos{\theta_{n-1}})&=\sqrt{\frac{2l'+1}{2}}\sqrt{\frac{(l'-m')!}{(l'+m')!}}P_{l'}^{m'}(\cos{\theta_{n-1}})\\
&=\sqrt{\frac{2l_{n-1}+n-2}{2}}\sqrt{\frac{(l_{n-1}-l_{n-2})!}{(l_{n-1}+l_{n-2}+n-3)!}}P_{l_{n-1}+\frac{n-3}{2}}^{l_{n-2}+\frac{n-3}{2}}(\cos{\theta_{n-1}}),
\end{split}
\end{equation*}
which implies
$$
g_{n-1}(\cos{\theta_{n-1}})=\sqrt{\frac{2l_{n-1}+n-2}{2}}\sqrt{\frac{(l_{n-1}-l_{n-2})!}{(l_{n-1}+l_{n-2}+n-3)!}}\left[\sin{\theta_{n-1}}\right]^{\frac{3-n}{2}}P_{l_{n-1}+\frac{n-3}{2}}^{l_{n-2}+\frac{n-3}{2}}(\cos{\theta_{n-1}}).
$$
It is obvious that $l'+m',l'-m'\in\mathbb{N}_0$, so
$$
g_{n-1}(\cos{\theta_{n-1}})\overset{(\ref{da+ma-m})}=\sqrt{\frac{2l_{n-1}+n-2}{2}}\sqrt{\frac{(l_{n-1}+l_{n-2}+n-3)!}{(l_{n-1}-l_{n-2})!}}\left[\sin{\theta_{n-1}}\right]^{\frac{3-n}{2}}P_{l_{n-1}+\frac{n-3}{2}}^{-\left(l_{n-2}+\frac{n-3}{2}\right)}(\cos{\theta_{n-1}}).
$$
Summarizing,
\be\label{Y_final}
Y_{\bm{l}}(\theta_{d},\cdots,\theta_2,\theta_1)=\frac{e^{il_1\theta_1}}{\sqrt{2\pi}}\left[\prod_{n=2}^{D-1}{}_n \overline{P}_{l_{n}}^{l_{n-1}}(\theta_n) \right],
\ee
where
\be\label{Poverline}
{}_j \overline{P}_{L}^{M}(\theta):=\sqrt{\frac{2L+j-1}{2}}\sqrt{\frac{(L+M+j-2)!}{(L-M)!}}\left[\sin{\theta}\right]^{\frac{2-j}{2}}P_{L+\frac{j-2}{2}}^{-\left(M+\frac{j-2}{2}\right)}(\cos{\theta}).
\ee

\subsubsection{The orthonormality of $Y_{\bm{l}}$}
The $Y_{\bm{l}}$ built above are eigenvectors of self-adjoint operators, so
$$
\int_{S^d} Y_{\bm{l}}Y^*_{\bm{l'}} d\alpha=0\quad\mbox{if}\quad \bm{l}\neq \bm{l'};
$$
in order to prove (\ref{ArmSfe})$_3$ it remains to show that
$$
\int_{S^d} Y_{\bm{l}}Y^*_{\bm{l}} d\alpha=1,
$$
and this is done recursively.

First of all, it is important to underline that from (\ref{Y_final}-\ref{Poverline}) it follows
$$
\int_{S^2}Y_{\bm{l}}Y_{\bm{l'}}^*d\alpha=\int_{0}^{2\pi}\frac{e^{i(l_1-l_1')\theta_1}}{\sqrt{2\pi}}d\theta_1\cdot\left[\prod_{n=2}^{d}\int_0^{\pi}{}_n \overline{P}_{l_{n}}^{l_{n-1}}(\theta_n){}_n \overline{P}_{l'_{n}}^{l'_{n-1}}(\theta_n)\sin^{n-1}{\theta_n}d\theta_n \right].
$$
While from
$$
\int_{0}^{2\pi}\frac{e^{i(l_1-l_1')\theta_1}}{{2\pi}}d\theta_1=\delta_{l_1}^{l_1'},
$$
it follows that, if $l_1\neq l'_1$, then (\ref{ArmSfe})$_3$ vanishes; otherwise, if $l_1=l'_1\geq 0$ (the case $l_1<0$ is essentially the same), then
\begin{equation*}
\begin{split}
\int_0^{\pi}{}_2 \overline{P}_{l_{2}}^{l_{1}}(\theta_2){}_2 \overline{P}_{l'_{2}}^{l_{1}}(\theta_2)\sin{\theta_2}d\theta_2\overset{(\ref{Poverline})}=&\sqrt{\frac{2l_2+1}{2}\frac{(l_2+l_1)!}{(l_2-l_1)!}}\sqrt{\frac{2l'_2+1}{2}\frac{(l'_2+l_1)!}{(l'_2-l_1)!}}\int_0^{\pi}P_{l_{2}}^{-l_{1}}(\theta_2){}{P}_{l'_{2}}^{-l_{1}}(\theta_2)\sin{\theta_2}d\theta_2\\
\overset{(\ref{da+ma-m})}=&\sqrt{\frac{2l_2+1}{2}\frac{(l_2-l_1)!}{(l_2+l_1)!}}\sqrt{\frac{2l'_2+1}{2}\frac{(l'_2-l_1)!}{(l'_2+l_1)!}}\int_0^{\pi}P_{l_{2}}^{l_{1}}(\theta_2){}{P}_{l'_{2}}^{l_{1}}(\theta_2)\sin{\theta_2}d\theta_2\\
\overset{x=\cos{\theta_2}}=&\sqrt{\frac{2l_2+1}{2}\frac{(l_2-l_1)!}{(l_2+l_1)!}}\sqrt{\frac{2l'_2+1}{2}\frac{(l'_2-l_1)!}{(l'_2+l_1)!}}\int_{-1}^{1} P_{l_{2}}^{l_{1}}(x){P}_{l'_{2}}^{l_{1}}(x)dx\\
\overset{(\ref{condition_ortho_Plm})}=\delta_{l_2}^{l'_2}
\end{split}
\end{equation*}
and if $l_2\neq l'_2$, then (\ref{ArmSfe})$_3$ vanishes.

In general, if $l_i=l'_i$ for $i\in\{1,\cdots,n-1\}$,
\begin{equation*}
\begin{split}
\int_0^{\pi}{}_2 \overline{P}_{l_{n}}^{l_{n-1}}(\theta_n){}_2 \overline{P}_{l'_{n}}^{l_{n-1}}(\theta_n)\sin{\theta_n}d\theta_n\overset{(\ref{Poverline})}=&\sqrt{\frac{2l_n+n-1}{2}\frac{(l_n+l_{n-1}+n-2)!}{(l_n-l_{n-1})!}}\sqrt{\frac{2l'_n+1}{2}\frac{(l'_n+l_{n-1}+n-2)!}{(l'_n-l_{n-1})!}}\\
&\cdot \int_0^{\pi}P_{l_{n}+\frac{n-2}{2}}^{-\left(l_{n-1}+\frac{n-2}{2}\right)}(\theta_n){}{P}_{l'_{n}+\frac{n-2}{2}}^{-\left(l_{n-1}+\frac{n-2}{2}\right)}(\theta_n)\sin{\theta_n}d\theta_n\\
\overset{(\ref{da+ma-m})}=&\sqrt{\frac{2l_n+n-1}{2}\frac{(l_n-l_{n-1})!}{(l_n+l_{n-1}+n-2)!}}\sqrt{\frac{2l'_n+1}{2}\frac{(l'_n-l_{n-1})!}{(l'_n+l_{n-1}+n-2)!}}\\
&\cdot \int_0^{\pi}P_{l_{n}+\frac{n-2}{2}}^{-\left(l_{n-1}+\frac{n-2}{2}\right)}(\theta_n){}{P}_{l'_{n}+\frac{n-2}{2}}^{-\left(l_{n-1}+\frac{n-2}{2}\right)}(\theta_n)\sin{\theta_n}d\theta_n\\
\overset{x=\cos{\theta_n}}=&\sqrt{\frac{2l_n+n-1}{2}\frac{(l_n-l_{n-1})!}{(l_n+l_{n-1}+n-2)!}}\sqrt{\frac{2l'_n+1}{2}\frac{(l'_n-l_{n-1})!}{(l'_n+l_{n-1}+n-2)!}}\\
&\cdot \int_{-1}^{1} P_{l_{n}+\frac{n-2}{2}}^{-\left(l_{n-1}+\frac{n-2}{2}\right)}(x){P}_{l'_{n}+\frac{n-2}{2}}^{-\left(l_{n-1}+\frac{n-2}{2}\right)}(x)dx\\
\overset{(\ref{condition_ortho_Plm})}=\delta_{l_n}^{l'_n};
\end{split}
\end{equation*}
and this proves (\ref{ArmSfe})$_3$.

\subsubsection{The $Y_{\bm{l}}$ seen as homogeneous polynomials}\label{homog_Y_l}
The next proposition uses the equality (see section \ref{GALF} for its proof)
\be\label{nuovascritturaPlm}
{}_h\overline{P}_{l}^{m}(\theta)=\left(\sin{\theta}\right)^m \widetilde{P}_{l+\frac{h-2}{2}}^{-\left(m+\frac{h-2}{2}\right)}\left(\cos{\theta}\right) \equiv \left(\sin{\theta}\right)^{m}\left\{\left[\cos{\theta}\right]^{l-m}+ \left[\cos{\theta}\right]^{l-m-2}+\cdots \right\},
\ee
which is true up to any multiplicative constant before every power of $\cos{\theta}$ and $\sin{\theta}$.
\begin{propo}
Every $D$-dimensional spherical harmonic $Y_{\bm{l}}$ can be written as a homogeneous polynomial of degree $l$ in the $t^h$ variables.
\proof
This proof is given by induction over $D$; if $D=3$ and $m\geq 0$ [the assumption $m<0$ is essentially equivalent, because of (\ref{da+ma-m})], then (\ref{nuovascritturaPlm}) implies
$$
{}_2\overline{P}_{l}^m\left(\theta_2\right)=\left(\sin{\theta_2}\right)^m\left[
\left(\cos{\theta_2}\right)^{l-m}+ \left(\cos{\theta_2}\right)^{l-m-2}+ \left(\cos{\theta_2}\right)^{l-m-4}
+\cdots
\right],
$$
so
\begin{equation*}
\begin{split}
&Y_l^m\left(\theta_2,\theta_1\right)={}_2\overline{P}_{l}^m\left(\theta_2\right)e^{im\theta_1}\\
&=\left(t^1+i t^2\right)^m\left[
\left(t^3\right)^{l-m}+ \left(t^3\right)^{l-m-2}\left(t^1t^1+t^2t^2+t^3t^3\right)+ \left(t^3\right)^{l-m-4}\left(t^1t^1+t^2t^2+t^3t^3\right)^2+\cdots
\right];
\end{split}
\end{equation*}
which means that the claim is true for $D=3$.

Let $D>3$ and assume that the claim is true for $D-1$, then there exists  $\widehat{P}_{l_{d-1},\cdots,l_1}$, a suitable homogeneous polynomial of degree $l_{d-1}$ in the $t_1,\cdots,t_{d}$ variables, such that 
$$
Y_{l_{d-1},\cdots,l_1}=\prod_{h=2}^{d-1}{}_h\overline{P}_{l_h}^{l_{h-1}}\left(\theta_h\right)\cdot e^{il_1\theta_1}=\widehat{P}_{l_{d-1},\cdots,l_1}\left(t_1,\cdots,t_{d}\right).
$$
On the other hand, the polar system of coordinates (\ref{cambiocoord}) depends on the dimension of the carrier space, and then, in $\mathbb{R}^D$,
$$
\prod_{h=2}^{d-1}{}_h\overline{P}_{l_h}^{l_{h-1}}\left(\theta_h\right)\cdot e^{il_1\theta_1}=\left(\sin{\theta_{d}} \right)^{l_{d-1}}\widehat{P}_{l_{d-1},\cdots,l_1}\left(t_1,\cdots,t_{d}\right),
$$
for the same $\widehat{P}$.

This,
\begin{equation*}
\begin{split}
{}_d\overline{P}_{l_d}^{l_{d-1}}\left(\theta_d\right)=\left(\sin{\theta_d}\right)^{l_{d-1}}&\left[
\left(\cos{\theta_d}\right)^{l-l_{d-1}}+ \left(\cos{\theta_d}\right)^{l-l_{d-1}-2}+ \left(\cos{\theta_d}\right)^{l-l_{d-1}-4}
+\cdots
\right]\\
=\left(\sin{\theta_d}\right)^{l_{d-1}}&\left[\left(t^D\right)^{l-l_{d-1}}+ \left(t^D\right)^{l-l_{D-1}-2}\left(t^1t^1+\cdots+t^Dt^D\right) \right.\\
&\left.+\left(t^D\right)^{l-l_{d-1}-4}\left(t^1t^1+\cdots+t^Dt^D\right)^2+\cdots \right]
\end{split}
\end{equation*}
and (\ref{explicitY}) imply the claim.
\endproof
\end{propo}
\subsubsection{The $Y_{\bm{l}}$ are a basis of $\mathcal{L}^2\left(S^d\right)$}\label{Y_lbasisL^2}
Let 
\begin{itemize}
\item $\mathcal{P}_l^D$ be the vector space of polynomials in the $x_1,\cdots,x_D$ variables of degree at most $l$;
\item $\mathcal{Q}_l^D$ be the vector space of homogeneous polynomials in the $x_1,\cdots,x_D$ variables of degree $l$;
\item $\mathcal{T}_l^D$ be the vector space of homogeneous harmonic polynomials in the $x_1,\cdots,x_D$ variables of degree $l$ (the $q\in\mathcal{Q}_l^D$  fulfilling $\Delta q=0$);
\item $\widetilde{\mathcal{P}}_l^D$, $\widetilde{\mathcal{Q}}_l^D$ and $\widetilde{\mathcal{T}}_l^D$ be the restriction to the sphere $S^d$ of $\mathcal{P}_l^D$, $\mathcal{Q}_l^D$ and $\mathcal{T}_l^D$, respectively;
\item \be\label{decompVVtilde}
\Omega_{l,D}:=\bigoplus_{m=0}^l \widetilde{\mathcal{T}}_m^D,\quad \widehat{\Omega}_{l,D}:=\bigoplus_{m=0}^l V_{m,D}.
\ee
\end{itemize}
The goal is to show that
\be\label{goalbasisL2}
\forall f\in\mathcal{L}^2\left(S^d\right),\forall \varepsilon>0\exists l\in\mathbb{N}_0,\exists g\in \widehat{\Omega}_{l,D}\quad\mbox{such that}\quad \left\|f-g\right\|_2<\varepsilon.
\ee
The density of $C^0\left(S^d\right)$ in $\mathcal{L}^2\left(S^d\right)$ implies that it is sufficient to show (\ref{goalbasisL2}) for a generic continuous function on the unit sphere; on the other hand, from the Stone-Weierstrass theorem it follows that the function $f$ can be replaced with a polynomial, without loss of generality. According to this and
$$
\mathcal{P}_{l}^D=\bigoplus_{m=0}^{l}\mathcal{Q}_m^D
$$
it remains to show that every homogeneous polynomial can be approximated by the harmonic homogeneous ones, and then that $\Omega_{l,D}\equiv \widehat{\Omega}_{l,D}$ $\forall l,D$.

In order to do this, let
$$
L:p(x_1,\cdots,x_D)\in \mathcal{P}_l^D\longrightarrow \left(x_1^2+\cdots+x_D^2\right)p(x_1,\cdots,x_D) \mathcal{P}_{l+2}^D,
$$
and define in this way an internal product in $\mathcal{P}_l^D$:
$$
\left\langle x_1^{n_1}\cdots x_D^{n_D},x_1^{m_1}\cdots x_D^{m_D}\right\rangle_l:=(n_1)!\cdots (n_D)!\quad\mbox{if}\quad n_1=m_1,\cdots,n_D=m_D,
$$
$0$ otherwise.

$L$ is linear and
$$
\left\langle L[p],q\right\rangle_{l+2}=\left\langle p,\Delta q\right\rangle_l \quad\forall p\in \mathcal{P}_l^D\quad\mbox{and}\quad q\in \mathcal{P}_{l+2}^D,
$$
which means that $L^*=\Delta$; then
$$
\mathcal{Q}_{l+2}^D=L\left(\mathcal{Q}_{l}^D\right)\oplus Ker\left(L^*\right)=r^2 \mathcal{Q}_{l}^D\oplus \mathcal{T}_{l+2}^D.
$$
This implies (the dimension of $\mathcal{Q}_{l}^D $ is the the number of ways to sample $l$ elements from a set of $D$ elements allowing for duplicates, but disregarding different orderings)
\begin{equation}\label{dimH}
\begin{split}
\dim\left(\mathcal{T}_{l}^D\right)&=\dim\left(\mathcal{Q}_{l}^D \right)-\dim\left(\mathcal{Q}_{l-2}^D \right)\\
&={{D+l-1}\choose{l}}-{{D+l-3}\choose{l-2}}={{D+l-3}\choose{l-2}}\frac{(D+2l-2)(D-1)}{l(l-1)}\\
&={{D+l-3}\choose{l-1}}\frac{D+2l-2}{l},
\end{split}
\end{equation}
and also that
$$
\mathcal{P}_{l}^D=\mathcal{T}_{l}^D\oplus r^2 \mathcal{T}_{l-2}^D\oplus r^4 \mathcal{T}_{l-4}^D\oplus\cdots\quad\Longrightarrow\quad \widetilde{\mathcal{Q}}_{l}^D=\widetilde{\mathcal{T}}_{l}^D\oplus \widetilde{\mathcal{T}}_{l-2}^D\oplus \widetilde{\mathcal{T}}_{l-4}^D\oplus\cdots,
$$
in other words, every homogeneous polynomial on the sphere is a linear combination of homogeneous harmonic polynomials.

Furthermore,
$$
h\in\mathcal{T}_l^D\quad\Longrightarrow\quad  p=r^l q,\quad \Delta p=0\quad\mbox{and with}\quad q\in\widetilde{\mathcal{T}}_l^D;
$$
this and (\ref{decomp_Lapla})$_1$ imply
$$
\bm{L}^2 q=l(l+D-2)q.
$$

Then, both $q\in\widetilde{\mathcal{T}}_l^D$ and $Y_{\bm{l}}$ are eigenfunctions of $\bm{L}^2$ with eigenvalue $l(l+D-2)$ and homogeneous polynomials in the $t^h$ variables of degree $l$; this and (\ref{decompVVtilde}) imply that $\Omega_{l,D}\equiv \widehat{\Omega}_{l,D}$ $\forall l,D$ is equivalent to the proof of the following
\begin{teorema}
\be\label{H=Y}
\widetilde{\mathcal{T}}_{\overline{l}}^D=V_{\overline{l},D}\quad\forall\overline{l}\in\mathbb{N}_0,\forall D\in\mathbb{N}.
\ee
\proof
This proof is by induction on the dimension $D$ of the carrier space $\mathbb{R}^D$. When $D=3$,
$$
\dim{\left(\widetilde{\mathcal{T}}_{\overline{l}}^3 \right)}=\dim{\left({\mathcal{T}}_{\overline{l}}^3 \right)}\overset{(\ref{dimH})}={{\overline{l}}\choose{\overline{l}-1}}\frac{2\overline{l}+1}{\overline{l}}=2\overline{l}+1,
$$
and
$$
V_{\overline{l},D}\overset{(\ref{defV_lD})}=span\left\{Y_{\bm{l}}:\overline{l}\equiv l\geq l_{d-1}\geq\cdots\geq l_2\geq \vert l_1\vert, l_i\in\mathbb{Z}\forall i \right\}=span\left\{Y_{\overline{l}}^m:|m|\leq \overline{l},m\in\mathbb{Z}\right\},
$$
so (\ref{H=Y}) is true when $D=3$.

Assume that it is true for $d$, this means that
\be\label{H=Y_D-1} 
\dim{V_{l_{d-1},d}}={{D+l_{d-1}-4}\choose{l_{d-1}-1}}\frac{D+2l_{d-1}-3}{l_{d-1}};
\ee
this, the hockey stick identity (see \cite{hockey} formula (2))
\begin{equation}\label{hockey}
\begin{split}
{{n+1}\choose {r+1}  }&=\sum_{i=r}^n {{i}\choose {r}}=\sum_{i=r}^n\frac{i!}{(i-r)! r!}\overset{m=i-r}=\sum_{m=0}^{n-r}\frac{(m+r)!}{m! r!}=\sum_{m=0}^{n-r}{{m+r}\choose m}\\
&\overset{n=a+r}\Longrightarrow \sum_{m=0}^a {{m+r}\choose {m}}={{a+r+1}\choose {r+1}}
\end{split}
\end{equation}
and (\ref{Y_final}) imply
\begin{equation*}
\begin{split}
\dim{V_{\overline{l},D}}=&\sum_{m=0}^{\overline{l}}\dim{V_{m,d}}\overset{(\ref{H=Y_D-1})}=\sum_{m=0}^{\overline{l}}{{D+m-4}\choose{m-1}}\frac{D+2m-3}{m}\\
=&\sum_{m=0}^{\overline{l}}\frac{(D+m-4)!}{(D-3)!(m-1)!}\frac{D+2m-3}{m}=\sum_{m=0}^{\overline{l}}\frac{(D+m-4)!}{(D-4)! m!} +2\sum_{m=0}^{\overline{l}}m\frac{(D+m-4)!}{(D-3)!m!}\\
=&\sum_{m=0}^{\overline{l}}{{D-4+m}\choose {m}}+2\sum_{m=1}^{\overline{l}}{{D-4+m}\choose{m-1}}=\sum_{m=0}^{\overline{l}}{{D-4+m}\choose {m}}+2\sum_{n=0}^{\overline{l}-1}{{D-3+n}\choose{n}}\\
\overset{(\ref{hockey})}=&{{\overline{l}+D-4+1}\choose {D-4+1}}+2{{\overline{l}-1+D-3+1}\choose {D-3+1}}={{D+\overline{l}-3}\choose {D-3}}+2{{D+\overline{l}-3}\choose {D-2}}\\
&=\frac{(D+\overline{l}-3)!}{(D-3)!\overline{l}!}+2\frac{(D+\overline{l}-3)!}{(D-2)!(\overline{l}-1)!}={{D+\overline{l}-3}\choose {\overline{l}-1}}\frac{2\overline{l}+D-2}{\overline{l}}=\dim{\widetilde{\mathcal{T}}_{\overline{l}}^D },
\end{split}
\end{equation*}
so the proof is finished.
\endproof
\end{teorema}
According to this last proof
\begin{itemize}
\item The spherical harmonics $Y_{\bm{l}}$ are the harmonic homogeneous polynomials on the unit sphere $S^d$;
\item The spherical harmonics are an orhonormal basis of $\mathcal{L}^2\left(S^d\right)$;
\item The collection of operators $\left\{L_{1,2},C_2,\cdots,C_{D}\right\}$ is a CSCO for the $\bm{L}^2$-eigenfunctions in $\mathcal{L}^2\left(S^d\right)$;
\item The dimension of $\mathcal{H}_{\Lambda,D}$ coincides with the one of $\mathcal{T}_{\Lambda}^{D+1}$ (and then also with the one of $V_{\Lambda,D+1}$), so
\be\label{dimHLD}
\dim \mathcal{H}_{\Lambda,D}={{D+\Lambda-2}\choose{\Lambda-1}}\frac{D+2\Lambda-1}{\Lambda}.
\ee
\item Every $V_{l,D}$ is the representation space of a $SO(D)$-irrep, the one with $\bm{L}^2\equiv l(l+D-2) I$, and
$$
V_{l,D}\quad\mbox{is isomorphic to}\quad \bigoplus_{m=0}^{l}V_{m,d}.
$$
\end{itemize}
\subsection{The associated Legendre function of first kind}\label{GALF}
In this section $L,l,h\in\mathbb{N}_0$ and the behavior of ${}_h\overline{P}_L^{l}\left(\theta\right)$ is investigated, in order to prove the regularity of $g_{n-1}(z)$ in (\ref{dagadf}).

The equations (\ref{da+ma-m}) and (\ref{Poverline}) imply that ${}_h\overline{P}_L^{l}\left(\theta\right)$ basically coincides (up to a multiplicative constant, depending on $L,l$ and $h$) with
$$
\left[\sin{\theta}\right]^{\frac{2-h}{2}}P_{L+\frac{h-2}{2}}^{l+\frac{h-2}{2}}\left(\cos{\theta}\right)
$$
where $P_r^s$ is the associated Legendre function of first kind, $L+\frac{h-2}{2}$ (and also $l+\frac{h-2}{2}$) is integer if and only if $h$ is even, while it is half-integer if and only if $h$ is odd; according to this, one has to analyze the following two cases.
\subsubsection{The case $h$ even} 
When $h$ is even, then from eq. (6) pag. 148 and eq. (17) pag. 151 in \cite{Bateman}
\begin{equation}\label{Rodrigues}
P_l^m(x)=(-1)^m(1-x^2)^{\frac{m}{2}}\frac{d^m P_l(x)}{dx^m},\quad P_l(x)=\frac{1}{2^l l!}\frac{d^l}{dx^l}(x^2-1)^l
\end{equation}
it follows (in the next equations there is not any multiplicative constant, depending on the indices of $P$, because they are not relevant in this case, except when that constant is $0$)
$$
P_{L+\frac{h-2}{2}}^{-(l+\frac{h-2}{2})}\left(\cos{\theta}\right)=\left(\sin{\theta}\right)^{l+\frac{h-2}{2}}\widetilde{P}_{L+\frac{h-2}{2}}^{l+\frac{h-2}{2}}\left(\cos{\theta}\right),
$$
where $\widetilde{P}_{L+\frac{h-2}{2}}^{l+\frac{h-2}{2}}\left(\cos{\theta}\right)$ is a polynomial of degree $L-l$ in $\cos{\theta}$ which does not contain any terms of degree $L-l-(2n+1)$, with $n\in\mathbb{N}_0$; so,
$$
{}_h\overline{P}_L^{l}\left(\theta\right)= \left(\sin{\theta}\right)^l \widetilde{P}_{L+\frac{h-2}{2}}^{l+\frac{h-2}{2}}\left(\cos{\theta}\right).
$$
{In addition, from (\ref{da+ma-m}) and (\ref{Rodrigues}) it follows that the highest coefficient multiplying a power of $\cos{\theta}$ in $P_{L}^{l}\left(\cos{\theta}\right)$, when $L\geq |l|$ and $L,l\in\mathbb{Z}$ is
$$
\frac{(2L)!}{2^L L!}\overset{L \leq \Lambda}\leq \frac{(2\Lambda)!}{2^{\Lambda} \Lambda!}< 2^{\Lambda}\left[(2\Lambda+1)!!\right]^2.
$$}
\subsubsection{The case $h$ odd}
In  {\cite{Bateman} eq. (7) pag. 122} there is another explicit expression of the associated Legendre function of first kind (pay attention to the different fonts $P$ and $\mathrm{P}$, while here $\bm{F}$ is the Gauss hypergeometric function)
\be\label{defassleg}
\begin{split}
P_L^{l}(z)&=\frac{2^l}{\Gamma(1-l)}{\left(\frac{z+1}{z-1}\right)^{\frac{l}{2}}}\bm{F}\left(1-l+L,-l-L,1-l,\frac{1-z}{2}\right)\\
&=\frac{2^l}{\Gamma(1-l)}\frac{1}{\left({z^2-1}\right)^{\frac{l}{2}}}\bm{F}\left(1-l+L,-l-L,1-l,\frac{1-z}{2}\right),
\end{split}
\ee
while in {\cite{Bateman} eq. (5) pag. 143} there is the following definition
\be\label{defPmaiusc}
\mathrm{P}_L^l(x):=e^{\frac{1}{2} i l \pi}P_{L}^l(x+i0).
\ee
So, putting together {(\ref{defassleg})} and {(\ref{defPmaiusc})},
\be\label{nuovadefP}
\begin{split}
\mathrm{P}_L^l(x)&= e^{\frac{1}{2} i l \pi} \frac{2^l}{\Gamma(1-l)}\frac{1}{(x^2-1)^{\frac{l}{2}}}\bm{F}\left(1-l+L,-l-L,1-l,\frac{1-x}{2}\right)\\
&= \frac{2^l}{\Gamma(1-l)}\frac{1}{(1-x^2)^{\frac{l}{2}}}\bm{F}\left(1-l+L,-l-L,1-l,\frac{1-x}{2}\right);
\end{split}
\ee
in this work there the $P$ font is always used, but this is only a stylistic choice, in fact it is always referring to this `real' function $\mathrm{P}$ of {(\ref{nuovadefP})}.

In addition, from \cite{Bateman} p. 161 (12)-(14) it follows
\begin{equation}\label{Legendreprop}
\begin{split}
\sqrt{1-x^2}P_L^l(x)&=\frac{1}{2L+1}\left[(L-l+1)(L-l+2)P_{L+1}^{l-1}(x)-(L+l-1)(L+l)P_{L-1}^{l-1}(x)
\right],\\
\sqrt{1-x^2}P_L^{l}(x)&=\frac{1}{2L+1}\left[
-P_{L+1}^{l+1}(x)+P_{L-1}^{l+1}(x)
\right],\\
xP_L^l(x)&=\frac{1}{2L+1}\left[(L-l+1)P_{L+1}^l(x)+(L+l)P_{L-1}^l(x)\right];
\end{split}
\end{equation}
so, if $L=l=\frac{1}{2}$, eq. (11) p. 101 in \cite{Bateman}
$$
\cos{az}=\bm{F}\left(
\frac{1}{2}a,-\frac{1}{2}a,\frac{1}{2},\left(\sin{z}\right)^2
\right),
$$
implies
\be\label{Leg121}
\begin{split}
P_{\frac{1}{2}}^{\frac{1}{2}}\left(\cos{\theta}\right)&=\frac{\sqrt{2}}{\Gamma\left(\frac{1}{2} \right)}\frac{1}{\sqrt{\sin{\theta}}}\bm{F}\left(1,-1,\frac{1}{2},\frac{1-\cos{\theta}}{2}\right)\\
&=\sqrt{\frac{2}{\pi}}\frac{1}{\sqrt{\sin{\theta}}} \bm{F}\left(1,-1,\frac{1}{2},\sin^2\frac{\theta}{2}\right)=\sqrt{\frac{2}{\pi}}\frac{\cos{\theta}}{\sqrt{\sin{\theta}}}.
\end{split}
\ee
If $L=-l=\frac{1}{2}$, then eq. (4) p. 101, eq. (18) p. 102 and eq. (3) p.105 in \cite{Bateman}
\begin{equation*}
\begin{split}
\bm{F}\left(-a,b,b,z\right)=(1+z)^a\quad&,\quad \bm{F}(a,b,c,z)=\bm{F}(b,a,c,z),\\
\bm{F}(a,b,c,z)=(1-z)^{-a}&\bm{F}\left(a,c-b,c,\frac{z}{z-1}\right),
\end{split}
\end{equation*}
imply
\begin{equation*}
\begin{split}
\bm{F}\left(2,0,\frac{3}{2},\frac{1-x}{2}\right)&=\left(1-\frac{1-x}{2} \right)^{-2}\bm{F}\left(2,\frac{3}{2},\frac{3}{2},\frac{\frac{1-x}{2}}{\frac{1-x}{2}-1}\right)\\
&= \left(\frac{1+x}{2} \right)^{-2}F\left[-(-2),\frac{3}{2},\frac{3}{2},-\left(-\frac{\frac{1-x}{2}}{\frac{-1-x}{2}}\right)\right]\\
&= \left(\frac{1+x}{2} \right)^{-2}\left(1 -\frac{\frac{1-x}{2}}{\frac{-1-x}{2}}\right)^{-2}\\
&= \left(\frac{1+x}{2} \right)^{-2}\left(\frac{1}{\frac{1+x}{2}}\right)^{-2}=1;
\end{split}
\end{equation*}
and then
\be\label{Leg122}
P_{\frac{1}{2}}^{-\frac{1}{2}}\left(\cos{\theta}\right)=\frac{2^{-\frac{1}{2}}}{\Gamma\left(\frac{3}{2}\right)}\sqrt{\sin{\theta}}\bm{F}\left(2,0,\frac{3}{2},\frac{1-\cos{\theta}}{2}\right)=\sqrt{\frac{2}{\pi}}\sqrt{\sin{\theta}}.
\ee
Once calculated these $P_{\frac{1}{2}}^{\pm \frac{1}{2}}$, (\ref{Legendreprop}) leads to
\be\label{Leg321}
\cos{\theta} P_{\frac{1}{2}}^{-\frac{1}{2}}\left(\cos{\theta}\right)=P_{\frac{3}{2}}^{-\frac{1}{2}}\left(\cos{\theta}\right)=\sqrt{\frac{2}{\pi}}\cos{\theta}\sqrt{\sin{\theta}}
\ee
and
\be\label{Leg322}
\sin{\theta}P_{\frac{1}{2}}^{-\frac{1}{2}}\left(\cos{\theta}\right)=\frac{1}{2}[2\cdot 3]P_{\frac{3}{2}}^{-\frac{3}{2}}\left(\cos{\theta}\right)=\sqrt{\frac{2}{\pi}}\sin{\theta}\sqrt{\sin{\theta}};
\ee
for completeness, according to (\ref{da+ma-m}),
\be\label{3su2positivi}
P_{\frac{3}{2}}^{\frac{3}{2}}\left(\cos{\theta}\right)= 2\sqrt{\frac{2}{\pi}} \sin{\theta}\sqrt{\sin{\theta}}  \quad,\quad P_{\frac{3}{2}}^{\frac{1}{2}}\left(\cos{\theta}\right)=2\sqrt{\frac{2}{\pi}} \cos{\theta}\sqrt{\sin{\theta}}.
\ee
At this point, in order to conclude the proof of the regularity of $g_{n-1}(z)$ in (\ref{dagadf}), it is necessary the following
\begin{propo}
Let $L$ and $l$ be half-integer and positive, $0\leq l\leq L$, then 
$$
P_L^{-l}\left(\cos{\theta}\right)=\left(\sin{\theta}\right)^l \widetilde{P}_L^{-l}\left(\cos{\theta}\right),
$$
where $\widetilde{P}_L^{-l}\left(\cos{\theta}\right)$ is a polynomial of degree $L-l$ in $\cos{\theta}$ which does not contain any term of degree $L-l-(2h+1)$, with $h\in\mathbb{N}_0$.
\proof
This is proved by induction over $L$. When $L=\frac{1}{2}$ and $L=\frac{3}{2} $, (\ref{Leg121})-(\ref{Leg322}) imply that the claim is true in these two cases. Let $L=\frac{1}{2}+n$,  with $2\leq n\in\mathbb{N}$ and assume that the claim is true for $n-1$, then (\ref{Legendreprop})$_2$ implies, if $n> n'+1\in\mathbb{N}$ and $n'\in\mathbb{N}_0$,
$$
\sin{\theta}P_{\frac{1}{2}+(n-1)}^{-\left[\frac{1}{2}+(n'-1)\right]}= P_{\frac{1}{2}+n}^{-\left[\frac{1}{2}+n'\right]}+ P_{\frac{1}{2}+(n-2)}^{-\left[\frac{1}{2}+n'\right]},
$$
then, from
$$
\sin{\theta}P_{\frac{1}{2}+(n-1)}^{-\left[\frac{1}{2}+(n'-1)\right]} \left(\cos{\theta}\right) =\left(\sin{\theta}\right)^{\frac{1}{2}+n'}\widetilde{P}_{\frac{1}{2}+(n-1)}^{-\left[\frac{1}{2}+(n'-1)\right]} \left(\cos{\theta}\right)
$$
and
$$
P_{\frac{1}{2}+(n-2)}^{-\left[\frac{1}{2}+n'\right]} \left(\cos{\theta}\right) = \left(\sin{\theta}\right)^{\frac{1}{2}+n'} \widetilde{P}_{\frac{1}{2}+(n-2)}^{-\left[\frac{1}{2}+n'\right]} \left(\cos{\theta}\right),
$$
it follows that also  
$$
P_{\frac{1}{2}+n}^{-\left[\frac{1}{2}+n'\right]}\left(\cos{\theta}\right)= \left(\sin{\theta}\right)^{\frac{1}{2}+n'} \widetilde{P}_{\frac{1}{2}+n}^{-\left[\frac{1}{2}+n'\right]} \left(\cos{\theta}\right),
$$
where $\widetilde{P}_{\frac{1}{2}+n}^{-\left[\frac{1}{2}+n'\right]}$ is a polynomial of degree $n-n' $ in $\cos{\theta}$ which does not contain any term of degree $n-n'-(2h+1)$, with $h\in\mathbb{N}_0$.

On the other hand, (\ref{Legendreprop})$_3$ implies
$$
\cos{\theta} P_{\frac{1}{2}+(n-1)}^{-\left[\frac{1}{2}+(n-1)\right]}\left(\cos{\theta}\right)= P_{\frac{1}{2}+n}^{-\left[\frac{1}{2}+(n-1)\right]} \left(\cos{\theta}\right) +0= P_{\frac{1}{2}+n}^{-\left[\frac{1}{2}+(n-1)\right]} \left(\cos{\theta}\right);
$$
so
$$
P_{\frac{1}{2}+n}^{-\left[\frac{1}{2}+(n-1)\right]} \left(\cos{\theta}\right)=\left(\sin{\theta}\right)^{\frac{1}{2}+(n-1)}\widetilde{P}_{\frac{1}{2}+n}^{-\left[\frac{1}{2}+(n-1)\right]} \left(\cos{\theta}\right),
$$
where $\widetilde{P}_{\frac{1}{2}+n}^{-\left[\frac{1}{2}+(n-1)\right]} \left(\cos{\theta}\right)$ is a polynomial in $\cos{\theta}$ of degree $1$.

Furthermore, (\ref{Legendreprop})$_1$ implies
$$
\sin{\theta} P_{\frac{1}{2}+(n-1)}^{-\left[\frac{1}{2}+(n-1)\right]} \left(\cos{\theta}\right) = P_{\frac{1}{2}+n}^{-\left[\frac{1}{2}+n\right]} \left(\cos{\theta}\right)=\left(\sin{\theta}\right)^{\frac{1}{2}+n}\widetilde{P}_{\frac{1}{2}+(n-1)}^{-\left[\frac{1}{2}+(n-1)\right]} \left(\cos{\theta}\right),
$$
so the claim is true also for $n$, because this last equality means that
$$
P_{\frac{1}{2}+n}^{-\left[\frac{1}{2}+\widetilde{n}\right]} \left(\cos{\theta}\right)=\left[\sin{\theta}\right]^{\frac{1}{2}+\widetilde{n}} \widetilde{P}_{\frac{1}{2}+n}^{-\left[\frac{1}{2}+\widetilde{n}\right]} \left(\cos{\theta}\right),
$$
where $\widetilde{P}_{\frac{1}{2}+n}^{-\left[\frac{1}{2}+\widetilde{n}\right]} \left(\cos{\theta}\right)$ is a polynomial of degree $n-\widetilde{n}$ in $\cos{\theta}$ which does not contain any term of degree $n-\widetilde{n}-(2h+1)$, with $h\in\mathbb{N}_0$.
\endproof
\end{propo}
It is important to underline that the hypothesis $l\geq 0$ in the last proof is not stringent, in fact the same result can be proved also when $l$ is negative, because of (\ref{da+ma-m}).

{In addition, from (\ref{Leg121})-(\ref{3su2positivi}) it turns out that the highest coefficient multiplying a power of $\cos{\theta}$ in $P_{L}^{l}\left(\cos{\theta}\right)$, when $\frac{3}{2}\geq L\geq |l|$ and $L,l\in\frac{\mathbb{Z}}{2}$ is always less or equal that $2L+1$; on the other hand, from (\ref{Legendreprop}) it follows
\begin{equation*}
\begin{split}
P_{L+1}^{l-1}(x)&=\frac{2L+1}{(L-l+1)(L-l+2)}\sqrt{1-x^2}P_L^l(x)+\frac{(L+l-1)(L+l)}{(L-l+1)(L-l+2)}P_{L-1}^{l-1}(x),\\
P_{L+1}^{l+1}(x)&=-\sqrt{1-x^2}(2L+1)P_L^l(x)+P_{L-1}^{l+1}(x),\\ 
P_{L+1}^{l}(x)&=\frac{2L+1}{L-l+1}xP_L^l(x)-\frac{L+l}{L-l+1}P_{L-1}^{l}(x);
\end{split}
\end{equation*}
and then that the highest coefficient multiplying a power of $\cos{\theta}$ in $P_{L+1}^{l'}\left(\cos{\theta}\right)$ is less then $[2(L+1)+1]^2$ times the sum of the highest coefficient multiplying a power of $\cos{\theta}$ in $P_{L}^{l''}\left(\cos{\theta}\right)$.

According to this, by recursion one has that the highest coefficient multiplying a power of $\cos{\theta}$ in $P_{L}^{l}\left(\cos{\theta}\right)$ is 
$$c
2^L \left[(2L+1)!!\right]^2\overset{L\leq \Lambda} \leq 2^{\Lambda}\left[(2\Lambda+1)!!\right]^2.
$$}
\subsection{The square-integrability of $\psi_{\bm{l},D}$}\label {squintpsi}
In this section there is the proof that that every $\psi_{\bm{l},D}$ is square-integrable and also the explicit calculation of $M_{l,D}$.

The integral
$$
\int_{\mathbb{R}^D}\left\vert\psi_{\bm{l},D}\right\vert^2 dx
$$
can be factorized in this way:
\begin{equation*}
\begin{split}
\int_{\mathbb{R}^D}\left\vert\psi_{\bm{l},D}\right\vert^2 dx=&\left|M_{l,D}\right|^2\left(\int_{0}^{+\infty}
\frac{\left[g_{0,l,D}(r)\right]^2}{r^d} r^d dr
\right)\\
&\cdot\left[\int_{S^d}\left\vert Y_{\bm{l}}\right\vert^2 \left(\sin^{d-1}{\theta_{d}} \sin^{d-2}{\theta_{d-1}} \cdots \sin{\theta_2}\right) d\theta_1 d\theta_2\cdots d\theta_{d}
 \right]\\
\overset{(\ref{ArmSfe})_2}=&\left|M_{l,D}\right|^2\int_{0}^{+\infty}
\left[g_{0,l,D}(r)\right]^2 dr.
\end{split}
\end{equation*}
So, proceeding as in {section 6.5} of \cite{FiorePisacane},
\be\label{valueM_l,D}
\int_{\mathbb{R}^D}\left\vert\psi_{\bm{l},D}\right\vert^2 dx=1\Longleftrightarrow M_{l,D}=\frac{\sqrt[8]{k_{l,D}}}{\sqrt[4]{\pi}}.
\ee
\subsection{The action of $t^h$ on $Y_{\bm{l}}$}\label{Appendix_defi}
First of all
\begin{defi}\label{defiABCDFG}
Let $L\geq |l|$ and $2\leq j\in\mathbb{N}$, then
\begin{equation}\label{ABCDFG}
\begin{split}
&A\left(L,l,j\right):= \sqrt{\frac{(L+l+j-1)(L+l+j)}{(2L+j-1)(2L+j+1)}}\quad,\quad
B\left(L,l,j\right):= -\sqrt{\frac{(L-l-1)(L-l)}{(2L+j-1)(2L+j-3)}},\\
&C\left(L,l,j\right):= -\sqrt{\frac{(L-l+2)(L-l+1)}{(2L+j-1)(2L+j+1)}}\quad,\quad
D\left(L,l,j\right):= \sqrt{\frac{(L+l+j-2)(L+l+j-3)}{(2L+j-1)(2L+j-3)}},\\
&F\left(L,l,j\right):= \sqrt{\frac{(L+l+j-1)(L-l+1)}{(2L+j-1)(2L+j+1)}}\quad,\quad
G\left(L,l,j\right):= \sqrt{\frac{(L-l)(L+l+j-2)}{(2L+j-1)(2L+j-3)}}.
\end{split}
\end{equation}
\end{defi}
They fulfill
\begin{equation}\label{relazioniABCDFG}
\begin{split}
&A(L,l,j)=D(L+1,l+1,j),\quad B(L,l,j)=C(L-1,l+1,j),\\
& F(L,l,j)=G(L+1,l,j),\quad F(L,l,j)A(L+1,l,j)=A(L,l,j)F(L+1,l+1,j),\\
&G(L,l,j)B(L-1,l,j)=B(L,l,j)G(L-1,l+1,j),
\end{split}
\end{equation}
but it is also important to point out something about the generalized associated Legendre functions $P_r^s$.

From (\ref{Legendreprop}) and (\ref{ABCDFG}), it follows
\be
\begin{split}
\left[\sin{\theta}\right]{}_j\overline{P}_L^l\left(\theta\right)&=
\sqrt{\frac{2L+j-1}{2}}\sqrt{\frac{(L+l+j-2)!}{(L-l)!}}\left[\sin{\theta} \right]^{\frac{2-j}{2}}\left[\sin{\theta}\right]P_{L+\frac{j-2}{2}}^{-(l+\frac{j-2}{2})}\left(\cos{\theta}\right)\\
&=\sqrt{\frac{2L+j-1}{2}}\sqrt{\frac{(L+l+j-2)!}{(L-l)!}}\left[\sin{\theta} \right]^{\frac{2-j}{2}}\frac{1}{2L+j-1}\\
&\cdot\left\{
(L+l+j-1)(L+l+j)P_{[(L+1)+\frac{j-2}{2}]}^{-[(l+1)+\frac{j-2}{2}]}(\cos{\theta})\right.\\
&\left.\quad -(L-l)(L-l-1)P_{[(L-1)+\frac{j-2}{2}]}^{-[(l+1)+\frac{j-2}{2}]}(\cos{\theta})
\right\}\\
&=\sqrt{\frac{2L+j+1}{2}}\sqrt{\frac{(L+l+j)!}{(L-l)!}}\left[\sin{\theta} \right]^{\frac{2-j}{2}}P_{[(L+1)+\frac{j-2}{2}]}^{-[(l+1)+\frac{j-2}{2}]}(\cos{\theta})\sqrt{\frac{(L+l+j-1)(L+l+j)}{(2L+j-1)(2L+j+1)}}\\
&-\sqrt{\frac{2L+j-3}{2}}\sqrt{\frac{(L+l+j-2)!}{(L-l-2)!}}\left[\sin{\theta} \right]^{\frac{2-j}{2}}P_{[(L-1)+\frac{j-2}{2}]}^{-[(l+1)+\frac{j-2}{2}]}(\cos{\theta})\sqrt{\frac{(L-l-1)(L-l)}{(2L+j-1)(2L+j-3)}}\\
&=\sqrt{\frac{(L+l+j-1)(L+l+j)}{(2L+j-1)(2L+j+1)}}{}_j\overline{P}_{L+1}^{l+1}(\theta)+\left[-\sqrt{\frac{(L-l-1)(L-l)}{(2L+j-1)(2L+j-3)}} \right]{}_j\overline{P}_{L-1}^{l+1}(\theta)\\
&=A(L,l,j) {}_j\overline{P}_{L+1}^{l+1}(\theta)+B(L,l,j){}_j\overline{P}_{L-1}^{l+1}(\theta);
\end{split}\label{az_sin1}
\ee
\be
\begin{split}
\hspace{-1cm}\left[\sin{\theta}\right]{}_j\overline{P}_L^l\left(\theta\right)&=
\sqrt{\frac{2L+j-1}{2}}\sqrt{\frac{(L+l+j-2)!}{(L-l)!}}\left[\sin{\theta} \right]^{\frac{2-j}{2}}\left[\sin{\theta}\right]P_{L+\frac{j-2}{2}}^{-(l+\frac{j-2}{2})}\left(\cos{\theta}\right)\\
&=\sqrt{\frac{2L+j-1}{2}}\sqrt{\frac{(L+l+j-2)!}{(L-l)!}}\left[\sin{\theta} \right]^{\frac{2-j}{2}}\frac{1}{2L+j-1}\\
&\cdot\left\{
-P_{[(L+1)+\frac{j-2}{2}]}^{-[(l-1)+\frac{j-2}{2}]}(\cos{\theta})+P_{[(L-1)+\frac{j-2}{2}]}^{-[(l-1)+\frac{j-2}{2}]}(\cos{\theta})
\right\}\\
&=-\sqrt{\frac{2L+j+1}{2}}\sqrt{\frac{(L+l+j-2)!}{(L-l+2)!}}\left[\sin{\theta} \right]^{\frac{2-j}{2}}P_{[(L+1)+\frac{j-2}{2}]}^{-[(l-1)+\frac{j-2}{2}]}(\cos{\theta})\sqrt{\frac{(L-l+2)(L-l+1)}{(2L+j-1)(2L+j+1)}}\\
&+\sqrt{\frac{2L+j-3}{2}}\sqrt{\frac{(L+l+j-4)!}{(L-l)!}}\left[\sin{\theta} \right]^{\frac{2-j}{2}}P_{[(L-1)+\frac{j-2}{2}]}^{-[(l-1)+\frac{j-2}{2}]}(\cos{\theta})\sqrt{\frac{(L+l+j-2)(L+l+j-3)}{(2L+j-1)(2L+j-3)}}\\
&=\left[-\sqrt{\frac{(L-l+2)(L-l+1)}{(2L+j-1)(2L+j+1)}}\right]{}_j\overline{P}_{L+1}^{l-1}(\theta)+\sqrt{\frac{(L+l+j-2)(L+l+j-3)}{(2L+j-1)(2L+j-3)}} {}_j\overline{P}_{L-1}^{l-1}(\theta)\\
&=C(L,l,j) {}_j\overline{P}_{L+1}^{l-1}(\theta)+D(L,l,j){}_j\overline{P}_{L-1}^{l-1}(\theta);
\end{split}\label{az_sin2}
\ee
\be 
\begin{split}
\left[\cos{\theta}\right]{}_j\overline{P}_L^l\left(\theta\right)&=\sqrt{\frac{2L+j-1}{2}}\sqrt{\frac{(L+l+j-2)!}{(L-l)!}}\left[\sin{\theta} \right]^{\frac{2-j}{2}}\left[\cos{\theta}\right]P_{L+\frac{j-2}{2}}^{-(l+\frac{j-2}{2})}\left(\cos{\theta}\right)\\
&=\sqrt{\frac{2L+j-1}{2}}\sqrt{\frac{(L+l+j-2)!}{(L-l)!}}\left[\sin{\theta} \right]^{\frac{2-j}{2}}\frac{1}{2L+j-1}\\
&\left\{
(L+l+j-1)P_{[(L+1)+\frac{j-2}{2}]}^{-[(l)+\frac{j-2}{2}]}(\cos{\theta})+(L-l)P_{[(L-1)+\frac{j-2}{2}]}^{-[(l)+\frac{j-2}{2}]}(\cos{\theta})
\right\}\\
&=\left[\sqrt{\frac{(L+l+j-1)(L-l+1)}{(2L+j-1)(2L+j+1)}} \right]{}_j\overline{P}_{L+1}^l(\theta)+\left[\sqrt{\frac{(L-l)(L+l+j-2)}{(2L+j-1)(2L+j-3)}}\right]{}_j\overline{P}_{L-1}^l(\theta)\\
&=F(L,l,j){}_j\overline{P}_{L+1}^l(\theta) + G(L,l,j){}_j\overline{P}_{L-1}^l(\theta).
\end{split}\label{az_cos}
\ee
These last relations are fundamental, in fact they are used in order to understand the action of a coordinate $t^h$ (seen as a multiplication operator) on a $D$-dimensional spherical harmonic $Y_{\bm{l}}$.
\begin{remark}
Let $t^{\pm}:=\frac{x^1\pm ix^2}{\sqrt{2}r}$ and $t^\nu:=\frac{x^\nu}{r}$, when $\nu\in\{1,2,\cdots, D\}$; obviously $t^+t^-+t^-t^+=\left(t^1\right)^2+\left(t^2\right)^2$, so {(\ref{az_sin1})-(\ref{az_cos})} can be used to write $t^hY_{\bm{l}}$ in terms of other $D$-dimensional spherical harmonics, for instance
$$
t^+ Y_{\bm{l}}=\sin{\theta_{d}}\sin{\theta_{d-1}}\cdots\sin{\theta_{2}} \left[\prod_{n=2}^{d}{}_n\overline{P}_{l_n}^{l_{n-1}}\left(\theta_n\right) \right]\frac{1}{\sqrt{2\pi}} e^{i(l_1+1)\theta_1};
$$
then in the product $\sin{(\theta_{2})}\cdot {}_{2}P_{l_2}^{l_1}\left(\theta_2\right)$ it is necessary to use {(\ref{az_sin1})}, because $t^+$ changes $e^{il_1\theta_1}$ to $e^{i(l_1+1)\theta_1}$, so
\begin{equation*}
\begin{split}
t^+ Y_{\bm{l}}=&\sin{\theta_{d}}\sin{\theta_{d-1}}\cdots\sin{\theta_{3}}\left[
A\left(l_2,l_1,2\right) {}_2P_{l_2+1}^{l_1+1}\left(\theta_1\right)+
B\left(l_2,l_1,2\right) {}_2P_{l_2-1}^{l_1+1}\left(\theta_1\right)
\right]\\
&\cdot\left[\prod_{n=3}^{d}{}_n\overline{P}_{l_n}^{l_{n-1}}\left(\theta_n\right) \right]\frac{1}{\sqrt{2\pi}} e^{i(l_1+1)\theta_1}
\end{split}
\end{equation*}
and so on with the remaining factors $\sin{\theta_j}\cdot {}_j\overline{P}_{l'_j}^{l'_{j-1}}\left(\theta_j\right)$.\label{remark1}
\end{remark}
Of course, this last procedure can be repeated also for $t^-$ and then for every $t^\nu$ with $\nu\in\{3,\cdots,D\}$, while the actions of $t^1$ and $t^2$ can be recovered from the ones of $t^+$ and $t^-$. According to this, let
\be\label{scalprodY}
R_{h,D}\left(\bm{l};\bm{l'}\right):=\left\langle Y_{\bm{l'}},
t^h Y_{\bm{l}}\right\rangle
\ee
and this definition implies that, in general
\begin{equation}\label{aztu}
t^{\nu}Y_{\bm{l}}=\sum_{
\substack{
{l'}_j:|l_j-{l'}_j|=1\\
\mbox{for }j=\nu-1,\cdots,d
}}R_{\nu,D}\left(\bm{l};\bm{l'}_\nu\right)
\cdot Y_{\bm{l'}_\nu},\quad\mbox{where}\quad\bm{l'}_\nu:=\left(l',{l'}_{d-1},\cdots,{l'}_{\nu-1},l_{\nu-2},\cdots,{l}_1\right).
\end{equation}

{Remark \ref{remark1}} and (\ref {scalprodY}) suggest that every $R_{\nu,D}$ can be written as a sum of elements, where every addend is a product of several $A,B,C,D,F,G$; it is important to note that there are some simple rules, reported in the the next lines, which help to calculate every $R_{\nu,D}$.

The first rule is that the generic term of a $R_{\nu,D}$ is always written in an `ordered' way, in fact the factors appear in this `order': 
$$
R_{\nu,D}(\cdots;\cdots)=\cdots+\cdots D(l_{j+2},l_{j+1},j+2)B(l_{j+1},l_j,j+1)A(l_j,l_{j-1},j) \cdots+\cdots
$$
in other words a factor having third argument $j+1$ is always right-multiplied by a factor having third argument $j$ and always left-multiplied by a factor having thirs argument $j+2$.
\begin{remark}
The other rules are these ones:
\begin{itemize}
\item Every $A(l_j,l_{j-1},j)$ is always left-multiplied by an $A(l_{j+1},l_j,j+1)$ or $B(l_{j+1},l_j,j+1)$;
\item Every $B(l_j,l_{j-1},j)$ is always left-multiplied by an $C(l_{j+1},l_j,j+1)$ or $D(l_{j+1},l_j,j+1)$;
\item Every $C(l_j,l_{j-1},j)$ is always left-multiplied by an $A(l_{j+1},l_j,j+1)$ or $B(l_{j+1},l_j,j+1)$;
\item Every $D(l_j,l_{j-1},j)$ is always left-multiplied by an $C(l_{j+1},l_j,j+1)$ or $D(l_{j+1},l_j,j+1)$;
\item In $R_{1,D}$ the first factor (from right to left) is $A(l_2,l_1,2)$, or $B(l_2,l_1,2)$, or $C(l_2,l_1,2)$, or $D(l_2,l_1,2)$;
\item In $R_{2,D}$ the first factor (from right to left) is $A(l_2,l_1,2)$, or $B(l_2,l_1,2)$, or $C(l_2,l_1,2)$, or $D(l_2,l_1,2)$;
\item If $\nu\geq 3$, in order to calculate $R_{\nu,D}$, it is better to start by using \emph{(\ref{az_cos})} with $\theta=\theta_{\nu-1}$, and then go `backward'.
\end{itemize}\label{remark2}
\end{remark}
\subsection{Proof of (\ref{propLij})$_1$}\label{proofLij1}

The {definition \ref{defiL}} (which uses the $R$ coefficients) allows to take the relations among the coordinates $t^h$ (seen as multiplication operators) and obtain from them some relations among the components $L_{h,j}$ of the $D$-dimensional angular momentum operator.

In particular,
$$
\left(t^1\right)^2+\left(t^2\right)^2+\cdots+\left(t^D\right)^2=t^+t^-+t^-t^++\left(t^3\right)^2+\cdots+\left(t^D\right)^2=1
$$
implies
\be\label{R^2_1}
\left[
\left(t^1\right)^2+\left(t^2\right)^2+\cdots+\left(t^D\right)^2
\right]Y_{\bm{l}}=Y_{\bm{l}};
\ee
but (\ref{aztu}) implies also that [here $Z\left(\bm{l},\bm{l'}\right)$ are suitable coefficients, which can be obtained from the $R$s]
\begin{equation}\label{R^2_2}
\left[
\left(t^1\right)^2+\left(t^2\right)^2+\cdots+\left(t^D\right)^2
\right]Y_{\bm{l}}=\sum_{\substack{j=1,\cdots, d\\ |l_j-{l'j}|\leq 2}
}Z\left(\bm{l},\bm{l'}\right)Y_{\bm{l'}};
\end{equation}
in addition, (\ref{R^2_1}) and (\ref{R^2_2}) imply $Z\left(\bm{l},\bm{l'}\right)=0$ if there is at least one $j$ such that $l_j\neq {l'}_j$.
On the other hand, it is obvious that
\be\label{R^2_3}
Z\left(\bm{l},\bm{l}\right)=:Z\left(\bm{l}\right)=1;
\ee
so only the $Z\left(\bm{l}\right)$ are relevant.
\begin{remark}
Equations {(\ref{az_sin1})-(\ref{az_cos})} imply that
\begin{itemize}
\item if $R_{h,D}\left(\cdots, l_j,l_{j-1},\cdots; \cdots, {l'}_j,{l'}_{j-1},\cdots\right) $ contains a factor $A(l_j,l_{j-1},j)$, then ${l'}_j=l_j+1$ and ${l'}_{j-1}=l_{j-1}+1$;
\item if $R_{h,D}\left(\cdots, l_j,l_{j-1},\cdots; \cdots, {l'}_j,{l'}_{j-1},\cdots\right) $ contains a factor $B(l_j,l_{j-1},j)$, then ${l'}_j=l_j-1$ and ${l'}_{j-1}=l_{j-1}+1$;
\item if $R_{h,D}\left(\cdots, l_j,l_{j-1},\cdots; \cdots, {l'}_j,{l'}_{j-1},\cdots\right) $ contains a factor $C(l_j,l_{j-1},j)$, then ${l'}_j=l_j+1$ and ${l'}_{j-1}=l_{j-1}-1$;
\item if $R_{h,D}\left(\cdots, l_j,l_{j-1},\cdots; \cdots, {l'}_j,{l'}_{j-1},\cdots\right) $ contains a factor $D(l_j,l_{j-1},j)$, then ${l'}_j=l_j-1$ and ${l'}_{j-1}=l_{j-1}-1$;
\item if $R_{h,D}\left(\cdots, l_j,l_{j-1},\cdots; \cdots, {l'}_j,{l'}_{j-1},\cdots\right) $ contains a factor $F(l_j,l_{j-1},j)$, then ${l'}_j=l_j+1$ and ${l'}_{j-1}=l_{j-1}$;
\item if $R_{h,D}\left(\cdots, l_j,l_{j-1},\cdots; \cdots, {l'}_j,{l'}_{j-1},\cdots\right) $ contains a factor $G(l_j,l_{j-1},j)$, then ${l'}_j=l_j-1$ and ${l'}_{j-1}=l_{j-1}$;
\end{itemize}
in other words, these $A,B,C,D,F,G$ express that an index is raising or lowering, as in \emph{{remark \ref{remark1}}}.

Furthermore
\begin{equation}\label{t-theta}
\begin{split}
1=&t^+t^-+t^-t^++\left(t^3\right)^2+\cdots+\left(t^D\right)^2\\
=&\left[\cos{\theta_{d}}\right]^2+\left[\sin{\theta_{d}}\right]^2\left\{
\left[\cos{\theta_{d-1}}\right]^2+\left[\sin{\theta_{d-1}}\right]^2
\left\{ \cdots \left\{\left[\cos{\theta_{2}}\right]^2+\left[\sin{\theta_{2}}\right]^2 \right\}\cdots
\right\}
\right\}
\end{split}
\end{equation}
implies
\begin{equation}
\begin{split}
\left\{\left[\cos{\theta_{2}}\right]^2+\left[\sin{\theta_{2}}\right]^2 \right\}Y_{\bm{l}}=&\left\{\left[F(l_{2},l_{1},2)\right]^2+\left[A(l_{2},l_{1},2)\right]^2+\left[C(l_{2},l_{1},2)\right]^2 \right.\\
&+\left.\left[G(l_{2},l_{1},2)\right]^2+\left[B(l_{2},l_{1},2)\right]^2+\left[D(l_{2},l_{1},2)\right]^2 \right\}\\
&\cdot Y_{\bm{l}}\\
=:& \left\{
Z_{1,2} \left(l_2\right)+Z_{2,2} \left(l_2\right)
\right\}Y_{\bm{l}};
\end{split}
\end{equation}
while {{remark \ref{remark2}}} implies
\begin{equation}
\begin{split}
&\left\{\left[\cos{\theta_{3}}\right]^2+\left[\sin{\theta_{3}}\right]^2\left\{\left[\cos{\theta_{2}}\right]^2+\left[\sin{\theta_{2}}\right]^2\right\} \right\}Y_{\bm{l}}=\\
&\left\{\left[F(l_{3},l_{2},3)\right]^2+\left[A(l_{3},l_{2},3)\right]^2
Z_{1,2} \left(l_2\right)
+\left[C(l_{3},l_{2},3)\right]^2
Z_{2,2} \left(l_2\right)
 \right.\\
&+\left.\left[G(l_{3},l_{2},3)\right]^2+\left[B(l_{3},l_{2},3)\right]^2
Z_{1,2} \left(l_2\right)+\left[D(l_{3},l_{2},3)\right]^2Z_{2,2} \left(l_2\right)
\right\}\\
&\cdot Y_{\bm{l}}\\
=:& \left\{
Z_{1,3} \left(l_3,l_2\right)+Z_{2,3} \left(l_3,l_2\right)
\right\}Y_{\bm{l}};
\end{split}
\end{equation}
and so on with the other elements of \emph{(\ref{t-theta})}, so
\begin{equation}
\begin{split}
&\left\{\left[\cos{\theta_{j}}\right]^2+\left[\sin{\theta_{j}}\right]^2\left\{\left[\cos{\theta_{j-1}}\right]^2+\left[\sin{\theta_{j-1}}\right]^2\left\{ 
\cdots\left\{\left[\cos{\theta_{2}}\right]^2+\left[\sin{\theta_{2}}\right]^2 \right\}
\right\}\right\} \right\}Y_{\bm{l}}=\\
&\left\{\left[F(l_{j},l_{j-1},j)\right]^2+\left[A(l_{j},l_{j-1},j)\right]^2
Z_{1,j-1} \left(l_{j-1},l_{j-2},\cdots,l_2\right)
+\left[C(l_{3},l_{2},3)\right]^2
Z_{2,j-1} \left(l_{j-1},l_{j-2},\cdots,l_2\right)
 \right.\\
&+\left.\left[G(l_{j},l_{j-1},j)\right]^2+\left[B(l_{j},l_{j-1},j)\right]^2
Z_{1,j-1} \left(l_{j-1},l_{j-2},\cdots,l_2\right)
+\left[D(l_{3},l_{2},3)\right]^2
Z_{2,j-1} \left(l_{j-1},l_{j-2},\cdots,l_2\right)
\right\}\\
&\cdot Y_{\bm{l}}\\
=:& \left\{
Z_{1,j} \left(l_{j},l_{j-1},\cdots,l_2\right)+Z_{2,j}\left(l_{j},l_{j-1},\cdots,l_2\right)
\right\}Y_{\bm{l}}.
\end{split}\label{defP_{1-2,j}}
\end{equation}
\label{remark3}
\end{remark}

It is important to underline that every $Z_{h,j}$ defined above does not depend on the dimension $D$ of the ambient space, and this is a direct consequence of the factorization in (\ref{Y_final}).

A crucial result of this section is the following
\begin{propo}
\be
\begin{split}
Z_{1,d} \left(\bm{l}\right)=\frac{l+d-1}{2l+d-1}\quad,\quad Z_{2,d} \left(\bm{l}\right)=\frac{l}{2l+d-1}.
\end{split}\label{P_1234}
\ee
\proof
The proof is by induction on the dimension $D$ of the carrier space $\mathbb{R}^D$. If $D=3$, then
\begin{equation*}
\begin{split}
t^+Y_{l_2,l_1}=A(l_2,l_1,2)Y_{l_2+1,l_1+1}+B(l_2,l_1,2)Y_{l_2-1,l_1+1},\\
t^-Y_{l_2,l_1}=C(l_2,l_1,2)Y_{l_2+1,l_1-1}+D(l_2,l_1,2)Y_{l_2-1,l_1-1},\\
t^3Y_{l_2,l_1}=F(l_2,l_1,2)Y_{l_2+1,l_1}+G(l_2,l_1,2)Y_{l_2-1,l_1};
\end{split}
\end{equation*}
and
\begin{equation*}
\begin{split}
\left[t^+t^-+t^-t^++\left(t^3\right)^2\right]Y_{l_2,l_1}=&\left\{\frac{1}{2}\left[A(l_{2},l_{1},2)\right]^2+\frac{1}{2}\left[B(l_{2},l_{1},2)\right]^2+\frac{1}{2}\left[C(l_{2},l_{1},2)\right]^2+\right.\\
&\left.\frac{1}{2}\left[D(l_{2},l_{1},2)\right]^2+\left[F(l_{2},l_{1},2)\right]^2+\left[G(l_{2},l_{1},2)\right]^2\right\}Y_{l_2,l_1};
\end{split}
\end{equation*}
so
\begin{equation*}
Z_{1,2} \left(l_2\right)=\frac{l_{2}+3-3}{2l_{2}+3-3}=\frac{1}{2},\quad Z_{2,2} \left(l_2\right)=\frac{l_{2}}{2l_{2}+3-3}=\frac{1}{2},
\end{equation*}
then (\ref{P_1234}) is true when $D=3$. Let $D>3$ and assume that (\ref{P_1234}) is true for $D-1$, from (\ref{defP_{1-2,j}}) it follows
\begin{equation}
\begin{split}
Z_{1,d} \left(l,l_{d-1},\cdots,l_2\right)=&\left[F(l,l_{d-1},d)\right]^2\\
&+\left[A(l,l_{d-1},d)\right]^2
Z_{1,d-1} \left(l_{d-1},l_{d-2},\cdots,l_2\right)
\\&+\left[C(l,l_{d-1},d)\right]^2
Z_{2,d-1} \left(l_{d-1},l_{d-2},\cdots,l_2\right)\\
=&\frac{(l+l_{d-1}+d-1)(l-l_{d-1}+1)}{(2l+d-1)(2l+D)}
\\
&+\frac{(l+l_{d-1}+d-1)(l+l_{d-1}+d)}{(2l+d-1)(2l+D)}\frac{l_{d-1}+d-2}{2l_{d-1}+d-2}\\
&+\frac{(l-l_{d-1}+2)(l-l_{d-1}+1)}{(2l+d-1)(2l+D)}\frac{l_{d-1}}{2l_{d-1}+d-2} \\
=&\frac{l+d-1}{2l+d-1},
\end{split}
\end{equation}
and
\begin{equation}
\begin{split}
Z_{2,d} \left(l,l_{d-1},\cdots,l_2\right)=&\left[G(l,l_{d-1},d)\right]^2\\
&+\left[B(l,l_{d-1},d)\right]^2
Z_{1,d-1} \left(l_{d-1},l_{d-2},\cdots,l_2\right)
\\&+\left[D(l,l_{d-1},d)\right]^2
Z_{2,d-1} \left(l_{d-1},l_{d-2},\cdots,l_2\right)\\
=&\frac{(l-l_{d-1})(l+l_{d-1}+d-2)}{(2l+d-1)(2l+D-4)}
\\
&+\frac{(l-l_{d-1}-1)(l-l_{d-1})}{(2l+d-1)(2l+D-4)}\frac{l_{d-1}+d-2}{2l_{d-1}+d-2}\\
&+\frac{(l+l_{d-1}+d-2)(l+l_{d-1}+D-4)}{(2l+d-1)(2l+D-4)}\frac{l_{d-1}}{2l_{d-1}+d-2} \\
=&\frac{l}{2l+d-1};
\end{split}
\end{equation}
so the proof is finished.
\endproof\label{propo1}
\end{propo}
It is interesting to note that, because of this last proposition, 
\be\label{P1+P2}
Z_{1,d} \left(\bm{l}\right)+Z_{2,d} \left(\bm{l}\right)=1,
\ee
which agrees with
\begin{equation}
Y_{\bm{l}}=\left[t^+t^-+t^-t^++\left(t^3\right)^2+\cdots+\left(t^D\right)^2\right]Y_{\bm{l}}=\left\{Z_{1,d} \left(\bm{l}\right)+Z_{2,d} \left(\bm{l}\right) \right\}Y_{\bm{l}}.
\end{equation}
Here comes the proof of (\ref{propLij})$_1$.
\begin{teorema}
The {{definition \ref{defiL}}} implies
\be \label{L^2}
\bm{L}^2 Y_{\bm{l}}:=\sum_{1\leq i<j\leq D}L_{h,j}^2Y_{\bm{l}}=l\left(l+D-2 \right)Y_{\bm{l}}.
\ee
\proof
This proof is by induction on the dimension $D$ of the carrier space; if $D=2$, then $\bm{L}^2Y_{l_1}=L_{1,2}^2Y_{l_1}=(l_1)^2Y_{l_1}$; so (\ref{L^2}) is true for $D=2$. Let $D>2$ and assume that (\ref{L^2}) is true for $D-1$, which means that
\be\label{L^2D-1}
\sum_{1\leq i<j\leq d}L_{h,j}^2Y_{\bm{l}}=l_{d-1}\left(l_{d-1}+D-3 \right)Y_{\bm{l}}.
\ee
From {remark \ref{remark3}}, {proposition \ref{propo1}} and {definition \ref{defiL}} it follows
\begin{equation}\label{L^2D}
\begin{split}
&\sum_{i=1}^{d} L_{h,D}^2Y_{\bm{l}}=\left(d_{l,l_{d-1},D}\right)^2Z_{2,d-1} \left(l_{d-1},\cdots,l_1\right)Y_{\bm{l}}\\
&  +\left(d_{l,l_{d-1}+1,D}\right)^2Z_{1,d-1} \left(l_{d-1},\cdots,l_1\right)Y_{\bm{l}}\\
=&\left\{\left[{(l+1)(l+D-3)-l_{d-1}(l_{d-1}+D-4)}\right]\frac{l_{d-1}}{2l_{d-1}+D-3}\right\}Y_{\bm{l}}\\
&+\left\{\left[{(l+1)(l+D-3)-(l_{d-1}+1)(l_{d-1}+D-3)}\right]\frac{l_{d-1}+D-3}{2l_{d-1}+D-3}\right\}Y_{\bm{l}}\\
=&\left[l\left(l+D-2\right)-l_{d-1}\left(l_{d-1}+D-3\right)\right]Y_{\bm{l}}.
\end{split}
\end{equation}
The proof can be now completed because
\be
\bm{L}^2 Y_{\bm{l}}=\sum_{1\leq h<j\leq D}L_{h,j}^2Y_{\bm{l}}=\sum_{1\leq h<j\leq d}L_{h,j}^2Y_{\bm{l}}+\sum_{j=1}^{d}L_{j,D}^2Y_{\bm{l}}\overset{(\ref{L^2D-1})\&(\ref{L^2D})}=l\left(l+D-2\right)Y_{\bm{l}}.
\ee
\endproof
\end{teorema}
\subsection{Proof of (\ref{propLij})$_2$}\label{proofLij2}
The {definition \ref{defiL}} is given by induction on the dimension $D$ of the carrier space $\mathbb{R}^D$, this means that, in order to prove (\ref{propLij})$_2$, it is sufficient to show that
\be\label{toproveL}
\begin{split}
\left[L_{h,D},L_{j,D}\right]&=i L_{h,j}\quad,\quad \left[L_{h,j},L_{j,D}\right]=\frac{1}{i}L_{h,D},\\
 \left[L_{h,j},L_{p,D}\right]&=0 \mbox{ if }D\neq h,j\mbox{ and }p\neq h,j.
\end{split}
\ee
\subsubsection{Proof of (\ref{toproveL})$_1$}\label{proofLDaLDb}
Let $h<j$, of course $[t^h,t^j]Y_{\bm{l}}=0$ $\forall h,j$, but this and (\ref{aztu}) can be used to obtain some informations about the action of $[L_{h,D},L_{j,D}]$ on a spherical harmonic $Y_{\bm{l}}$.

It is important to point out that
\begin{remark}\label{remark4}
Let $1\leq h<j\leq d$, then $t^ht^jY_{l_{d-1},l_{d-2},\cdots,l_{1}}$ can be written as a linear combination of $(D-1)$-dimensional spherical harmonics $Y_{{l'}_{d-1},\cdots,{l'}_1}$ with, in principle, $|l_h-{l'}_h|\leq 2$, $\forall h\leq d-1$. 

More precisely, $t^h\cdot$ on $Y_{l_{d-1},l_{d-2},\cdots,l_{1}}$, `modifies' only the integers $l_{d-1},\cdots,l_{h-1}$,  while $t^j\cdot$ `modifies' only $l_{d-1},\cdots,l_{j-1}$, then the `modified' integers from the action of $t^ht^j\cdot$, as the ones from the action of $t^jt^h\cdot$ are $l_{d-1},\cdots,l_{h-1}$ and, in particular, $|l_p-{l'}_p|\leq 2$ if $p\in\{d-1,\cdots,j-1\}$, while $|l_p-{l'}_p|= 1$ if $p\in\{j-2,\cdots,h-1\}$.

Then
\begin{equation}\label{commtatb}
0=\left[
t^h,t^j
\right]Y_{{}_{d}\bm{l}}=\sum_{
\substack{
|{l'}_{p_1}-l_{p_1}|\leq 2\\
p_1=d-1,\cdots,j-1\\
|{l'}_{p_2}-l_{p_2}|=1\\
p_2=j-2,\cdots,h-1
}}Q_{D,h,j}\left({}_{d}\bm{l},{}_{d}\bm{l'}_h
\right)Y_{
{}_{d}\bm{l'}_h
},
\end{equation}
where
$$
{}_{d}\bm{l}:=\left(l_{d-1},\cdots,l_{1}\right)\quad\mbox{and}\quad{}_{d}\bm{l'}_h:=\left({l'}_{d-1},\cdots,{l'}_{h-1},l_{h-2},\cdots,l_{1}\right).
$$
\end{remark}
Anyway, the {definition \ref{defiL}} implies that the action of $L_{h,D}$ on $Y_{\bm{l}}$ is similar to the action of the coordinate $t^h$ on $Y_{{}_{d}\bm{l}}$, the only difference is given by the presence of the $d$ coefficients; so
\begin{equation}\label{commLaLb}
\begin{split}
\left[
L_{h,D},L_{j,D}
\right]Y_{\bm{l}}
=\sum_{
\substack{
|{l'}_{p_1}-l_{p_1}|\leq 2\\
p_1=d-1,\cdots,j-1\\
|{l'}_{p_2}-l_{p_2}|=1\\
p_2=j-2,\cdots,h-1
}}\widetilde{Q}_{D,h,j}\left(\bm{l},\bm{l'}_h
\right) Y_{
\bm{l'}_h
},
\end{split}
\end{equation}
where
$$
\bm{l'}_h:=\left(l,{l'}_{d-1},\cdots,{l'}_{h-1},l_{h-2},\cdots,l_{1}\right).
$$
It is necessary to prove the following
\begin{propo}
$$
\widetilde{Q}_{D,h,j}\left(\bm{l},\bm{l'}_h\right)=0
$$
if there exists at least one $p\in\{d-1,\cdots,j-1\}$ such that $|l_p-{l'}_p|=2$.
\proof
First of all, if $l_{d-1}\neq{l'}_{d-1}$, for example ${l'}_{d-1}=l_{d-1}+2$ (the case ${l'}_{d-1}=l_{d-1}-2$ is similar), then
\begin{equation*}
\widetilde{Q}_{D,h,j}\left(\bm{l},\bm{l'}_h\right)=-d_{l,l_{d-1}+1,D}d_{l,l_{d-1}+2,D}{Q}_{D,h,j}\left(\bm{l},\bm{l'}_h\right),
\end{equation*}
but ${Q}_{D,h,j}\left(\bm{l},\bm{l'}_h\right)=0$ because of (\ref{commtatb}), and this implies $\widetilde{Q}_{D,h,j}\left(\bm{l},\bm{l'}_h\right)=0$.

Furthermore, if $j\leq d-1$ and $l_{d-1}={l'}_{d-1}$, while $l'_{d-2}=l_{d-2}+2$ (the case $l'_{d-2}=l_{d-2}-2$ is similar), then it must be
\begin{equation*}
\begin{split}
\left\langle Y_{\bm{l'}_h},  L_{h,D}L_{j,D}Y_{\bm{l}}\right\rangle=&\left[d_{l,l_{d-1},D}B\left(l_{d-1},l_{d-2},d-1\right)d_{l,l_{d-1},D}A\left(l_{d-1}-1,l_{d-2}+1,d-1\right)\right.\\
&\left.+d_{l,l_{d-1}+1,D}A\left(l_{d-1},l_{d-2},d-1\right)d_{l,l_{d-1}+1,D}B\left(l_{d-1}+1,l_{d-2}+1,d-1\right)\right]g\left(\bm{l},\bm{l'}_h
\right)\\
=:&\widetilde{g}\left(l,l_{d-1},l_{d-2}\right)g\left(\bm{l},\bm{l'}_h
\right),
\end{split}
\end{equation*}
for a certain function $g$ and, similarly,
\begin{equation*}
\begin{split}
\left\langle Y_{\bm{l'}_h},L_{j,D}L_{h,D}Y_{\bm{l}}\right\rangle=&\left[d_{l,l_{d-1},D}B\left(l_{d-1},l_{d-2},d-1\right)d_{l,l_{d-1},D}A\left(l_{d-1}-1,l_{d-2}+1,d-1\right)\right.\\
&\left.+d_{l,l_{d-1}+1,D}A\left(l_{d-1},l_{d-2},d-1\right)d_{l,l_{d-1}+1,D}B\left(l_{d-1}+1,l_{d-2}+1,d-1\right)\right]g\left(\bm{l},\bm{l'}_h
\right)\\
=:&\widetilde{g}\left(l,l_{d-1},l_{d-2}\right)g\left(\bm{l},\bm{l'}_h
\right),
\end{split}
\end{equation*}
for the same function $g$, because $Q_{D,h,j}\left(\bm{l},\bm{l'}_h\right)=0$; so, also in this case $
\widetilde{Q}_{D,h,j}\left(\bm{l},\bm{l'}_h\right)=0.
$

In general, if there is a $p\in\{d-1,\cdots,j+1\}$ such that ${l'}_{p-1}=l_{p-1}+2$ (the case ${l'}_{p-1}=l_{p-1}-2$ is similar), while $l_q={l'}_q$ $\forall q\geq p$, then
\begin{equation*}
\begin{split}
\left\langle Y_{\bm{l'}_h},L_{h,D}L_{j,D}Y_{\bm{l}}\right\rangle=&g_1\left(l,l_{d-1},\cdots,l_p\right)\\
&\cdot\left[A\left(l_{p},l_{p-1},p\right)B\left(l_{p}+1,l_{p-1}+1,p\right)+B\left(l_{p},l_{p-1},p\right)A\left(l_{p}-1,l_{p-1}+1,p\right)\right]\\
&\cdot g_2\left(\bm{l},\bm{l'}_h
\right),
\end{split}
\end{equation*}
and
\begin{equation*}
\begin{split}
\left\langle Y_{\bm{l'}_h}, L_{j,D}L_{h,D}Y_{\bm{l}}\right\rangle=&g_1\left(l,l_{d-1},\cdots,l_p\right)\\
&\cdot\left[A\left(l_{p},l_{p-1},p\right)B\left(l_{p}+1,l_{p-1}+1,p\right)+B\left(l_{p},l_{p-1},p\right)A\left(l_{p}-1,l_{p-1}+1,p\right)\right]\\
&\cdot g_2\left(\bm{l},\bm{l'}_h
\right),
\end{split}
\end{equation*}
for the same functions $g_1$ (because $l_q={l'}_q$ $\forall q\geq p$) and $g_2$ [because $Q_{D,h,j}\left(\bm{l},\bm{l'}_h\right)=0$]; so, also in this case, $
\widetilde{Q}_{D,h,j}\left(\bm{l},\bm{l'}_h\right)=0.
$

Finally, if ${l'}_{j-1}=l_{j-1}+2$ (the case ${l'}_{j-1}=l_{j-1}-2$ is similar), ${l'}_{j-2}=l_{j-2}+1$ (also here, the case ${l'}_{j-2}=l_{j-2}-1$ is similar), while $l_q={l'}_q$ $\forall q\geq j$, then
\begin{equation*}
\begin{split}
\left\langle Y_{\bm{l'}_h},L_{h,D}L_{j,D}Y_{\bm{l}}\right\rangle=&g_3\left(l,l_{d-1},\cdots,l_j\right)\\
&\cdot\left[F\left(l_{j-1},l_{j-2},j-1\right)A\left(l_{j-1}+1,l_{j-2},j-1\right)\right]\\
&\cdot g_4\left(\bm{l},\bm{l'}_h
\right),
\end{split}
\end{equation*}
\begin{equation*}
\begin{split}
\left\langle Y_{\bm{l'}_h},L_{j,D}L_{h,D}Y_{\bm{l}}\right\rangle=&g_3\left(l,l_{d-1},\cdots,l_j\right)\\
&\cdot \left[A\left(l_{j-1},l_{j-2},j-1\right)F\left(l_{j-1}+1,l_{j-2}+1,j-1\right)\right]\\
&\cdot g_4\left(\bm{l},\bm{l'}_h
\right),
\end{split}
\end{equation*}
for the same functions $g_3$ (because $l_q={l'}_q$ $\forall q\geq j$) and $g_4$ [because $Q_{D,h,j}\left(\bm{l},\bm{l'}_h\right)=0$]; so, because of (\ref{relazioniABCDFG}), $
\widetilde{Q}_{D,h,j}\left(\bm{l},\bm{l'}_h\right)=0.
$
\endproof
\end{propo}
The last proof implies
\begin{equation}\label{az_LaLb}
\left[L_{h,D},L_{j,D}\right]Y_{
\bm{l}
}
=\sum_{
\substack{
|{l'}_p-l_p|=1\\
p=j-2,\cdots,h-1
}}\widetilde{Q}_{D,h,j}\left(\bm{l},\bm{l'}_h\right)Y_{
\bm{l'}_h
},
\end{equation}
and from now on assume that $l'_p=l_p$ $\forall p\geq j-1$, otherwise $\widetilde{Q}_{D,h,j}\left(\bm{l},\bm{l'}_h\right)=0$.

It is necessary to further investigate about these 
$$
\widetilde{Q}_{D,h,j}\left(\bm{l},\bm{l'}_h\right)\quad\mbox{when }|{l'}_p-l_p|=1, p\in\{j-2,\cdots,h-1\}.
$$
In order to do this, let
\begin{equation*}
\begin{split}
T_1^1(l_{p},l_{p-1},p)&:=A(l_p,l_{p-1},p)G(l_p+1,l_{p-1}+1,p)-F(l_p,l_{p-1},p)B(l_p+1,l_{p-1},p),\\
T_2^1(l_{p},l_{p-1},p)&:=B(l_p,l_{p-1},p)F(l_p-1,l_{p-1}+1,p)-G(l_p,l_{p-1},p)A(l_p-1,l_{p-1},p),\\
T_3^1(l_{p},l_{p-1},p)&:=C(l_p,l_{p-1},p)G(l_p+1,l_{p-1}-1,p)-F(l_p,l_{p-1},p)D(l_p+1,l_{p-1},p),\\
T_4^1(l_{p},l_{p-1},p)&:=D(l_p,l_{p-1},p)F(l_p-1,l_{p-1}-1,p)-G(l_p,l_{p-1},p)C(l_p-1,l_{p-1},p);
\end{split}
\end{equation*}
they fulfill
\begin{equation}
\begin{split}
T_1^1(l_{p},l_{p-1},p)&=\frac{\sqrt{(l_p+l_{p-1}+p-1)(l_{p}-l_{p-1})}}{2l_{p}+p-1}=\frac{d_{l_{p},l_{p-1}+1,p+1}}{2l_{p}+p-1},\\
T_2^1(l_{p},l_{p-1},p)&=-\frac{\sqrt{(l_p+l_{p-1}+p-1)(l_{p}-l_{p-1})}}{2l_{p}+p-1}=-\frac{d_{l_{p},l_{p-1}+1,p+1}}{2l_{p}+p-1},\\
T_3^1(l_{p},l_{p-1},p)&=-\frac{\sqrt{(l_p+l_{p-1}+p-2)(l_{p}-l_{p-1}+1)}}{2l_{p}+p-1}=-\frac{d_{l_{p},l_{p-1},p+1}}{2l_{p}+p-1},\\
T_4^1(l_{p},l_{p-1},p)&=\frac{\sqrt{(l_p+l_{p-1}+p-2)(l_{p}-l_{p-1}+1)}}{2l_{p}+p-1}=\frac{d_{l_{p},l_{p-1},p+1}}{2l_{p}+p-1}.
\end{split}\label{valueT_1}
\end{equation}

Similarly, for $n\geq 2$, let
\begin{equation}\label{Tn}
\begin{split}
T_1^n(l_{p+n-1},l_p,l_{p-1},p):=&\left[A(l_{p+n-1},l_{p+n-2},p+n-1)\right]^2 T_1^{n-1}(l_{p+n-2},l_p,l_{p-1},p)\\
&+\left[C(l_{p+n-1},l_{p+n-2},p+n-1)\right]^2 T_2^{n-1}(l_{p+n-2},l_p,l_{p-1},p)\\
=&\frac{d_{l_{p},l_{p-1}+1,p+1}}{2l_{p+n-1}+p+n-2},\\
T_2^n(l_{p+n-1},l_p,l_{p-1},p):=&\left[B(l_{p+n-1},l_{p+n-2},p+n-1)\right]^2 T_1^{n-1}(l_{p+n-2},l_p,l_{p-1},p)\\
&+\left[D(l_{p+n-1},l_{p+n-2},p+n-1)\right]^2 T_2^{n-1}(l_{p+n-2},l_{p-1},p)\\
=&-\frac{d_{l_{p},l_{p-1}+1,p+1}}{2l_{p+n-1}+p+n-2},\\
T_3^n(l_{p+n-1},l_p,l_{p-1},p):=&\left[A(l_{p+n-1},l_{p+n-2},p+n-1)\right]^2 T_3^{n-1}(l_{p+n-2},l_p,l_{p-1},p)\\
&+\left[C(l_{p+n-1},l_{p+n-2},p+n-1)\right]^2 T_4^{n-1}(l_{p+n-2},l_p,l_{p-1},p)\\
=&-\frac{d_{l_{p},l_{p-1},p+1}}{2l_{p+n-1}+p+n-2},\\
T_4^n(l_{p+n-1},l_p,l_{p-1},p):=&\left[B(l_{p+n-1},l_{p+n-2},p+n-1)\right]^2 T_3^{n-1}(l_{p+n-2},l_p,l_{p-1},p)\\
&+\left[D(l_{p+n-1},l_{p+n-2},p+n-1)\right]^2 T_4^{n-1}(l_{p+n-2},l_p,l_{p-1},p)\\
=&\frac{d_{l_{p},l_{p-1},p+1}}{2l_{p+n-1}+p+n-2}.
\end{split}
\end{equation}
Assume (witout loss of generality) that ${l'}_{j-2}=l_{j-2}-1$, then
\begin{equation}\label{diff1}
\begin{split}
\widetilde{Q}_{D,h,j}\left(\bm{l},\bm{l'}_h\right)=\left\{\left(d_{l,l_{d-1}+1,D}
\right)^2\right.\left\{ \left[A(l_{d-1},l_{d-2},d-1)\right]^2\right.&\left. \left\{\left[A(l_{d-2},l_{d-3},D-3)\right]^2\left\{
\cdots 
\right\}\right.\right.\\
&\left.\left.
+\left[C(l_{d-2},l_{d-3},D-3)\right]^2\left\{
\cdots  \right\}\right\}\right.\\
+\left[C(l_{d-1},l_{d-2},d-1)\right]^2 &\left\{\left[B(l_{d-2},l_{d-3},D-3)\right]^2\left\{
\cdots  \right\}\right.\\
&\left.\left.+\left[D(l_{d-2},l_{d-3},D-3)\right]^2\left\{
\cdots  \right\}\right\}
\right\}\\
+\left(d_{l,l_{d-1},D}
\right)^2\left\{ \left[B(l_{d-1},l_{d-2},d-1)\right]^2\right.&\left. \left\{\left[A(l_{d-2},l_{d-3},D-3)\right]^2\left\{
\cdots 
\right\}\right.\right.\\
&\left.\left.
+\left[C(l_{d-2},l_{d-3},D-3)\right]^2\left\{
\cdots  \right\}\right\}\right.\\
+\left[D(l_{d-1},l_{d-2},d-1)\right]^2& \left\{\left[B(l_{d-2},l_{d-3},D-3)\right]^2\left\{
\cdots  \right\}\right.\\
&\left.\left.\left.+\left[D(l_{d-2},l_{d-3},D-3)\right]^2\left\{
\cdots  \right\}\right\}
\right\}\right\}\\
\cdot R_{h,j-1}\left(\bm{l}\vert_{j,h},\widetilde{\bm{l'}}\vert_{j,h}\right),
\end{split}
\end{equation}
where
$$
\bm{l}\vert_{j,h}:=\left(l_{j-2},\cdots,l_{h-1},l_{h-2}\right)\quad,\quad \widetilde{\bm{l'}}\vert_{j,h}:=\left({l}_{j-2}-1,\cdots,{l'}_{h-1},l_{h-2}\right).
$$
\begin{remark}\label{remark5}
The $\left\{\cdots\right\}$ in \emph{(\ref{diff1})} is such that
\begin{itemize}
\item every $\left[A(l_{h},l_{h-1},h)\right]^2$ is always left-multiplied by $\left[A(l_{h+1},l_{h},h+1)\right]^2$ or $\left[B(l_{h+1},l_{h},h+1)\right]^2$;
\item every $\left[B(l_{h},l_{h-1},h)\right]^2$ is always left-multiplied by $\left[C(l_{h+1},l_{h},h+1)\right]^2$ or $\left[D(l_{h+1},l_{h},h+1)\right]^2$;
\item every $\left[C(l_{h},l_{h-1},h)\right]^2$ is always left-multiplied by $\left[A(l_{h+1},l_{h},h+1)\right]^2$ or $\left[B(l_{h+1},l_{h},h+1)\right]^2$;
\item every $\left[D(l_{h},l_{h-1},h)\right]^2$ is always left-multiplied by $\left[C(l_{h+1},l_{h},h+1)\right]^2$ or $\left[D(l_{h+1},l_{h},h+1)\right]^2$;
\item the most `internal' term of $\left\{\cdots\right\}$ is  $T_p\left(l_{j-1},l_{j-2},j\right)$, with $p\in\{1,2,3,4\}$;
\item every $T_1^1\left(l_{j-1},l_{j-2},j\right)$ and $T_2^1\left(l_{j-1},l_{j-2},j\right)$ are always left-multiplied by $\left[A(l_{j},l_{j-1},j)\right]^2$ or $\left[B(l_{j},l_{j-1},j)\right]^2$;
\item every $T_3^1\left(l_{j-1},l_{j-2},j\right)$ and $T_4^1\left(l_{j-1},l_{j-2},j\right)$ are always left-multiplied by $\left[C(l_{j},l_{j-1},b)\right]^2$ or $\left[D(l_{j},l_{j-1},j)\right]^2$.
\end{itemize}
\end{remark}
This and (\ref{Tn}) imply
\begin{equation}\label{diff2}
\begin{split}
\widetilde{Q}_{D,h,j}\left(\bm{l},\bm{l'}_h\right)=\left\{\left(d_{l,l_{d-1}+1,D}
\right)^2\right.\left\{\right.&\left[A(l_{d-1},l_{D-3},D-2)\right]^2 T_3^{D-j-1}\left(l_{D-3},l_{j-1},l_{j-2},j-1\right)
\\
&\left.+\left[C(l_{d-1},l_{D-3},D-2)\right]^2 T_4^{D-j-1}\left(l_{D-3},l_{j-1},l_{j-2},j-1\right)\right\}\\
+\left(d_{l,l_{d-1},D}
\right)^2\left\{\right.& \left[B(l_{d-1},l_{D-3},D-2)\right]^2 T_3^{D-j-1}\left(l_{D-3},l_{j-1},l_{j-2},j-1\right)\\
&\left.\left.+\left[D(l_{d-1},l_{D-3},D-2)\right]^2 T_4^{D-j-1}\left(l_{D-3},l_{j-1},l_{j-2},j-1\right)\right\}\right\}\\
&\cdot R_{h,j-1}\left(\bm{l}\vert_{j,h},\widetilde{\bm{l'}}\vert_{j,h}\right)\\
\overset{(\ref{Tn})}=\left[ 
-\left(d_{l,l_{d-1}+1,D}
\right)^2+\left(d_{l,l_{d-1},D}
\right)^2
\right]&\frac{d_{l_{j-1},l_{j-2},j}}{2l_{d-1}+D-3} R_{h,j-1}\left(\bm{l}\vert_{j,h},\widetilde{\bm{l'}}\vert_{j,h}\right)\\
=d_{l_{j-1},l_{j-2},j}R_{h,j-1}\left(\bm{l}\vert_{j,h},\widetilde{\bm{l'}}\vert_{j,h}\right);
\end{split}
\end{equation}
and this proves the following
\begin{teorema}
The equation \emph{(\ref{diff1})}, using the rules of {{remark \ref{remark5}}}, becomes
\begin{equation}\label{valueQ}
\widetilde{Q}_{D,h,j}\left(\bm{l},\bm{l'}_h\right)=d_{l_{j-1},l_{j-2},j} R_{h,j-1}\left(\bm{l}\vert_{j,h},\widetilde{\bm{l'}}\vert_{j,h}\right).
\end{equation}
\end{teorema}
The same job can be done with the assumption ${l'}_{j-2}=l_{j-2}+1$, in this case the result is an equation like (\ref{diff1}), but with $T_1$ and $T_2$ instead of $T_3$ and $T_4$, respectively; and in this case it turns out that
\begin{equation}\label{valueQ2}
\widetilde{Q}_{D,h,j}\left(\bm{l},\bm{l'}_h\right)=-d_{l_{j-1},l_{j-2}+1,j}\cdot  R_{h,j-1}\left(\bm{l}\vert_{j,h},\widehat{\bm{l'}}\vert_{j,h}\right),
\end{equation}
where $\widehat{\bm{l'}}:=\left({l}_{j-2}+1,\cdots,{l'}_{h-1},l_{h-2}\right)$.

Finally, definition \ref{defiL} and (\ref{valueQ}-\ref{valueQ2}) imply
\be\label{commDjDw} 
\left[L_{h,D},L_{j,D}\right]=iL_{h,j}.
\ee
\subsubsection{Proof of (\ref{toproveL})$_2$ and (\ref{toproveL})$_3$}\label{proofLx} 
Let $1\leq h<j\leq d$, the {definition \ref{defiL}} implies that the action of $L_{j,D}$ in $\mathbb{R}^{D}$ is the same of $t^j$ in $\mathbb{R}^{d}$, the only difference is given by the $\frac{1}{i}d_{l,l'_{d-1},D}$ coefficients and their signs (here $l'_{d-1}=l_{d-1}$ or $l'_{d-1}=l_{d-1}+1$), but anyway the action of $L_{h,j}$ on a $Y_{\bm{l}}$ does not change the indices $l$ and $l_{d-1}$.

According to this and (\ref{commLhjxp}), from
$$
\left[L_{h,j},x^j\right]=\frac{1}{i} x^h\quad\mbox{it follows}\quad \left[L_{h,j},L_{j,D}\right]=\frac{1}{i}L_{h,D},
$$
and from
$$
\left[L_{h,j},x^p\right]=0\quad \mbox{if}\quad p\neq h,j\quad\mbox{it follows}\quad \left[L_{h,j},L_{p,D}\right]=0 \mbox{ if }D\neq h,j\mbox{ and }p\neq h,j.
$$
\subsection{On the action of `projected' coordinate operators $\overline{x}^h$}\label{scalprod}
{The behavior (\ref{pure_psi})-(\ref{statopsi}) of a generic $\psi_{\bm{l},D}$ and the expression of the integration measure $d\bm{x}$ of $\mathbb{R}^D$ in spherical coordinates
$$
d\bm{x}=r^{D-1}\sin^{d-1}{(\theta_{d})} \sin^{d-2}{(\theta_{d-1})} \cdots \sin{(\theta_2)} dr d\theta_1 d\theta_2\cdots d\theta_{d}
$$
allow to factorize the scalar product $\left\langle\psi_{\bm{l'},D},\overline{x}^{h}\psi_{\bm{l},D}\right\rangle_{\mathbb{R}^D}$
in this way:
$$
\left\langle\psi_{\bm{l'},D},\overline{x}^{h}\psi_{\bm{l},D}\right\rangle=\left\langle f_{0,l',D},r f_{0,l,D}\right\rangle_{R^+} \cdot \left\langle Y_{\bm{l'}}, t^{h}Y_{\bm{l}}\right\rangle_{S^d},
$$
where
\be\label{psixpsi_r}
\left\langle f_{0,l',D},r f_{0,l,D}\right\rangle_{\mathbb{R}^+}:=M_{l,D}M_{l',D}\int_0^{+\infty}re^{-\sqrt{k_{l,D}}\left(r-\widetilde{r}_{l,D}\right)^2}e^{-\sqrt{k_{l',D}}\left(r-\widetilde{r}_{l',D}\right)^2}dr,
\ee
while the value of the `angular' scalar product
$$
\left\langle Y_{\bm{l'}}, t^{h}Y_{\bm{l}}\right\rangle_{S^d}=\int_{S^d}Y_{\bm{l'}}^*t^hY_{\bm{l}}\left[\sin^{d-1}{(\theta_{d})} \sin^{d-2}{(\theta_{d-1})} \cdots \sin{(\theta_2)}\right] d\theta_1 d\theta_2\cdots d\theta_{d}
$$
is
$$
\left\langle Y_{\bm{l'}}, t^{h}Y_{\bm{l}}\right\rangle_{S^d}\equiv R_{h,D}\left(\bm{l},\bm{l'}\right),
$$
according to sections \ref{D-dimsa} and \ref{Appendix_defi}.

On the other hand, as for {section 6.6} in \cite{FiorePisacane}, 
$$
\left\langle f_{0,l\pm 1,D},r f_{0,l,D}\right\rangle_{\mathbb{R}^+}=M_{l,D}M_{l\pm1,D}\hspace{0.2cm}e^{-\frac{\sqrt{k_{l,D}k_{l\pm1,D}}\left(\widetilde{r}_{l,D}-\widetilde{r}_{l\pm1,D} \right)^2}{2\left(\sqrt{k_{l,D}}+\sqrt{k_{l\pm1,D}}\right)}}\sqrt{\frac{2\pi}{\sqrt{k_{l,D}}+\sqrt{k_{l\pm1,D}}}}\hspace{0.2cm}\widehat{r}_{l,l\pm1,D};
$$
with
\be\label{widehatr}
\widehat{r}_{l,l\pm1,D}=\frac{\sqrt{k_{l,D}}\widetilde{r}_{l,D}+\sqrt{k_{l\pm1,D}}\widetilde{r}_{l\pm1,D}}{\sqrt{k_{l,D}}+\sqrt{k_{l\pm1,D}}}.
\ee
Then, in order to calculate $\left\langle f_{0,l\pm 1,D},r f_{0,l,D}\right\rangle_{\mathbb{R}^+}$ at leading orders in $1/\sqrt{k}$, the following steps are needed.

First of all
\begin{equation}\label{rtilde}
\begin{split}
\widetilde{r}_{l,D}=&1+\frac{b(l,D)}{2k}-3\frac{\left[b(l,D)\right]^2}{4k^2}+9\frac{\left[b(l,D)\right]^3}{8k^3}-\frac{27\left[b(l,D)\right]^4}{16k^4}+O\left(k^{-5}\right);
\end{split}
\end{equation}
while
\begin{equation}\label{sqrtk}
\begin{split}
\sqrt{k_{l,D}}=&\sqrt{2k}+\frac{3}{2\sqrt{2k}}b(l,D)-\frac{9}{8}\frac{\left[b(l,D)\right]^2}{2k\sqrt{2k}}+\frac{27}{16}\frac{\left[b(l,D)\right]^3}{4k^2\sqrt{2k}}\\
&-\frac{405}{128}\frac{\left[b(l,D)\right]^4}{8k^3\sqrt{2k}}+O\left(k^{-4}\right),\\
\end{split}
\end{equation}
implies
\begin{equation}\label{sqrtkk}
\begin{split}
\sqrt{k_{l,D}k_{l\pm 1,D}}=&2k+\frac{3}{2}\left[b(l,D)+b(l\pm1,D)\right]-\frac{9}{8}\frac{\left[b(l,D)-b(l\pm1,D)\right]^2}{2k}\\
&+\frac{27}{16}\frac{\left[b(l,D)\right]^3+\left[b(l\pm1,D)\right]^3-\left[b(l,D)\right]^2\left[b(l\pm1,D)\right]}{4k^2}\\
&-\frac{\left[b(l\pm1,D)\right]^2\left[b(l,D)\right]}{4k^2}+O\left(k^{-3}\right),
\end{split}
\end{equation}
\begin{equation*}
\begin{split}
\sqrt{k_{l,D}}+\sqrt{k_{l\pm1,D}}=&2\sqrt{2k}+\frac{3}{2}\frac{b(l,D)+b(l\pm1,D)}{\sqrt{2k}}-\frac{9}{8}\frac{\left\{\left[b(l,D)\right]^2+\left[b(l\pm1,D)\right]^2\right\}}{2k\sqrt{2k}}\\
&+\frac{27}{16}\frac{\left\{\left[b(l,D)\right]^3+\left[b(l\pm1,D)\right]^3\right\}}{4k^2\sqrt{2k}}-\frac{405}{128}\frac{\left\{\left[b(l,D)\right]^4+\left[b(l\pm1,D)\right]^4\right\}}{8k^3\sqrt{2k}}\\
&+O\left(k^{-3}\right),\end{split}
\end{equation*}
\begin{equation}\label{1sukk}
\begin{split}
\frac{1}{\sqrt{k_{l,D}}+\sqrt{k_{l\pm1,D}}}=&\frac{1}{2\sqrt{2k}}-\frac{3}{8}\frac{\left[b(l,D)+b(l\pm1,D)\right]}{(2k)^{\frac{3}{2}}}\\
&+\frac{9}{16}\frac{\left[b(l,D)\right]^2+\left[b(l\pm1,D)\right]^2+b(l,D)b(l\pm1,D)}{(2k)^{\frac{5}{2}}}+O\left(k^{-3}\right),\\
\end{split}
\end{equation}
\begin{equation*}
\begin{split}
\sqrt{\sqrt{k_{l,D}}+\sqrt{k_{l\pm1,D}}}=&\sqrt{2}\sqrt[4]{2k}+\frac{3\sqrt{2}}{8}\frac{b(l,D)+b(l\pm1,D)}{(2k)^{\frac{3}{4}}}\\
&-\frac{45\sqrt{2}}{128}\frac{\left[b(l,D)\right]^2+\left[b(l\pm1,D)\right]^2+\frac{2}{5}b(l,D)b(l\pm1,D)}{(2k)^{\frac{7}{4}}}\\
&-\frac{567\sqrt{2}}{1024}\frac{\left[b(l,D)\right]^3+\left[b(l+1,D)\right]^2+\frac{1}{3}\left[b(l,D)\right]^2b(l\pm1,D)+\frac{1}{3}b(l,D)\left[b(l\pm1,D)\right]^2}{(2k)^{\frac{11}{4}}}\\
&+O\left(k^{-3}\right),\\
\end{split}
\end{equation*}
\begin{equation}\label{1susqrtkperk}
\begin{split}
\frac{1}{\sqrt{\sqrt{k_{l,D}}+\sqrt{k_{l\pm1,D}}}}=&\frac{1}{\sqrt{2}\sqrt[4]{2k}}-\frac{3\sqrt{2}}{16}\frac{b(l,D)+b(l\pm1,D)}{(2k)^{\frac{5}{4}}}\\
&+\frac{63\sqrt{2}}{256}\frac{\left[b(l,D)\right]^2+\left[b(l\pm1,D)\right]^2+\frac{6}{7}b(l,D)b(l\pm1,D)}{(2k)^{\frac{9}{4}}}+O\left(k^{-3}\right).\\
\end{split}
\end{equation}
So, from (\ref{valueM_l,D}) and (\ref{sqrtkk}), it follows
\begin{equation*}
\begin{split}
\sqrt{\pi}M_{l,D}M_{l\pm1,D}=&\sqrt[8]{k_{l,D}k_{l\pm1,D}}=\sqrt[4]{2k}+\frac{3}{8}\frac{\left[b(l,D)+b(l\pm1,D)\right]}{(2k)^{\frac{3}{4}}}\\
&-\frac{63}{128}\frac{\left[b(l,D)\right]^2+\left[b(l\pm1,D)\right]^2-\frac{2}{7}b(l,D)b(l\pm1,D)}{(2k)^{\frac{7}{4}}}\\
&+\frac{945}{1024}\frac{\left[b(l,D)\right]^3+\left[b(l\pm1,D)\right]^3-\frac{1}{5}\left[b(l,D)\right]^2b(l\pm1,D)}{(2k)^{\frac{11}{4}}}\\
&-\frac{\frac{1}{5}\left[b(l\pm1,D)\right]^2b(l,D)}{(2k)^{\frac{11}{4}}}+O\left(k^{-3}\right),
\end{split}
\end{equation*}
and then
\begin{equation}\label{valueMMsqrtpi}
\begin{split}
M_{l,D}M_{l\pm1,D}\sqrt{\frac{2\pi}{\sqrt{k_{l,D}}+\sqrt{k_{l\pm1,D}}}}\overset{(\ref{1susqrtkperk})}=1-\frac{9}{64}\frac{\left[b(l,D)-b(l\pm1,D)\right]^2}{4k^2}+O\left(k^{-3}\right).
\end{split}
\end{equation}
Furthermore, from
\begin{equation*}
\begin{split}
\sqrt{k_{l,D}}\widetilde{r}_{l,D}\overset{(\ref{rtilde})\& (\ref{sqrtk})}=&\sqrt{2k}+\frac{5b(l,D)}{2\sqrt{2k}}-\frac{21}{8}\frac{\left[b(l,D)\right]^2}{2k\sqrt{2k}}+\frac{81}{16}\frac{\left[b(l,D)\right]^3}{4k^2\sqrt{2k}}+O\left(k^{-3}\right),\\
\end{split}
\end{equation*}
it follows
\begin{equation*}
\begin{split}
\sqrt{k_{l,D}}\widetilde{r}_{l,D}+\sqrt{k_{l+1,D}}\widetilde{r}_{l+1,D}=&2\sqrt{2k}+\frac{5\left[b(l,D)+b(l\pm1,D)\right]}{2\sqrt{2k}}-\frac{21}{8}\frac{\left\{\left[b(l,D)\right]^2+\left[b(l\pm1,D)\right]^2\right\}}{2k\sqrt{2k}}\\
&+\frac{81}{16}\frac{\left\{\left[b(l,D)\right]^3+\left[b(l\pm1,D)\right]^3\right\}}{4k^2\sqrt{2k}}+O\left(k^{-3}\right);
\end{split}
\end{equation*}
then the last equalities and (\ref{widehatr}) imply
\begin{equation*}
\begin{split}
\widehat{r}_{l,l\pm1,D}=&1+\frac{1}{2}\frac{b(l,D)+b(l\pm1,D)}{2k}-\frac{9}{8}\frac{\left[b(l,D)\right]^2+\left[b(l+1,D)\right]^2+\frac{2}{3}b(l,D)b(l\pm1,D)}{4k^2}\\
&+O\left(k^{-3}\right);
\end{split}
\end{equation*}
Similarly,
\begin{equation}\label{differenzartilde}
\begin{split}
\left(\widetilde{r}_{l,D}-\widetilde{r}_{l\pm1,D} \right)^2\overset{(\ref{rtilde})}=&\frac{\left[b(l,D)-b(l\pm1,D)\right]^2}{4k^2}\\
&-\frac{6\left\{\left[b(l,D)\right]^3+\left[b(l\pm1,D)\right]^3-\left[b(l,D)\right]^2b(l\pm1,D)\left[b(l\pm1,D)\right]^2b(l,D) \right\}}{8k^3}\\
&+O\left(k^{-4}\right),\\
\end{split}
\end{equation}
\begin{equation*}
\begin{split}
&\frac{\sqrt{k_{l,D}k_{l\pm1,D}}\left(\widetilde{r}_{l,D}-\widetilde{r}_{l\pm1,D} \right)^2}{2\left(\sqrt{k_{l,D}}+\sqrt{k_{l\pm1,D}}\right)}\overset{(\ref{sqrtkk}),(\ref{1sukk})\&(\ref{differenzartilde})}=\frac{1}{4}\frac{\left[b(l,D)-b(l\pm1,D)\right]^2}{(2k)^{\frac{3}{2}}}\\
&-\frac{21}{16}\frac{\left[b(l,D)\right]^3+\left[b(l\pm1,D)\right]^3-\left[b(l,D)\right]^2b(l\pm1,D)-\left[b(l\pm1,D)\right]^2b(l,D)}{(2k)^{\frac{5}{2}}}+O\left(k^{-3}\right),
\end{split}
\end{equation*}
which implies
\begin{equation}\label{value_exponential}
\begin{split}
e^{-\frac{\sqrt{k_{l,D}k_{l\pm1,D}}\left(\widetilde{r}_{l,D}-\widetilde{r}_{l\pm1,D} \right)^2}{2\left(\sqrt{k_{l,D}}+\sqrt{k_{l\pm1,D}}\right)}}=&1-\frac{1}{4}\frac{\left[b(l,D)-b(l\pm1,D)\right]^2}{(2k)^{\frac{3}{2}}}\\
&+\frac{21}{16}\frac{\left[b(l,D)\right]^3+\left[b(l\pm1,D)\right]^3-\left[b(l,D)\right]^2b(l\pm1,D)-\left[b(l\pm1,D)\right]^2b(l,D)}{(2k)^{\frac{5}{2}}}\\
&+O\left(k^{-3}\right).\\
\end{split}
\end{equation}
So, according to the above equalities,
\be\label{valuescalprod}
\begin{split}
\left\langle f_{0,l\pm 1,D},r f_{0,l,D}\right\rangle_{\mathbb{R}^+}=&1+\frac{1}{2}\frac{\left[b(l,D)+b(l\pm1,D)\right]}{2k}-\frac{1}{4}\frac{\left[b(l,D)-b(l\pm1,D) \right]^2}{(2k)^{\frac{3}{2}}}\\
&-\frac{81}{64}\frac{\left[b(l,D)\right]^2+\left[b(l\pm1,D)\right]^2+\frac{54}{5}b(l,D)b(l\pm1,D)}{4k^2}+O\left(k^{-\frac{5}{2}}\right).
\end{split}
\ee

Furthermore, (\ref{defc}) and this last equality imply
\be\label{diffc}
\left(c_{l+1,D}\right)^2-\left(c_{l,D}\right)^2=\frac{b(l+1,D)-b(l-1,D)}{2k}+O\left(\frac{1}{k^2}\right)=\frac{2l+D-2}{k}+O\left(\frac{1}{k^2}\right).
\ee

Similarly, the scalar product $\left\langle\psi_{\bm{l'},D},t^{h}\psi_{\bm{l},D}\right\rangle_{\mathbb{R}^D}$ can be factorized, obtaining
$$
\left\langle\psi_{\bm{l'},D},t^{h}\psi_{\bm{l},D}\right\rangle=\left\langle f_{0,l',D}, f_{0,l,D}\right\rangle_{R^+} \cdot \left\langle Y_{\bm{l'}}, t^{h}Y_{\bm{l}}\right\rangle_{S^d},
$$
and also in this case
$$
\left\langle Y_{\bm{l'}}, t^{h}Y_{\bm{l}}\right\rangle_{S^d}\equiv R_{h,D}\left(\bm{l},\bm{l'}\right)
$$
does not vanish if $l'=\pm 1$. On the other hand, as for the previous `radial' scalar product, 
\begin{equation*}
\begin{split}
\left\langle f_{0,l\pm 1,D}, f_{0,l,D}\right\rangle_{R^+}=&M_{l,D}M_{l\pm 1,D}\int_0^{+\infty}e^{-\frac{\sqrt{k_{l,D}}}{2}\left(r-\widetilde{r}_{l,D}\right)^2}e^{-\frac{\sqrt{k_{l\pm 1,D}}}{2}\left(r-\widetilde{r}_{l\pm 1,D}\right)^2}dr\\
\simeq & M_{l,D}M_{l\pm 1,D}\int_{-\infty}^{+\infty}e^{-\frac{\sqrt{k_{l,D}}}{2}\left(r-\widetilde{r}_{l,D}\right)^2}e^{-\frac{\sqrt{k_{l\pm 1,D}}}{2}\left(r-\widetilde{r}_{l\pm 1,D}\right)^2}dr,
\end{split}
\end{equation*}
with
\begin{equation*}
\begin{split}
&\int_{-\infty}^{+\infty}e^{-\frac{\sqrt{k_{l,D}}}{2}\left(r-\widetilde{r}_{l,D}\right)^2}e^{-\frac{\sqrt{k_{l\pm 1,D}}}{2}\left(r-\widetilde{r}_{l\pm 1,D}\right)^2}dr\\
=&e^{-\frac{\sqrt{k_{l,D}}\widetilde{r}_{l,D}^2+\sqrt{k_{l\pm 1,D}}\widetilde{r}_{l\pm 1,D}^2}{2}}\int_{-\infty}^{+\infty}e^{-r^2\frac{\sqrt{k_{l,D}}+\sqrt{k_{l\pm 1,D}}}{2}+2r\frac{\sqrt{k_{l,D}}\widetilde{r}_{l,D}+\sqrt{k_{l\pm 1,D}}\widetilde{r}_{l\pm 1,D}}{2}}dr\\
=&e^{-\frac{\sqrt{k_{l,D}}\widetilde{r}_{l,D}^2+\sqrt{k_{l\pm 1,D}}\widetilde{r}_{l\pm 1,D}^2}{2}+\frac{\left(\sqrt{k_{l,D}}\widetilde{r}_{l,D}+\sqrt{k_{l\pm 1,D}}\widetilde{r}_{l\pm 1,D} \right)^2}{2\left(\sqrt{k_{l,D}}+\sqrt{k_{l\pm 1,D}}\right) }}\int_{-\infty}^{+\infty}e^{-\frac{\sqrt{k_{l,D}}+\sqrt{k_{l\pm 1,D}}}{2}\left(r-\widehat{r}_{l,l\pm1,D}\right)^2}dr\\
=&e^{-\frac{\sqrt{k_{l,D}k_{l\pm 1,D}}}{2\left(\sqrt{k_{l,D}}+\sqrt{k_{l\pm 1,D}}\right)}\left(\widetilde{r}_{l,D}-\widetilde{r}_{l\pm 1,D}\right)^2}  \int_{-\infty}^{+\infty}e^{-\frac{\sqrt{k_{l,D}}+\sqrt{k_{l\pm 1,D}}}{2}\left(r-\widehat{r}_{l,l\pm1,D}\right)^2}dr\\
=& e^{-\frac{\sqrt{k_{l,D}k_{l\pm 1,D}}}{2\left(\sqrt{k_{l,D}}+\sqrt{k_{l\pm 1,D}}\right)}\left(\widetilde{r}_{l,D}-\widetilde{r}_{l\pm 1,D}\right)^2}\sqrt{\frac{2\pi}{\sqrt{k_{l,D}}+\sqrt{k_{l\pm 1,D}}}},
\end{split}
\end{equation*}
then
\begin{equation}\label{valueflscalarefl+1}
\begin{split}
\left\langle f_{0,l\pm 1,D}, f_{0,l,D}\right\rangle_{R^+}=&M_{l,D}M_{l\pm 1,D}e^{-\frac{\sqrt{k_{l,D}k_{l\pm 1,D}}}{2\left(\sqrt{k_{l,D}}+\sqrt{k_{l\pm 1,D}}\right)}\left(\widetilde{r}_{l,D}-\widetilde{r}_{l\pm 1,D}\right)^2}\sqrt{\frac{2\pi}{\sqrt{k_{l,D}}+\sqrt{k_{l\pm 1,D}}}}\\
\overset{(\ref{valueMMsqrtpi})\&(\ref{value_exponential})}=&1+O\left(\frac{1}{k^{\frac{3}{2}}}\right).
\end{split}
\end{equation}}
\subsection{The algebraic relations fulfilled by $\overline{L}_{h,j}$ and $\overline{x}^s$}
\subsubsection{Proof of (\ref{commx-ax-b})} \label{proofcommx-ax-b}
The proofs of {section \ref{proofLDaLDb}} can be used here to calculate $\left[\overline{x}^h,\overline{x}^j\right]\psi_{\bm{l},D}$ when $h<j$ and $l<\Lambda$; this is possible because {definition \ref{defiL}} implies that the action of $L_{h,D+1}$ in $\mathbb{R}^{D+1}$ is very similar to the one of $t^h$ (and also of $\overline{x}^h$) in $\mathbb{R}^D$. In fact, the only difference is the replacement of $-\frac{1}{i}d_{l_D,l+1,D+1}$ and $\frac{1}{i}d_{l_D,l,D+1}$ with $c_{l+1,D}$ and $c_{l,D}$, respectively.

Then it must be $l'_p=l_p$ $\forall p\geq j-1$ and
\begin{equation}
\left[\overline{x}^h;\overline{x}^j\right]\psi_{
\bm{l},D}=\sum_{
\substack{
|{l'}_h-l_h|=1\\
h=j-2,\cdots,h-1
}}\widehat{Q}_{D,h,j}\left(\bm{l},\bm{l'}_h\right)\psi_{
\bm{l'}_h,D
}.
\end{equation}
If ${l'}_{j-2}=l_{j-2}-1$, then
\begin{equation}\label{diff}
\begin{split}
\widehat{Q}_{D,h,j}\left(\bm{l},\bm{l'}_h\right)=\left\{\left(c_{l+1,D}
\right)^2\left\{\right.\right.&\left[A(l,l_{d-1},D-1)\right]^2 T_3^{D-j}\left(l_{d-1},l_{j-1},l_{j-2},j-1\right)
\\
&\left.+\left[C(l,l_{d-1},D-1)\right]^2 T_4^{D-j}\left(l_{d-1},l_{j-1},l_{j-2},j-1\right)\right\}\\
+\left(c_{l,D}
\right)^2\left\{\right.& \left[B(l,l_{d-1},D-1)\right]^2 T_3^{D-j}\left(l_{d-1},l_{j-1},l_{j-2},j-1\right)\\
&\left.\left.+\left[D(l,l_{d-1},D-1)\right]^2 T_4^{D-j}\left(l_{d-1},l_{j-1},l_{j-2},j-1\right)\right\}\right\}\\
&\cdot R_{h,j-1}\left(\bm{l}\vert_{j,h},\widetilde{\bm{l'}}\vert_{j,h}\right).
\end{split}
\end{equation}
The equations (\ref{Tn}) and (\ref{diffc}) imply
\begin{equation}
\begin{split}
\widehat{Q}_{D,h,j}\left(\bm{l},\bm{l'}_h\right)&=\left[-\left(c_{l+1,D}
\right)^2+ \left(c_{l,D}
\right)^2\right]\frac{d_{l_{j-1},l_{j-2},j}}{2l+D-2} R_{h,j-1}\left(\bm{l}\vert_{j,h},\widetilde{\bm{l'}}\vert_{j,h}\right)\\
&=-\frac{d_{l_{j-1},l_{j-2},j}}{k(\Lambda)}R_{h,j-1}\left(\bm{l}\vert_{j,h},\widetilde{\bm{l'}}\vert_{j,h}\right);
\end{split}
\end{equation}
similarly, if ${l'}_{j-2}=l_{j-2}+1$, then
\begin{equation}
\widehat{Q}_{D,h,j}\left(\bm{l},\bm{l'}_h\right)
=\frac{d_{l_{j-1},l_{j-2}+1,j}}{k(\Lambda)}R_{h,j-1}\left(\bm{l}\vert_{j,h},\widehat{\bm{l'}}\vert_{j,h}\right);
\end{equation}
and then, when $l<\Lambda$,
$$
\left[\overline{x}^h,\overline{x}^j\right]\psi_{\bm{l},D}=-i\frac{\overline{L}_{h,j}}{k(\Lambda)}\psi_{\bm{l},D}.
$$
On the other hand, if $l=\Lambda$, the only difference is that $c_{\Lambda+1,D}=0$ and then (of course, the calculations of {section \ref{proofLDaLDb}} are used also here)
$$
\left[\overline{x}^h,\overline{x}^j\right]\psi_{\Lambda,l_{d-1},\cdots,l_1,D}=i\frac{\left(c_{\Lambda,D}
\right)^2}{2\Lambda+D-2}\overline{L}_{h,j}\psi_{\Lambda,l_{d-1},\cdots,l_1,D}.
$$
According to this,
$$
\left[\overline{x}^h,\overline{x}^j\right]=
i\left[-\frac{I}{k(\Lambda)}+\left(\frac{1}{k(\Lambda)}+\frac{\left(c_{\Lambda,D}
\right)^2}{2\Lambda+D-2} \right)\widehat{P}_{\Lambda,D}
\right]
\overline{L}_{h,j}.
$$
\subsubsection{Proof of (\ref{R^2value})} \label{proofR^2value}
The proofs of {section \ref{proofLij1} can be used here to calculate the value of $\mathcal{R}^2\psi_{\bm{l},D}$; in fact it is easy to see that, when $l<\Lambda$,
\begin{equation*}
\begin{split}
\mathcal{R}^2 \psi_{\bm{l},D}=&\left[\left(c_{l+1,D}\right)^2Z_{1,d} \left(\bm{l}\right)+\left(c_{l,D}\right)^2Z_{2,d} \left(\bm{l}\right)\right]\psi_{\bm{l},D}\\
\overset{(\ref{P1+P2})}=&\left\{1+\frac{[b(l,D)+b(l+1,D)]Z_{1,d} \left(\bm{l}\right)} {2k(\Lambda)}+\frac{[b(l,D)+b(l-1,D)]Z_{2,d} \left(\bm{l}\right)}{2k(\Lambda)}+O\left(\frac{1}{k^2} \right)\right\} \psi_{\bm{l},D}\\
=& \left\{1+\frac{b(l,D)+[b(l+1,D)]\frac{l+D-2}{2l+D-2}+[b(l-1,D)]\frac{l}{2l+D-2}}{2k(\Lambda)}+O\left(\frac{1}{k^2} \right)\right\}\psi_{\bm{l},D}.
\end{split}
\end{equation*}
On the other hand, if $l=\Lambda$, $c_{\Lambda+1,D}=0$; so
$$
\mathcal{R}^2 \psi_{\Lambda,l_{d-1},\cdots,l_1,D}=\left[\left(c_{\Lambda,D}\right)^2\frac{\Lambda}{2\Lambda+D-2}\right]\psi_{\Lambda,l_{d-1},\cdots,l_1,D}.
$$
And then, [up to $O\left(\frac{1}{k^2}\right)$]
\begin{equation*}
\begin{split}
\mathcal{R}^2 \psi_{\bm{l},D}=&\left\{1+\frac{b(l,D)+[b(l+1,D)]\frac{l+D-2}{2l+D-2}+[b(l-1,D)]\frac{l}{2l+D-2}}{2k(\Lambda)}\right.\\
&\left.-\left[\left(1+\frac{b(\Lambda,D)+b(\Lambda+1,D)}{2k(\Lambda)} \right)\frac{\Lambda+D-2}{2\Lambda+D-2} \right]\widehat{P}_{\Lambda,D} \right\}\psi_{\bm{l},D}.
\end{split}
\end{equation*}
\subsection{The product of two $D$-dimensional spherical harmonics}\label{subsprodYY}
First of all, it is important to summarize that in section \ref{GALF} it has been shown that (in the following equations there is not any multiplicative constant, depending on the indices of $P$, because they are not relevant also in this case, except when that constant is $0$)
$$
P_l^{-m}\left(\cos{\theta}\right)=\left(\sin{\theta}\right)^{m}\widetilde{P}_l^{-m}\left(\cos{\theta}\right),
$$
where $0\leq m\leq l$, $\widetilde{P}_l^{-m}\left(\cos{\theta}\right)$ is a polynomial of degree $l-m$ in $\cos{\theta}$ which does not contain any term of degree $l-m-(2n+1)$, with $n\in\mathbb{N}_0$; so, coming back to ${}_jP_{L}^{l}(\theta)$,
\be\label{explicitP_l^m}
{}_h\overline{P}_{l}^{m}(\theta)=\left(\sin{\theta}\right)^m \widetilde{P}_{l+\frac{h-2}{2}}^{-\left(m+\frac{h-2}{2}\right)}\left(\cos{\theta}\right) = \left(\sin{\theta}\right)^{m}\left\{\left[\cos{\theta}\right]^{l-m}+ \left[\cos{\theta}\right]^{l-m-2}+\cdots \right\}.
\ee
It is now possible to calculate the product of two spherical harmonics $Y_{\bm{l'}}$ and $Y_{\bm{l}}$; first of all, $e^{i{l'}_1\theta_1} e^{i{l}_1\theta_1}= e^{i(l_1+{l'}_1)\theta_1}$, then
\begin{equation}\label{prodPP}
\begin{split}
&{}_2 \overline{P}_{{l'}_2}^{{l'}_1}\left(\theta_2\right) {}_2 \overline{P}_{{l}_2}^{{l}_1}\left(\theta_2\right)e^{i(l_1+{l'}_1)\theta_1}\\
& \overset{(\ref{explicitP_l^m})}=\left(\sin{\theta_2}\right)^{{l'}_1}\left[
\left(\cos{\theta}\right)^{{l'}_2-{l'}_1}+ \left(\cos{\theta}\right)^{{l'}_2-{l'}_1-2}+ \left(\cos{\theta}\right)^{{l'}_2-{l'}_1-4}+\cdots
\right] {}_2 \overline{P}_{{l}_2}^{{l}_1}\left(\theta_2\right)e^{i(l_1+{l'}_1)\theta_1}\\
&\overset{(\ref{Legendreprop})}=\left[ {}_2\overline{P}_{l_2+{l'}_2}^{l_1+{l'}_1}\left(\theta_2\right)+ {}_2\overline{P}_{l_2+{l'}_2-2}^{l_1+{l'}_1}\left(\theta_2\right)+ {}_2\overline{P}_{l_2+{l'}_2-4}^{l_1+{l'}_1}\left(\theta_2\right)+\cdots
\right] e^{i(l_1+{l'}_1)\theta_1}.
\end{split}
\end{equation}
Similarly,
\be
\begin{split}
&\left[
{}_3 \overline{P}_{{l'}_3}^{{l'}_2}\left(\theta_3\right) {}_3 \overline{P}_{{l}_3}^{{l}_2}\left(\theta_3\right)
\right]{}_2\overline{P}_{l_2+{l'}_2}^{l_1+{l'}_1} \left(\theta_2\right)\\
&\overset{(\ref{explicitP_l^m})}=\left(\sin{\theta_3}\right)^{{l'}_2}\left[
\left(\cos{\theta_3}\right)^{{l'}_3-{l'}_2}+\left(\cos{\theta_3}\right)^{{l'}_3-{l'}_2-2}+\left(\cos{\theta_3}\right)^{{l'}_3-{l'}_2-4}+\cdots
\right]{}_2\overline{P}_{l_2+{l'}_2}^{l_1+{l'}_1}\left(\theta_2\right)\\
&\overset{(\ref{Legendreprop})}=
\left[
{}_3 \overline{P}_{{l}_3+{l'}_3}^{{l}_2+ {l'}_2}\left(\theta_3\right)+ {}_3 \overline{P}_{{l}_3+{l'}_3-2}^{{l}_2+ {l'}_2}\left(\theta_3\right)+ {}_3 \overline{P}_{{l}_3+{l'}_3-2}^{{l}_2+ {l'}_2}\left(\theta_3\right)+\cdots
\right] {}_2\overline{P}_{l_2+{l'}_2}^{l_1+{l'}_1}\left(\theta_2\right).
\end{split}
\ee
Furthermore, in order to calculate 
$$
{}_3 \overline{P}_{{l'}_3}^{{l'}_2}\left(\theta_3\right) {}_3 \overline{P}_{{l}_3}^{{l}_2}\left(\theta_3\right)
{}_2\overline{P}_{l_2+{l'}_2-2}^{l_1+{l'}_1} \left(\theta_2\right),
$$
the formula (\ref{Legendreprop})$_2$ must be used ${l'}_2-1$ times and then $1$ time the formula (\ref{Legendreprop})$_1$ in correspondence of $\sin{\theta_3}\cdot$, while the formula (\ref{Legendreprop})$_3$ must be used in correspondence of $\cos{\theta_3}\cdot$; then
\be
\begin{split}
&{}_3 \overline{P}_{{l'}_3}^{{l'}_2}\left(\theta_3\right) {}_3 \overline{P}_{{l}_3}^{{l}_2}\left(\theta_3\right)
{}_2\overline{P}_{l_2+{l'}_2-2}^{l_1+{l'}_1} \left(\theta_2\right)\\
&\overset{(\ref{explicitP_l^m})}=\left(\sin{\theta_3}\right)^{{l'}_2}\left[
\left(\cos{\theta_3}\right)^{{l'}_3-{l'}_2}+\left(\cos{\theta_3}\right)^{{l'}_3-{l'}_2-2}+ \left(\cos{\theta_3}\right)^{{l'}_3-{l'}_2-4}+\cdots
\right] {}_3 \overline{P}_{{l}_3}^{{l}_2}\left(\theta_3\right) {}_2\overline{P}_{l_2+{l'}_2-2}^{l_1+{l'}_1} \left(\theta_2\right)\\
&\overset{(\ref{Legendreprop})}=\left[
{}_3 \overline{P}_{{l}_3+{l'}_3}^{{l}_2+{l'}_2-2}\left(\theta_3\right)+ {}_3 \overline{P}_{{l}_3+{l'}_3-2}^{{l}_2+{l'}_2-2}\left(\theta_3\right)+ {}_3 \overline{P}_{{l}_3+{l'}_3-4}^{{l}_2+{l'}_2-2}\left(\theta_3\right)
+\cdots
\right] {}_2\overline{P}_{l_2+{l'}_2-2}^{l_1+{l'}_1} \left(\theta_2\right),
\end{split}
\ee
and so on with the other angles and factors. 

According to this,
\be\label{YY}
Y_{\bm{l'}} Y_{\bm{l}}=\sum_{l''=0}^{l+l'}\sum_{l''_{d-1}=0}^{{l}_{d-1}+{l'}_{d-1}}\cdots \sum_{l''_{2}=0}^{{l}_{2}+{l'}_{2}}\gamma_{\bm{l''}}Y_{\bm{l''}},\quad\mbox{where}\quad\bm{l''}:=\left(l'',l''_{d-1},\cdots,l''_2,{l}_1+{l'}_1\right);
\ee
so, this last equation describes the action of the generic multiplication operator $Y_{\bm{l'}} \cdot $ on the Hilbert space of $D$-dimensional spherical harmonics.

Furthermore, from section \ref{homog_Y_l} and the fact that the $t^h$ commute it follows
\begin{equation*}
\begin{split}
Y_{\bm{l}}&=\sum_{
\substack{
\bm{\alpha}\in\left(\mathbb{N}_0\right)^D\\
\|\bm{\alpha}\|_1=l
}}c_ {\bm{l}} ^{\bm{\alpha}}\left(t^1\right)^{\alpha_1} \left(t^2\right)^{\alpha_2}\cdots \left(t^D\right)^{\alpha_D}\\
&= \sum_{
\substack{
\bm{\alpha}\in\left(\mathbb{N}_0\right)^D\\
\|\bm{\alpha}\|_1=l
}}
c_ {\bm{l}} ^{\bm{\alpha}}
\frac{(\alpha_1)! (\alpha_2)!\cdots (\alpha_D)!}{l!}
\sum_{h}
N\left(h,\bm{\alpha},t_1,t_2,\cdots,t_D\right),
\end{split}
\end{equation*}
where $c_ {\bm{l}} ^{\bm{\alpha}}$ is a suitable constant and $N\left(h,\bm{\alpha},t_1,t_2,\cdots,t_D\right)$ is the ordered monomial obtained applying $\pi_h$ (the permutation with ripetition of $l$ objects with $\alpha_1$ identical objects of type $1$, $\alpha_2$ identical objects of type $2$,..., $\alpha_D$ identical objects of type $D$) to the monomial $\left(t^1\right)^{\alpha_1} \left(t^2\right)^{\alpha_2}\cdots \left(t^D\right)^{\alpha_D}$.

Inspired by this, define the fuzzy approximations $\widehat{Y}_{\bm{l}}$ of the spherical harmonics $Y_{\bm{l}}$ as
\begin{equation}\label{fuzzyY}
\widehat{Y}_{\bm{l}}:= \sum_{
\substack{
\bm{\alpha}\in\left(\mathbb{N}_0\right)^D\\
\|\bm{\alpha}\|_1=l
}}
c_ {\bm{l}} ^{\bm{\alpha}}
\frac{(\alpha_1)! (\alpha_2)!\cdots (\alpha_D)!}{l!}
\sum_{h}
N\left(h,\bm{\alpha},\overline{x}^1,\overline{x}^2,\cdots,\overline{x}^D\right).
\end{equation}

It is also important to underline that
\begin{remark}\label{remark8}
From $|A|,|B|,|C|,|D|,|F|,|G|\leq 1$ it follows
\begin{equation}\label{abs_scY}
\begin{split}
\left\vert\sin{\theta} {}_j\overline{P}_L^l\left(\theta\right)\right\vert&\leq \left\vert {}_j\overline{P}_{L+1}^{l+1}\left(\theta\right)\right\vert+\left\vert {}_j\overline{P}_{L-1}^{l+1}\left(\theta\right)\right\vert,\\
\left\vert\sin{\theta} {}_j\overline{P}_L^l\left(\theta\right)\right\vert &\leq \left\vert {}_j\overline{P}_{L+1}^{l-1}\left(\theta\right)\right\vert+\left\vert {}_j\overline{P}_{L-1}^{l-1}\left(\theta\right)\right\vert,\\
\left\vert\cos{\theta} {}_j\overline{P}_L^l\left(\theta\right)\right\vert &\leq \left\vert {}_j\overline{P}_{L+1}^{l}\left(\theta\right)\right\vert+\left\vert {}_j\overline{P}_{L-1}^{l}\left(\theta\right)\right\vert.
\end{split}
\end{equation}
This, the recursive procedures of {{section \ref{GALF}}} and the calculations of this section, imply that the product ${}_j\overline{P}_{L'}^{l'} \left(\theta\right){}_j\overline{P}_L^l \left(\theta\right) $ when $\Lambda\geq L'\geq l'\geq0$ and $\Lambda\geq L\geq l\geq0$ is the sum of (at most) $\Lambda 2^{\Lambda}$ terms ${}_j\overline{P}_{L''}^{l''} \left(\theta\right)$, and some of them may have the same indices. Furthermore, (\ref{abs_scY}) implies that the product of ${}_j\overline{P}_L^l$ by $\sin{\theta}$ and $\cos{\theta}$ returns coefficients that are bounded by $1$, while in (\ref{ArmSfe}) every ${}_j \overline{P}_L^l(\theta)$ contains a normalization constant which is less or equal than $(2\Lambda)!$.
\end{remark}
\subsection{Some proofs about convergence}\label{convergproof}
{Let $\varphi\in\mathcal{H}_{\Lambda,D}$, with $\|\varphi\|_2=1$, and
$$
\varphi=\sum_{\substack{
0\leq l\leq \Lambda\\
l_{h-1}\leq l_h\mbox{ for }h=d,\cdots, 3\\
|l_1|\leq l_2
}}
\varphi_{\bm{l}}\psi_{\bm{l},D}
$$
be the decomposition of $\varphi$ in that orthonormal basis of $\mathcal{H}_{\Lambda,D}$; of course, $\|\varphi\|_2=1$ implies $\left\vert \varphi_{\bm{l}}\right\vert\leq 1$. 

According to these statements,
\begin{equation*}
\begin{split}
\left\|
\left(
\overline{x}^h-t^h
\right)
\varphi
\right\|_2\leq &
\sum_{\substack{
0\leq l\leq \Lambda\\
l_{h-1}\leq l_h\mbox{ for }h=d,\cdots, 3\\
|l_1|\leq l_2
}}
|\varphi_{\bm{l}}|\left\|\left(\overline{x}^h-t^h\right)\psi_{\bm{l},D}\right\|_2\\
\overset{|\varphi_{\bm{l}}|\leq 1}\leq & \sum_{\substack{
0\leq l\leq \Lambda\\
l_{h-1}\leq l_h\mbox{ for }h=d,\cdots, 3\\
|l_1|\leq l_2
}}
\left\|\left(\overline{x}^h-t^h\right)\psi_{\bm{l},D}\right\|_2\\
\overset{*}\leq & [\dim{\mathcal{H}_{\Lambda,D}}]^2\left(
\frac{b(l+1,D)+2b(l,D)+b(l-1,D)}{4k}
\right)\\
\overset{\symbol{35}}\leq & [\dim{\mathcal{H}_{\Lambda,D}}]^2\frac{b(\Lambda,D)}{k};
\end{split}
\end{equation*}
where the $*$ inequality follows from the fact that the sum is of $\dim{\mathcal{H}_{\Lambda,D}}$ elements, that both $\overline{x}^h\psi_{\bm{l},D}$ and $t^h\psi_{\bm{l},D}$ can be written as the linear combination of (at most) $\dim{\mathcal{H}_{\Lambda,D}}$ elements, that $|A|,|B|,|C|,|D|,|F|,|G|\leq 1$ , that $\left\langle \psi_{\bm{l'},D},\overline{x}^h \psi_{\bm{l},D}\right\rangle_{\mathbb{R}^D}$ and $\left\langle \psi_{\bm{l'},D},t^h \psi_{\bm{l},D}\right\rangle_{\mathbb{R}^D}$ do not vanish if $l'=l\pm 1$, and the values of the corresponding `radial' scalal product are
\begin{equation*}
\begin{split}
\left\langle f_{0,l\pm 1,D},r f_{0,l,D}\right\rangle_{\mathbb{R}^+}\overset{(\ref{valuescalprod})}=&1+\frac{1}{2}\frac{\left[b(l,D)+b(l\pm1,D)\right]}{2k}+O\left(\frac{1}{k^{\frac{3}{2}}}\right)\\
\mbox{and}\quad \left\langle f_{0,l\pm,D},f_{0,l,D}\right\rangle_{\mathbb{R}^+}\overset{(\ref{valueflscalarefl+1})}=&1+O\left(\frac{1}{k^{\frac{3}{2}}}\right);
\end{split}
\end{equation*}
while the $\symbol{35}$ inequality follows from (\ref{definizioni1})$_1$.

So, if
$$
k\left(\Lambda\right)\geq \Lambda \left[\dim{\mathcal{H}_{\Lambda,D}}\right]^2b(\Lambda,D), \mbox{ then } \left\|
\left(
\overline{x}^h-t^h
\right)
\varphi
\right\|_2\overset{\Lambda\rightarrow+\infty}\longrightarrow 0.
$$}

In {section \ref{subsprodYY}} there is the product between two generic $D$-dimensional spherical harmonics and the in section \ref{homog_Y_l} it is shown that every $D$-dimensional spherical harmonic is a homogeneous polynomial in the $t^h$ variables, this suggested the definition (\ref{fuzzyY}).

Those $\widehat{Y}_{\bm{l}}$ are the fuzzy spherical harmonics, they are elements of $B[{\cal L}^2(S^d)]$ and, in particular,
\begin{remark}\label{remark7}
The action of $\widehat{Y}_{\bm{l}}$ on $Y_{\bm{l'}}$ can be obtained through the following replacements to $Y_{\bm{l}}\cdot Y_{\bm{l'}}$:
\begin{itemize}
\item replace every $A\left(l, l_{d-1},D-1\right)$ with $c_{l+1,D}A\left(l, l_{d-1},D-1\right)$;
\item replace every $B\left(l, l_{d-1},D-1\right)$ with $c_{l,D}B\left(l, l_{d-1},D-1\right)$;
\item replace every $C\left(l, l_{d-1},D-1\right)$ with $c_{l+1,D}C\left(l, l_{d-1},D-1\right)$;
\item replace every $D\left(l, l_{d-1},D-1\right)$ with $c_{l,D}D\left(l, l_{d-1},D-1\right)$;
\item replace every $F\left(l, l_{d-1},D-1\right)$ with $c_{l+1,D}F\left(l, l_{d-1},D-1\right)$;
\item replace every $G\left(l, l_{d-1},D-1\right)$ with $c_{l,D}G\left(l, l_{d-1},D-1\right)$.
\end{itemize}
\end{remark}
On the other hand, the fuzzy analogs of the vector space $B(S^d)$ are defined as
\be
{\cal C}_{\Lambda,D}:=span_{\mathbb{C}}\left\{\widehat{Y}_{\bm{l}}:2\Lambda\geq l\equiv l_d\geq \cdots\geq l_2\geq |l_1|,l_i\in\mathbb{Z}\forall i \right\}
\subset\A_{\Lambda,D}\subset B[{\cal L}^2(S^d)],
\label{def_CLambda3D}
\ee
and here the highest $l$ is $2\Lambda$ because $\widehat{Y}_{2\Lambda,2\Lambda,\cdots,2\Lambda}$ is the 
`highest' multiplying operator acting nontrivially on $\mathcal{H}_{\Lambda,D}$ (it does 
not annihilate $\psi_{\Lambda,\Lambda,\cdots,-\Lambda,D}$).

So
\be
{\cal C}_{\Lambda,D}=\bigoplus\limits_{l=0}^{2\Lambda} V_{l,D}\hspace{0.1cm}\label{deco2}
\ee
is the decomposition
of ${\cal C}_{\Lambda,D}$ into irreducible components under $O(D)$; furthermore, $V_{l,D}$ is trace-free
for all $l>0$, i.e. its projection on the single component $V_{0,D}$ is zero and it is easy to see that (\ref{deco2}) becomes the
decomposition of $B(S^d),C(S^d)$ in the limit $\Lambda\to\infty$. 

However, the fuzzy analog of $f\in B(S^d)$ is
\be
\hat f_\Lambda:=\sum_{l=0}^{2\Lambda}
\sum_{\substack{
l_{d-1}\leq l\\
l_{h-1}\leq l_h \mbox{ for }h=d-1,\cdots,3\\
|l_1|\leq l_2
}}
f_{\bm{l}}\widehat{Y}_{\bm{l}}
\in\mathcal{A}_{\Lambda,D}\subset B[{\cal L}^2(S^2)].
\ee
\subsubsection{Proof of Proposition \ref{propoD}}
Let $\phi,f\in B\left(S^d\right)$, then
\be 
\begin{split}
(f -\hat f_\Lambda)\phi=&\sum_{l=0}^{\Lambda}\sum_{\substack{
l_{d-1}\leq l\\
l_{h-1}\leq l_h \mbox{ for }h=d-1,\cdots,3\\
|l_1|\leq l_2
}}
Y_{\bm{l}} \chi_{\bm{l}}+\sum_{l>\Lambda}
\sum_{\substack{
l_{d-1}\leq l\\
l_{h-1}\leq l_h \mbox{ for }h=d-1,\cdots,3\\
|l_1|\leq l_2
}}
Y_{\bm{l}} (f\phi)_{\bm{l}}
, 
\end{split}
\ee
where  
$$
\chi_{\bm{l}}:=(f\phi)_{\bm{l}}-(\hat f_\Lambda \phi)_{\bm{l}}\quad ,\quad (f\phi)_{\bm{l}}= \langle Y_{\bm{l}},f\phi\rangle\quad \mbox{and}\quad\left(\hat f_\Lambda \phi\right)_{\bm{l}}=\langle Y_{\bm{l}},\hat{f}_\Lambda\phi\rangle;
$$
in particular
\be 
\begin{split}
\chi_{\bm{l}}&=\left\langle Y_{\bm{l}},\left( f-\hat{f}_{\Lambda}\right)\phi\right\rangle=\left\langle Y_{\bm{l}},\sum_{l'=0}^{2\Lambda}
\sum_{\substack{
l'_{d-1}\leq l'\\
l'_{h-1}\leq l'_h \mbox{ for }h=d-1,\cdots,3\\
|l'_1|\leq l'_2
}}
f_{\bm{l'}}\left(Y_{\bm{l'}}-\widehat{Y}_{\bm{l'}}\right)
\phi\right\rangle \\
&=
\sum_{l'=0}^{2\Lambda}
\sum_{\substack{
l'_{d-1}\leq l'\\
l'_{h-1}\leq l'_h \mbox{ for }h=d-1,\cdots,3\\
|l'_1|\leq l'_2
}}
\sum_{l''=0}^{+\infty}
\sum_{\substack{
l''_{d-1}\leq l''\\
l''_{h-1}\leq l''_h \mbox{ for }h=d-1,\cdots,3\\
|l''_1|\leq l''_2
}}
f_{\bm{l'}} \phi_{\bm{l''}} \:\: \left\langle Y_{\bm{l}},\left(Y_{\bm{l'}}-\widehat{Y}_{\bm{l'}}\right)Y_{\bm{l''}}\right\rangle.
\end{split}\label{chi}
\ee

On the other hand,
\be 
\begin{split}
 \Vert(f -\hat f_\Lambda)\phi\Vert^2=&\sum_{l=0}^{\Lambda}
\sum_{\substack{
l_{d-1}\leq l\\
l_{h-1}\leq l_h \mbox{ for }h=d-1,\cdots,3\\
|l_1|\leq l_2
}}
 |\chi_{\bm{l}}|^2+\sum_{l>\Lambda}
\sum_{\substack{
l_{d-1}\leq l\\
l_{h-1}\leq l_h \mbox{ for }h=d-1,\cdots,3\\
|l_1|\leq l_2
}}
|(f\phi)_{\bm{l}}|^2,
\end{split}      \label{normffh}
\ee
the second sum goes to zero as $\Lambda\to\infty$, it remains to show that the first sum
does as well.

{The sum over $l'$ in (\ref{chi}) consists of at most $\dim{\mathcal{H}_{2\Lambda,D}}$ elements, as for the one over $l''$ (because $0\leq l\leq \Lambda$),; the equality (\ref{YY}) can be applied in this case, and it implies that both $Y_{\bm{l'}}Y_{\bm{l''}}$ and $\widehat{Y}_{\bm{l'}}Y_{\bm{l''}}$ can be written as a linear combination of, at most, $\dim{\mathcal{H}_{2\Lambda,D}}$ basis elements, then the sum in (\ref{chi}) is made up by at most $[\dim{\mathcal{H}_{2\Lambda,D}}]^3$ non-vanishing addends, while the one over $l$ in (\ref{normffh}) is of at most $\dim{\mathcal{H}_{\Lambda,D}}$ elements.

In addition, the fact that in (\ref{ArmSfe}) every ${}_j \overline{P}_L^l(\theta)$ contains a normalization constant which is less or equal than $(2\Lambda)!$, that the highest coefficient multiplying a power of $\cos{\theta}$ in $P_{L}^{l}\left(\cos{\theta}\right)$ is less or equal than
$$
2^{\Lambda}\left[(2\Lambda+1)!!\right]^2,
$$
that $|A|,|B|,|C|,|D|,|F|,|G|\leq 1$, that $\left\langle \psi_{\bm{l'},D},\overline{x}^h \psi_{\bm{l},D}\right\rangle_{\mathbb{R}^D}$ and $\left\langle \psi_{\bm{l'},D},t^h \psi_{\bm{l},D}\right\rangle_{\mathbb{R}^D}$ do not vanish if $l'=l\pm 1$, and the values of the corresponding `radial' scalal product are
\begin{equation*}
\begin{split}
\left\langle f_{0,l\pm 1,D},r f_{0,l,D}\right\rangle_{\mathbb{R}^+}\overset{(\ref{valuescalprod})}=&1+\frac{1}{2}\frac{\left[b(l,D)+b(l\pm1,D)\right]}{2k}+O\left(\frac{1}{k^{\frac{3}{2}}}\right)\\
\mbox{and}\quad \left\langle f_{0,l\pm,D},f_{0,l,D}\right\rangle_{\mathbb{R}^+}\overset{(\ref{valueflscalarefl+1})}=&1+O\left(\frac{1}{k^{\frac{3}{2}}}\right)
\end{split}
\end{equation*}
imply
$$
\sum_{l=0}^{\Lambda}
\sum_{\substack{
l_{d-1}\leq l\\
l_{h-1}\leq l_h \mbox{ for }h=d-1,\cdots,3\\
|l_1|\leq l_2
}}
 |\chi_{\bm{l}}|^2\leq[\dim{\mathcal{H}_{\Lambda,D}}] \left\{[\dim{\mathcal{H}_{2\Lambda,D}}]^3 \left[(2\Lambda)!\right]^D 2^{\Lambda D}\left[(2\Lambda+1)!!\right]^{2D}\frac{\Lambda b(\Lambda,D)}{k(\Lambda)} \right\}^2
$$}
So, if
$$
k\left(\Lambda\right)\geq \Lambda^2 [\dim{\mathcal{H}_{2\Lambda,D}}]^3 \left[(2\Lambda)!\right]^D 2^{\Lambda D}\left[(2\Lambda+1)!!\right]^{2D}b(\Lambda,D)\sqrt{\dim{\mathcal{H}_{\Lambda,D}}},
$$
then
\bea
\Vert(f -\hat f_\Lambda)\phi\Vert^2\le \left\| f\right\|^2\left\| \phi\right\|^2  
\, \frac{1}{\Lambda^2}+\sum_{l>\Lambda}
\sum_{\substack{
l_{d-1}\leq l\\
l_{h-1}\leq l_h \mbox{ for }h=d-1,\cdots,3\\
|l_1|\leq l_2
}}
|(f\phi)_{\bm{l}}|^2
\:
\stackrel{\Lambda\to\infty}{\longrightarrow}0,       \label{flimitD}
\eea
i.e. $\widehat{f}_\Lambda\to f\cdot$ strongly for all $f\in B\big(S^d\big)$, as claimed.

The replacement $f\mapsto fg$, implies that $\widehat{(fg)}_\Lambda\to (fg)\cdot$ strongly for all $f,g\in B\big(S^d\big)$, while from (\ref{flimitD}) it follows
\be\label{finale1}
 \Vert(f -\hat f_\Lambda)\phi\Vert^2\leq \left\| f\right\|^2\left\| \phi\right\|^2  
\, \frac{1}{\Lambda^2}+\left\| f\phi\right\|^2\le \left(
\frac{\left\| f\right\|^2}{\Lambda^2}
+\left\| f\right\|_\infty^2\right)\left\| \phi\right\|^2,
\ee
with
\be\label{finale2}
\sqrt{\frac{\left\| f\right\|^2}{\Lambda^2}
+\left\| f\right\|_\infty^2}\leq \sqrt{\left\| f\right\|^2
+\left\| f\right\|_\infty^2}\leq \left\| f\right\|
+\left\| f\right\|_\infty,
\ee
and then
\be
\Vert \hat f_\Lambda\phi\Vert 
%=\Vert (\hat f_\Lambda\!-\! f)\phi+f\phi\Vert 
\le \Vert (\hat f_\Lambda\!-\! f)\phi\Vert +\Vert f\phi\Vert 
\le \Vert (\hat f_\Lambda\!-\! f)\phi\Vert +\Vert f\Vert_\infty \Vert\phi\Vert\overset{(\ref{finale1})\&(\ref{finale2})} \le 
\left(\left\| f\right\|+2 \Vert f\Vert_\infty\right) \Vert\phi\Vert, 
%        \label{normffh}
\ee
i.e. the operator norms $\Vert \hat f_\Lambda\Vert_{op}$ of the $ \hat f_\Lambda$ are bounded  uniformly in
$\Lambda$: \ $\Vert \hat f_\Lambda\Vert_{op}\le \left\| f\right\|+2\Vert f\Vert_\infty$.
Therefore, as claimed, (\ref{flimitD}) implies
\bea
\Vert(fg -\hat f_\Lambda\hat g_\Lambda)\phi\Vert
&\le & \Vert (f -\hat f_\Lambda)g\phi\Vert+\Vert \hat f_\Lambda(g-\hat g_\Lambda)\phi\Vert \nn[6pt]
&\le & \Vert (f -\hat f_\Lambda)(g\phi)\Vert+\Vert \hat f_\Lambda\Vert_{op} \: \: \Vert(g -\hat g_\Lambda)\phi\Vert
\stackrel{\Lambda\to\infty}{\longrightarrow}0.      \label{flimit'3D}
\eea


\begin{thebibliography}{9}
\bibitem{FiorePisacane}
G.Fiore, F.Pisacane, \emph{Fuzzy circle and new fuzzy sphere through confining potentials and energy cutoffs}, Journal of Geometry and Physics \textbf{132} (2018), 423-451,\\
DOI: 10.1016/j.geomphys.2018.07.001.
\bibitem{FiorePisacanePOS18}
G.Fiore, F.Pisacane, 
{\it New fuzzy spheres through confining potentials and energy cutoffs}, 
Proceedings of Science Volume 318,
%- Corfu Summer Institute 2017 "Schools and Workshops on Elementary Particle Physics and Gravity" (CORFU2017) - Workshop on Testing Fundamental Physics Principles;
PoS(CORFU2017)184.
%DOI: 10.22323/1.318.0184
\bibitem{Madore} J. Madore, \emph{The Fuzzy sphere}, Classical and Quantum Gravity,  Volume 9,  Number 1, 1992.

\bibitem{HopdeWNic} 
J. Hoppe, \emph{Quantum theory of a massless relativistic surface and a two-dimensional bound state problem},  PhD thesis, MIT 1982; B. de Wit, J. Hoppe, H. Nicolai,
%\emph{On the quantum mechanics of supermembranes},
Nucl. Phys. {\bf B305} (1988), 545.

\bibitem{Snyder}
H. S. Snyder, \emph{Quantized Space-Time},
Phys. Rev. {\bf 71} (1947), 38.
\bibitem{Shubin} M.A. Shubin, S.I. Andersson, \emph{Pseudodifferential operators and spectral theory}, Springer, 2001.
\bibitem{Bateman} H. Bateman, \emph{Higher transcendental functions 1}, McGraw-Hill Book Company, 1953.
\bibitem{Dolan:2003kq}
   B.~P.~Dolan and D.~O'Connor,
   \emph{A Fuzzy three sphere and fuzzy tori},
   JHEP {\bf 0310} (2003) 060
   doi:10.1088/1126-6708/2003/10/060
   [hep-th/0306231].
   \bibitem{Ramgoolam}
S. Ramgoolam,
%\emph{On spherical harmonics for fuzzy spheres in diverse dimensions}, 
Nucl. Phys. {\bf B610} (2001), 461-488
%hep-th/0105006
;
%\emph{Higher dimensional geometries related to fuzzy odd-dimensional spheres}, 
  JHEP10(2002)064;
%hep-th/0207111
and references therein.
\bibitem{Castelino} J. Castelino, S. Lee, W. Taylor, \emph{Longitudinal $5$-branes as $4$-spheres in Matrix theory}, Nuclear Physics B, Volume 526, Issues 1–3, 24 August 1998, Pages 334-350.
\bibitem{Medina-Connor}J.~Medina and D.~O'Connor,
   %``Scalar field theory on fuzzy S**4,''
   JHEP {\bf 0311} (2003) 051
   doi:10.1088/1126-6708/2003/11/051
   [hep-th/0212170].
   \bibitem{Dolan:2003th}
   B.~P.~Dolan, D.~O'Connor and P.~Presnajder,
   \emph{Fuzzy complex quadrics and spheres},
   JHEP {\bf 0402} (2004) 055
   doi:10.1088/1126-6708/2004/02/055
   [hep-th/0312190].
   \bibitem{Ste16}
H. Steinacker, \emph{Emergent gravity on covariant quantum spaces in the IKKT model},
J. High Energy Physics 2016: 156.  %arXiv:1606.00769. 
\bibitem{Ste17}
M. Sperling, H. Steinacker,  
%\emph{Covariant 4-dimensional fuzzy spheres, matrix models and higher spin},
J. Phys. A: Math. Theor. {\bf 50} (2017), 375202.
% arXiv:1704.02863
hys. {\bf 180} (1996), 429-438.
%hep-th/9602115
\bibitem{Morimoto} M. Morimoto, \emph{Analytic Functionals on the Sphere}, American Mathematical Society, 1998.
\bibitem{Stegun} M. Abramowitz M., I.A. Stegun, \emph{Handbook of mathematical functions With Formulas, Graphs, and Mathematical Tables}, National Bureau of Standards, 1972.
\bibitem{hockey} C. H. Jones, \emph{Generalized Hockey Stick Identities and $N$-Dimensional Block Walking}, Fibonacci Quarterly {\bf 34}(3), 280-288, 1996.
\end{thebibliography}
\end{document}